\documentclass[prd,aps,nofootinbib,nobibnotes,superscriptaddress,preprintnumbers]{revtex4}

\usepackage{amsmath}
\usepackage{amsfonts}
\usepackage{graphicx}
\usepackage[]{dsfont}
\usepackage{maths}
\usepackage{sistyle}
\usepackage{enumitem}
\usepackage{textcomp}
\usepackage{color}
\usepackage{hyperref}
\usepackage{tensor}
\usepackage{tcolorbox}
\usepackage{varwidth}

\usepackage{float}

\usepackage{physics}

\usepackage{tikz}

\newcommand{\be}{\begin{equation}}
\newcommand{\ee}{\end{equation}}
\newcommand{\ba}{\begin{eqnarray}}
\newcommand{\ea}{\end{eqnarray}}

\newcommand{\nn}{{\nonumber}}

\newcommand{\mn}{{\mu \nu}}
\newcommand{\ab}{{\alpha \beta}}

\newcommand{\cl}{{\cal{L}}}

\newcommand{\nb}{{\omega_c^{-1}}}
\newcommand{\pgw}{{\cal{P}}_{\rm GW}}

\newcommand{\Id}{1}
\newcommand{\re}{\mathrm{Re}}

\def\la{\langle}
\def\ra{\rangle}
\newcommand{\mean}[1]{\la{#1}\ra}

\newcommand{\mat} [4] {\left ( \begin{array}{cc}{#1}&{#2}\\{#3}&{#4} \end{array} \right ) }

\newcommand{\p}{\partial}
\newcommand{\na}{\nabla}

\newcommand{\f}{\frac}
\newcommand{\tl}{\tilde}

\renewcommand{\a}{\alpha} \renewcommand{\b}{\beta}   
\renewcommand{\d}{\delta}  \newcommand{\eps}{\epsilon} 
 \renewcommand{\th}{\theta}   
    \renewcommand{\l}{\lambda}
\let\m=\mu    \let\n=\nu   \let\r=\rho \let\om=\omega
 \newcommand{\s}{\sigma}  \renewcommand{\t}{\tau}    
\let\G=\Gamma \let\D=\Delta  \let\Th=\Theta \let\L=\Lambda 
\let\Si=\Sigma \let\Om=\Omega

\newcommand{\SO}{\mathrm{SO}}

\newcommand{\sscr}{\scriptscriptstyle\rm}

\newcommand{\cL}{{\cal L}}

\newcommand{\eqons}{\,\hat{=}\,}

\begin{document}

\begin{center}

{\bf\Large An Introduction to Gravitational Wave Theory}\\

 \date{\today}

\end{center}

\vspace{0.5cm}
\begin{center}
{\large S.~Speziale$^{(1)}$ and D.A.~Steer$^{(2)}$}
\vspace{0.5cm}
\\
{\it (1) Aix Marseille Univ., Univ.~de Toulon, CNRS, CPT, UMR 7332, 13288 Marseille, France 
\vspace{0.2cm}
\\
(2) Laboratoire de Physique de l'ENS, Universit\'e Paris Cit\'e, Ecole Normale Sup\'erieure, Universit\'e
PSL, Sorbonne Universit\'e, CNRS, 75005 Paris }\\
 \vspace{0.2cm}
 (9 October, 2025)
 \end{center}

\vspace{1cm}

{\small \noindent
Introduction to the theoretical foundations of gravitational waves: from general relativity to detection and binary system waveforms.
Lecture notes prepared for the MaNiTou summer school on gravitational waves. 
Draft chapter for the CNRS contemporary Encyclopaedia Sciences to be published by ISTE.} 

\tableofcontents


\newpage
\section{Overview and characteristic scales}
\label{sec:overview}

\subsection{Aims}
\label{sec:Aims}

Indirect evidence of gravitational waves (GWs) has been available for many years (see e.g.~\cite{Will:2014kxa} for a review).
Direct detection on the other hand had to wait until 2015 \cite{LIGOScientific:2016aoc}, that is nearly 100 years after GWs were predicted to exist in general relativity (GR). These detections by the LVK collaboration, consisting of the network of LIGO \cite{LIGOScientific:2014pky}, Virgo \cite{VIRGO:2014yos} and more recently KAGRA \cite{KAGRA:2020tym} GW interferometers, are ongoing with new GWs signals being observed on a weekly basis \cite{LIGOGracedb}. 
In the future more sensitive detectors on earth, together with ones working in different frequency bands such as the Laser Interferometer Space Antenna (LISA) \cite{LISA:2024hlh} as well as Pulsar Timing Arrays (PTAs), will lead to new observations of the universe, potential new discoveries, and unprecendented tests of general relativity, cosmology and astrophysics.

\vspace{0.2cm}

The aim of these lectures 
The aim of these first two chapters
is to present the basic introductory material required to understand GWs. We will address some of following questions: 
\begin{itemize}
\item What are GWs?  How do they emerge from GR? How does one deal with the symmetries (diffeomorphism invariance) of GR to fix gauges and coordinates, and what do they imply for the stress energy tensor of GWs? 
\item To what GW frequencies $f_{\text{GW}}$ are current and future GW detectors sensitive? 
Why are those detectors designed to be sensitive to particular GW frequency ranges?
\item For a source consisting of two {\it bound} compact binaries objects (such as black holes) of masses $m_1$ and $m_2$ at some distance $R$ from an observer, what is the characteristic frequency, amplitude etc of the GWs emitted?  Up to what distances $R$ can such sources be detected?
\item Using the quadrupole formula (which we derive) what is the waveform of the emitted GWs and how does it depend for example on the ellipticity of the bound orbit?
\item What sources correspond to the GW events detected by LVK?  Are there other possible GW sources? We give an example of compact binary sources on unbound orbits and discuss the GW memory effect. 
\item If we consider sources on cosmological distance scales, how are their amplitude, frequency e.t.c.~affected by the cosmological expansion?
\end{itemize}

\subsection{On wave-like solutions and relativity}

Gravitational waves are a natural expectation from GR, simply because it
is a relativistic theory of gravity. To understand why, let us first take a step back to {\it non-relativistic} Newtonian gravity. 
When the famous apple drops on Newton's head, the mass distribution of the Earth changes, and so does the gravitational field created.  In Newton's time, this variation, however negligible, was assumed to be the effect of some {\it instantaneous} ``action at a distance''.   After the discovery that the {\it speed of light is finite}, and that all effects in our universe appear to follow this causal limitation, it seems natural to expect that also the variations of the gravitational field will {\it not} be felt instantaneously in the whole universe, but will rather be propagated at the speed of light --- or less.\footnote{We will see that Einstein's relativistic theory of gravity, GR, predicts that --- {whatever their wavelength} ---  these variations propagate at exactly the speed of light, and if some future experiment shows that they propagate at a lesser speed, then this would be an explicit violation of GR.}
{\it The propagation of this perturbation of the gravitational field is intuitively what we call a gravitational wave.} 

Conceptually, a gravitational wave is similar to a water wave or an electromagnetic (EM) wave. However, while those propagate a modification in the depth of water or the intensities of the electromagnetic field, a GW propagates a modification of the structure of spacetime itself.
As for producing one, it is natural to expect that the Earth emits GWs when
orbiting the sun, thus carrying away energy and making the orbit decay, just like a charged particle emits EM waves when moving
on an accelerating trajectory.

In practise, however, understanding GWs is very subtle for a number of reasons. First of all, the concept itself of propagation makes reference to a background spacetime, and in GR there is no fixed background structure. Splitting the dynamical spacetime into a reference background and a perturbation on top of it is a delicate process in which potential ambiguities have to be dealt with. In fact many decades passed before a consensus was reached, to the point that, famously, Einstein himself initially doubted the physical existence of GW, for reasons that we will briefly review and clarify below. Secondly, if we think of GWs as waves propagating in a medium, this medium is extraordinarily rigid: the waves go as fast as possible and have very {\it tiny} amplitudes. To give an idea, the power emitted by the Earth-Sun system in the form of GW is around 200W! This rigidity has to do with the weakness of the gravitational coupling constant. One may think that gravity is strong when for e.g.~trying to beat a high jump record, or when skydiving, but this strength is ridiculously small compared to the much much stronger electro-magnetic force that dominates our daily life. These two points --- background independence and weakness of the signal --- are typical issues that one has to face when studying GWs.

Indeed only very massive and energetic objects can produce GWs of amplitudes that are actually detectable.  Amongst the most massive and compact astrophysical objects known are black holes (BH), neutron stars (NS) and white dwarfs (WD). 
The GW sources detected to date by the LVK collaboration are all `compact binary 
systems' made of a bound pair of BH and/or NS  on closed orbits. As a result of the energy lost through GW emission, the two bodies making up the bound system approach closer to each other, inspiralling inwards, and eventually merging into one final object.In fact the GW signals detected by the LVK collaboration correspond to the last moments in the life of these systems including their merger --- they are known as `{\it compact binary coalescences}' (CBC's).  For comparison with the earth-sun system mentioned above, the energy emitted in GWs by the very first detected GW event GW150914 \cite{LIGOScientific:2016aoc}, which  was due to the coalescence of two BHs of masses $m_1\sim 36 M_\odot$ and $m_1\sim 29 M_\odot$, was almost $10^{48}$Joules in 0.2 seconds.  

The direct detection of GWs can be used to test many aspects of gravity, for instance in the strong field regime, see e.g.~\cite{Yunes:2013dva}, as well as to probe cosmology, as will be discussed later. Indeed, the gravitational interaction is so weak that the universe is almost completely transparent to a gravitational wave. As a consequence, one can potentially collect pristine information about any cosmological era through GWs, and in particular through the detection and characterisation of a stochastic gravitational wave background. Sources relevant to cosmology include primordial GWs produced during inflation but there are also potential new sources to be discovered, such as 
primordial black holes, cosmic strings, and other exotic objects, see e.g.~\cite{Caprini:2018mtu,Caprini:2024ofd} for reviews.

Our aim in these lectures
 is not to provide an introduction to the broad set of fascinating GW sources, confirmed or hypothetical,
 nor to the many creative ideas to detect them that have been proposed, investigated and realised in 
 practise; but only to provide an introduction to the field, and to that end, we decided to focus on the most common type of sources, and most common type of detectors: CBCs and laser interferometers. In the rest of this overview section we review the characteristic properties of GWs emitted by CBCs and the relevant frequency bands of laser interferometers, in particular explaining why LVK detectors are sensitive to the merger of stellar mass BH, whilst LISA for example to that of supermassive BHs. The rest of the chapter will present the theoretical derivation of GWs from GR.

\subsection{Detectors and GW frequencies}
\label{ssec:detectors}

The LVK interferometers and future LISA detector are essentially {\it Michelson-Morley interferometers}, designed to be as sensitive as possible to time-varying changes in the separation between two freely falling test-masses --- mirrors in the case of interferometers. The invariant distance between the test masses varies when a GW passes (see Section \ref{sec:detectionofGWs}), leading to a change in the observed interference pattern in the detector. 
\begin{itemize}
\item The LVK interformeters are on earth (in Livingston and Handford in the USA, in Pisa in Europe, and in Kamioka in Japan) and have a typical arm length $L\sim3$km. They are sensitive to GWs with frequency of order
\be
10 \text{Hz}  \lesssim f_{\rm GW}  \lesssim 5 \text{kHz}  \qquad \text{(LVK)}.
\ee

\item The LISA interferometer \cite{LISA:2024hlh} was adopted by ESA on the 25th january 2024, and should be operational in 2037. The distance between the spacecraft which make up arms of LISA is $L \sim  2.5 \cdot10^6$km. LISA will be sensitive to GWs with frequencies in the range
\be
10^{-4} \text{Hz}  \lesssim f_{\rm GW}  \lesssim 1 \text{Hz}  \qquad \text{(LISA)}.
\ee

\item There are plans to build new interferometers on earth beyond LVK. These include the Einstein Telescope in Europe \cite{Sathyaprakash:2012jk} and Cosmic Explorer in the USA \cite{Reitze:2019iox}, both of which should have $L\sim 10$km, and
\be
\text{few Hz}  \lesssim f_{\rm GW}  \lesssim 10^4 \text{Hz}  \qquad \text{(ET, CE...)}.
\ee

\item An alternative to interferometers are PTAs which  search for GWs by exploiting the variation in distance $L \sim 10^{17}$km between the earth and a typical distant galactic pulsar due to GWs. Pulsars emit EM pulses with extreme regularity $\Delta t$, typically of the order of milliseconds. If GWs are present, then as the EM pulses propagate from the pulsar to the earth, the observed $\Delta t_{obs}$ will be modulated. PTA experiments searching for these modulations are sensitive to GWs in the frequency band of the inverse year,
\be
10^{-7} \text{Hz}  \lesssim f_{\rm GW}  \lesssim 10^{-9} \text{Hz}  \qquad \text{(PTA)}.
\ee
In 2023 different PTA experiments presented strong evidence for the existence of a stochastic GW background \cite{EPTA:2023fyk, NANOGrav:2023gor,  Reardon:2023gzh, Xu:2023wog}.

\item Prior to the success of interferometers, there was an effort pioneered by Weber in the 60's to use resonant bars to detect GWs, building material bars whose acoustic modes would resonate at a frequency as near as possible to that expected from the optimal sources \cite{Weber1960}. These experiments would typically have a narrow-band sensitivity around $10^3$Hz. In spite of constant experimental evolution throughout the 90's, no observation has occurred in this way.
\end{itemize}
Table \ref{tab:GWdetectors} summarises the different characteristics of the existing experiments, and in particular the ratio of their characteristic size $L$ to the GW wavelength $\lambda_{\rm{GW}} = c/f_{\rm{GW}}$.

\begin{table}[ht]
\centering
\begin{tabular}{|c||c|c|c|c|}
\hline  \textbf{} & \rm{Characteristic detector} &  {GW frequency} & {$f_{\rm GW}L$}  &{$L$} vs  {$\lambda_{\rm GW}$} \\ 
&  {size (km)} & {detectability range (Hz)} & & \\
\hline \hline
LVK  & $\sim 1$  & $\sim 10 - 10^4$   & $f_{\rm GW}L \ll 1$ & $L\ll \lambda_{\rm GW}$ \\ \hline
LISA & $\sim 10^6$    & $\sim 10^{-4} - 1$ & $f_{\rm GW}L \sim 1$ & $L \sim  \lambda_{\rm GW}$ \\ \hline
PTA  & $\sim 10^{17}$ & $\sim 10^{-9} - 10^{-7}$ & $f_{\rm GW}L \gg 1$ & $L\gg \lambda_{\rm GW}$ \\ \hline
\end{tabular}
\caption{Characteristics of different GW detectors and the corresponding GW wavelength.}
\label{tab:GWdetectors}
\end{table}

\subsection{Compact binary systems: orders of magnitude and characteristic scales}
\label{ssec:binaries-characteristic}

\begin{figure}[ht]\centering
  \includegraphics[width=12cm]{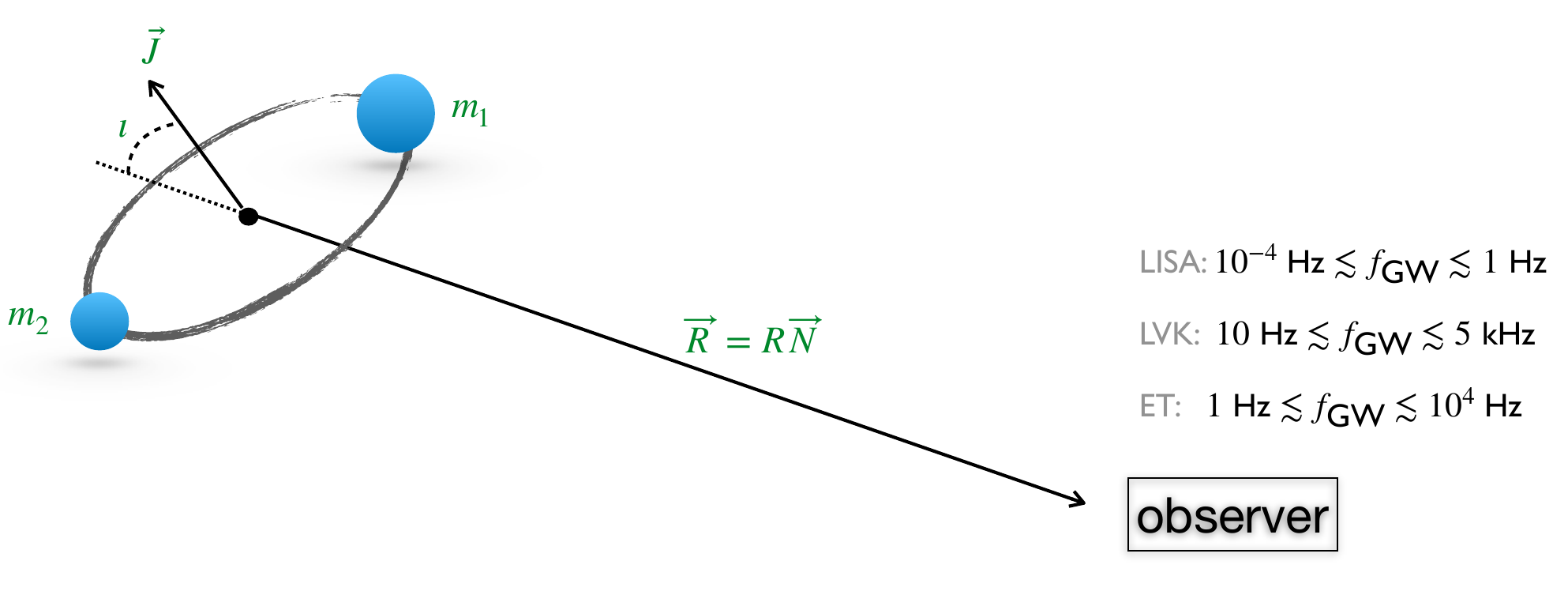}
  \caption{\small\emph{Sketch of a binary system of masses $m_{1,2}$ with conserved orbital angular momentum $\vec{J}$ and inclination $\iota$, at a distance $R$ from different detectors (LVK, ET and LISA). The approximate frequency bands of each detector are indicated.}} 
    \label{fig:characteristic-GW}
\end{figure}

LVK and LISA were conceived in order to be sensitive to the particular range of frequencies that are not only within experimental reach, but also that are likely to constitute a rich source according to the known astrophysical data.  Amongst those GW sources are compact binary systems.
We now focus on orders of magnitude and characteristic scales for such compact binary system, consisting of two masses $m_{1,2}$ at a distance $R$ from the detectors, see figure \ref{fig:characteristic-GW}. The expressions given here will be derived later in section \ref{sec:CBCs}. Furthermore, the expansion of the universe, neglected here, is considered in Section \ref{sec:cosmology}.

\begin{figure}[ht]\centering
  \includegraphics[width=10cm]{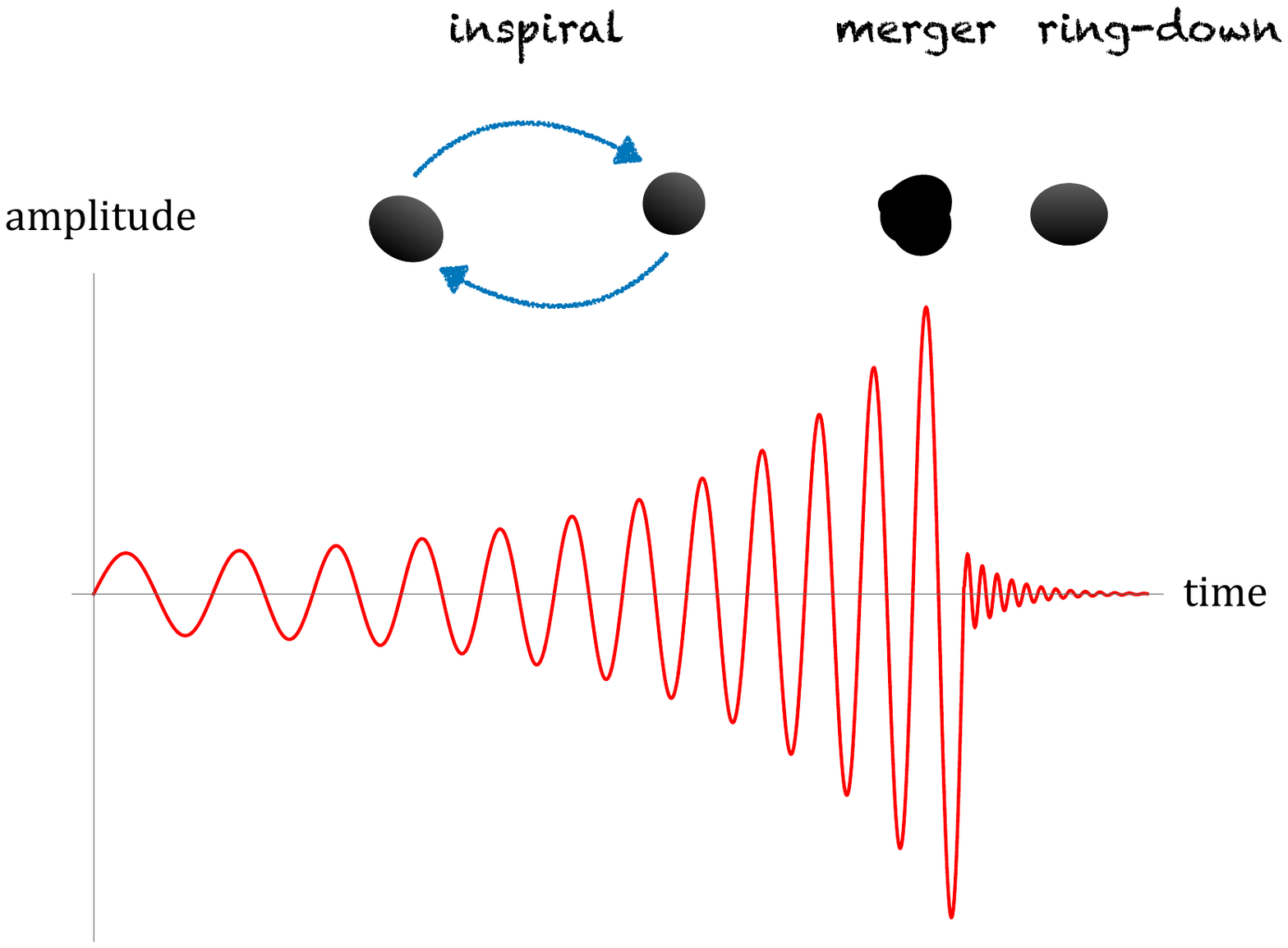}
  \caption{\small\emph{Qualitative behaviour of the inspiral, merger and ringdown phases of a CBC with their corresponding gravitational waveform as a function of time. There is no analytic method that can reproduce this signal entirely, and different approximation schemes are used. Linearized GR and the PN expansion that will be explained here can be used for the initial inspiral phase, and extrapolated to get a first estimate of the merging amplitude.
  }} 
    \label{fig:waveform}
\end{figure}

As shown in figure \ref{fig:waveform}, as a consequence of GW emission, the two masses $m_{1,2}$ approach each other --- the {\it inspiral phase} --- until they merge --- the {\it merger phase} --- and form a single object. This object will keep radiating GWs, in
the so-called {\it ringdown phase}, until it settles down to an equilibrium state (which for BHs is expected to be represented by the Kerr or Schwarzschild solutions, according to theoretical and numerical evidence) after which no further emission occurs. The typical corresponding waveform, related to the GW amplitude, is shown in figure \ref{fig:waveform} as a function of time.

\begin{itemize}
\item The {\it inspiral phase} can be understood with perturbation theory (the ``post-Newtonian (PN) expansion'' of the Einstein equations) presented below, more details in \cite{Thorne:1980ru,Blanchet:2006jqj,poisson}.  
\item  The {\it merger phase} generally requires numerical relativity, or other techniques such as effective one-body techniques, see e.g.~\cite{DeruelleUzan2018} for an introduction.  These techniques can also applied to the late inspiral phase, in particular to understand accurately the GW signal as the merger is reached.
\item  The {\it ringdown phase} can also be approached with perturbative methods, namely BH perturbation theory, see e.g.~\cite{Kokkotas:1999bd}.
\end{itemize}

\subsubsection{The chirp signal}
\label{sss:chirpsignal}
During the inspiral phase the GW frequency increases with time according to the well-known {\it chirp} signal. Using the dominant quadrupolar mode contribution for point masses $m_1$ and $m_2$ (with spins set to zero), and assuming circular orbits, the time dependence of the frequency is given by 
\be
{ f_{\rm GW} = \frac{1}{\pi} \left(\frac{G{\cal{M}}}{c^3}\right)^{-5/8} \left( \frac{5}{256 \tau} \right)^{3/8}}
\label{eq:fchirp}
\ee
see Eq.~\eqref{eq:chirpf}, section \ref{subsec:circular}.
Here the {\it chirp mass} is
\be
{\cal{M}} \equiv \frac{(m_1 m_2)^{3/5}}{(m_1+m_2)^{1/5}}
\label{eq:chirpmass}
\ee
and 
\be
\tau = t-t_c
\ee 
is the time to coalescence, with $t_c$ the coalescence time. Clearly Eq.~\eqref{eq:fchirp} will break down before $\tau=0$ where formally  $f_{\rm GW}$ diverges. We thus define a ``merger time'' $t_{\rm merger}<t_c$ up to which
Eq.~\eqref{eq:fchirp} is assumed valid, and whose meaning we now discuss.

\subsubsection{Merger frequency}
\label{sss:mergerfreq}

We now assume that the two objects are Schwarzschild BHs, and that merger occurs at the innermost stable circular orbit (ISCO) namely a distance $a=6GM/c^2$ with $m=m_1 + m_2$. It then follows from Keplers laws (see Sec.~\ref{sec:CBCs}) together with Eq.~\eqref{eq:fchirp} that
\be
{f_{\text{merger}} = \frac{1}{6^{3/2}\pi} \left( \frac{c^3}{Gm} \right).}
\label{eq:fmerger}
\ee
(Note given a length scale $a$ and a mass $m$, $\sqrt{Gm/{a^3}}$ has dimensions of frequency.  Setting $a=6Gm/c^2$ gives, modulo factors of $2\pi$, Eq.~\eqref{eq:fmerger}.)
\begin{itemize}
\item For a binary neutron stars ({\bf BNS}) system, with say $m_{1,2}\sim 1.4 M_\odot$ then Eq.~\eqref{eq:fmerger} gives
\be
f_{\text{merger}} \simeq 1.5\text{kHz} \qquad (\text{BNS})
\ee
This is in the upper part of the LVK frequency band. 
\item 
For a stellar mass binary black hole ({\bf{BBH}}) system with for instance $m_{1,2}\sim 35 M_\odot$,
\be
f_{\text{merger}} \simeq 60 \text{Hz} \qquad (\text{stellar mass BBH}).
\ee
This is right in the frequency band of LVK. 
\item 
For a massive black hole binary ({\bf{MBHB}}) system with for instance $m_{1,2}\sim 10^6 M_\odot$
\be
f_{\text{merger}} \simeq 10^{-3} \text{Hz} \qquad (\text{supermassive Binary BHs})
\ee
which is in the frequency band of LISA. 
\item Notice that PTA frequencies do not correspond to the merger frequency of any know astrophysical system. Rather, they correspond to the inspiral phase of super MBHB at times much before merger, as can be seen from Eq.~\eqref{eq:fmerger}. Hence these are on broad orbits, with periods of the order of years.
\end{itemize}

Detailed figures for the Ligo-Virgo, CE and ET sensitivities as a function of frequency can be found for instance in \cite{Maggiore:2024cwf}. LISA sensitivities can be found for instance in \cite{LISA:2024hlh}.


\subsubsection{Time to merger}

If GWs emitted during the inspiral enter the frequency band of a given detector at frequency $f_{\rm low}$, then it is straightforward to integrate Eq.~\eqref{eq:fchirp} from $f_{\rm low}$ to $f_{\rm merger}$ to find the total duration of the GW signal as will be observed by the experiment. Assuming $f_{\rm merger} \gg f_{\rm low}$ for simplicity, one finds
that the total duration of the signal is
\be
{ T \sim 10^{-3} f_{\rm low}^{-8/3} \left( \frac{c^3}{G{\cal{M}}} \right)^{5/3}}
\label{eq:T}
\ee
\begin{itemize}
\item For BNS entering the LVK band with $f_{\rm low} \sim 20$Hz, this gives $T\sim 4$ minutes.
\item  For BNS entering the ET band with $f_{\rm low} \sim 1$Hz, then $T\sim 5$ days. \\
(This implies for example that effects of the rotation of the earth cannot be neglected when calculating the GW properties in more detail, see e.g.~\cite{Iacovelli:2022bbs} and references within. Also one might expect other GW signals to be produced in such a long period, overlapping with the BNS one. This makes data analysis more complex \cite{Samajdar:2021egv}.)
\item For stellar mass BHs, with say $m_{1,2}\sim 35 M_\odot$ entering the LVK band with $f_{\rm low} \sim 20$Hz, then $T\sim 0.1$ seconds.
\item For stellar mass BHs, with say $m_{1,2}\sim 35 M_\odot$ entering the ET band with $f_{\rm low} \sim 1$Hz, then $T\sim 300$ seconds.
\item For MBHB with $m_{1,2}\sim 10^6 M_\odot$ entering the LISA band with $f_{\rm low} \sim 10^{-4}$Hz, then $T\sim 1$ month.\\
(The orbital motion of LISA will thus also be non-negligible and e.g.~Doppler effects must considered. Furthermore other LISA sources will overlap with the MBHB signal.)
\end{itemize}

\subsubsection{Amplitude and distance}
The dimensionless amplitude of the GW signal scales with distance $R$ to the source and GW frequency $f_{\rm GW}$ as
\be
{ h \sim \frac{4}{R}  \left( \frac{G{\cal{M}}}{c^2}  \right)^{5/3} \left(\frac{\pi f_{\rm GW}}{c}\right)^{2/3}}.
\label{eq:hamplitude}
\ee
As an example, consider say stellar mass BBH with $m_{1,2}\sim 35 M_\odot$ for which $f_{\rm merger}\sim 60$Hz. In order to generate (at merger) a signal with amplitude $h\sim 10^{-21}$, which is accessible to LVK, requires
\be
R \sim 400 \, \text{Mpc}
\ee
which is of the order of cosmological scales (for comparison, the observable universe has a scale of $c/H_0\sim$Gpc, where $H_0$ is the Hubble constant).

Clearly from Eq.~\eqref{eq:hamplitude}, given ${\cal{M}}$ and $f_{\rm GW}$, the more sensitive a detector, namely the smaller $h$ can be detected, the further one can detect a given GW source. The ``detection volume'' of LVK has been steadily increasing with the different observing runs of LVK obviously leading to increasing numbers of detected GW events.

Notice that if such a GW signal is detected, then from the time dependence of the GW frequency one can directly obtain chirp mass Eq.~\eqref{eq:fchirp}. With that, from the amplitude one can obtain the distance through Eq.~\eqref{eq:hamplitude}. Distance measurements can thus directly be obtained with GW observations from binaries, hence their name {\it standard sirens} \cite{Schutz:1986gp,Holz:2005df}. This should be contrasted with the case of EM observations ({\it standard candles}) for which the determination of the distance is particularly difficult. See Sec.~\ref{sec:cosmology} for more information about GWs as distance indicators and their use in cosmology.

\subsubsection{Distance between objects at merger}

When GWs are emitted with frequency $f_{\rm GW}$, the two bodies in the compact binary are separated by a characteristic scale
\be
{ r \sim \left( \frac{Gm}{f_{\rm GW}^2} \right)^{1/3}}
\label{eq:cc}
\ee
(see also the discussion after Eq.~\eqref{eq:fmerger}). Since, from Eq.~\eqref{eq:fchirp}, the GW frequency increases during inspiral, the distance $r$ between the two bodies decreases. The minimum distance is at the merger frequency $f_{\rm merger}$. For example, for stellar mass BBH with $m_{1,2}\sim 35 M_\odot$ and $f_{\rm merger}\sim 60$Hz then from Eq.~\eqref{eq:cc} $r \sim {\cal{O}}(100)$km.

A distance $r \sim {\cal{O}}(100)$km is tiny compared to the characteristic size of a star. Some of the most dense stars  in the universe --- for instance WDs --- have a size $\sim 10^3$km. Main sequence stars have a size which can go up to millions of km. Thus if GW signals are seen from objects which reach minimum approach distances $\sim {\cal{O}}(100)$km, those objects cannot be stars as they would already have collided. We must be dealing with BH (or possibly NS) for which the minimum distance will be determined by the Schwarzschild radius.

\subsection{Roadmap}

Having gone through the overview and discussed these orders of magnitude, the remainder of this chapter aims to derive formal results on GWs starting from Einstein's equations.  Many introductions and reviews on GWs already exist, see for instance \cite{poisson,Maggiore1,Maggiore2,Blanchet:2006jqj,AnderssonBook,DeruelleUzan2018} to mention a few. It is a rich and intricate topic, and each of these reviews tends to have a different angle on it, whose mutual compatibility may not always be clear to somebody entering the field. 
We have strived at presenting the material in a way that allows one to understand how the different approaches relate to one another. We have also strived to 
spend time on some of the subtleties and delicate conceptual aspects of GR and GWs which are often left to the side in gravitational wave introductions, such as gauge dependencies, asymptotic charges and memory effects, and which are becoming more and more relevant as theoretical research and experiments advance into more accurate comparisons. 
In these lecture notes, 
a reader will therefore find discussions of questions of such as coordinate invariance, diffeomorphisms and gauge invariance, spin and helicity, the controversies about the stress energy tensor of GWs, GW memory effects, and Noether charges. 
The hope is that
although most of the more advanced material is not needed for a first introduction, its inclusion here will stimulate the reader, and provide a useful reference for delving further into the topic.

\section{Einstein's equations: general covariance,  Noether's theorem and gauge transformations}
\label{sec:linearised theory}

\subsection{Einstein's equations and general covariance}
\label{ssec:gencov}
Einstein's great discovery about gravitation  was that it can be understood as the manifestation of the curvature of spacetime. 
In Wheeler's words, spacetime tells matter how to move, matter tells spacetime how to bend.\footnote{While pictorially charming, this statement is not exactly true: spacetime can be extraordinarily bent even in the absence of matter, as black hole solutions show.} 
Understanding gravity as a dynamical spacetime has changed profoundly our understanding of inertia. 
If we go back to Galilean relativity, an inertial observer is defined as one moving on a straight line at constant velocity. Special relativity introduces a non-trivial mixing of space and time, but leaves this notion unaffected: Inertial observers are still moving on a straight line, even though they are now related by Poincar\'e transformations as opposed to Galilean transformations, so to account for the experimental invariance of the speed of light. But in a curved spacetime, straight lines may no longer exist.
The notion that encompasses them is the one of geodesics, which describe free-falling observers.
Constant motion on a straight line is simply the flat-spacetime version of free falling. The understanding offered by general relativity thus has the merit of not only explaining gravity, but also explaining the origin of inertia. On the other hand, it changes the perspective on it radically:
You reading these notes at your desk are inertial in Newton's terms, but accelerated in Einstein's, since you are being held by the ground against Earth's gravitational attraction and not following a geodesic. 

Another profound consequence of a dynamical spacetime metric is that the field equations of gravity and matter are {\it covariant under general coordinate transformation}, as we will review below, introducing a new paradigm that goes under the name of {\it principle of general covariance}. In a curved spacetime, there are no more preferred Cartesian coordinates, no more Poincar\'e transformations relating inertial observers, and familiar physical concepts such as time evolution and energy become surprisingly subtle.
These aspects of GR are often glossed over in lectures aiming at introducing gravitational waves, where one can blissfully rely on the background spacetime introduced by the weak field approximation and ignore most of them. However we believe they are important in order to better appreciate some of the properties of gravitational waves, provide an understanding that is more conceptual and less application-driven, and mostly because frankly who'd need yet another introduction to GWs if we didn't attempt something different? So 
we will briefly review these aspects below, and use them as benchmark to discuss some conceptual aspects of gravitational waves.
For instance, the lack of preferred clocks in a curved spacetime is relieved in the weak field approximation, where one can use the flat Minkowski background to introduce a  class of Cartesian observers, and select their proper time as preferred time. But the lack of well-defined notion of energy density is a subtlety that persists also in the weak-field approximation, and has to be dealt with.

Let us start by recalling Einstein's equations
\be\label{EE1}
G_{\m\n}+\L g_{\m\n} = \f{8\pi G}{c^4} T_{\m\n}, 
\ee
where $G_{\m\n}:=R_{\m\n}-\f 12Rg_{\m\n}$ is the Einstein tensor with $R_{\m\n}$ the Ricci tensor and $R$ the Ricci scalar, and $T_{\m\n}$ is the (symmetric) stress-energy tensor of matter.
We use the definitions and conventions of \cite{poisson}, in particular mostly-plus convention for the spacetime metric $g_{\m\n}$.
The constants $G/{c^4}$ and $\L$ are respectively the relativistic gravitational coupling constant and the cosmological constant. 
The first can be determined from local gravitational experiments to be
\be
\f{8\pi G}{c^4}\simeq 10^{-43}\, {\tt kg}^{-1}{\tt m}^{-1}{\tt s}^{2}.
\label{eq:tiny}
\ee
This value is stupendously small, and it is the origin of the `rigidity' of spacetime mentioned in the overview section. The smallness of this parameter has, on the other hand, a positive side: the gravitational force is so weak that many of the observed phenomena, and virtually all solar system experiments, can be studied using the {\it weak field approximation}, namely a perturbative expansion around the Minkowski metric. 
This is quite helpful because Einstein's equations are non-linear and it is in general very difficult to find exact solutions. 
Strong gravity effects occur only near very compact objects, and to study them one has to resort to numerical techniques, or be able to push the perturbative treatment to high orders. 

The cosmological constant $\L$ can be determined from the observed acceleration of the  expansion of the universe assuming homogeneity and isotropy on large scales, and turns then out to be $\L\simeq10^{-52}\,{\tt m}^{-2}$. This coupling constant can also be interpreted as a sort of averaged `vacuum energy' density, often referred to as \emph{dark energy} since it is not associated to visible matter, and whose value is $\rho_{DE}={\L c^2}/G \sim 10^{-28}$kg/m$^3$.
The presence of $\L$ affects the propagation of gravitational waves on cosmological distances, but it can be ignored for a first understanding of the perturbative treatment. We will set $\L=0$ for now, and restore it below in Section \ref{sec:cosmology} when discussing cosmological effects.

Analysis of the 10 field equations in Eq.~\eqref{EE1} shows (see subsection \ref{ssec:dofcount}) that: 
four are redundant, because of the Bianchi identities;
four are elliptic, hence describe gravitational degrees of freedom constrained by the sources; 
two are hyperbolic, hence contain independent degrees of freedom.
This three-sided structure is a common feature to Maxwell and Yang-Mills theories, with the role of the Gauss constraint generating gauge transformation replaced by the so-called Hamiltonian and vector constraints generating diffeomorphisms, and it is our first indication that coordinate transformations are a gauge symmetry.
A second indication comes from Noether's theorem, but before talking about it, 
let us review how coordinate transformations act.

Recall that a tensor is a quantity that transforms homogeneously under general coordinate transformations $x^\m\to x'{}^\m(x^\n)$. For instance a scalar field, a vector field and the metric transform respectively as
\be\label{gdiffeo0}
\phi'(x')=\phi(x), \qquad v'{}^\m(x')=\f{\p x'{}^\m}{\p x^\n}v^\n(x), \qquad g'_{\m\n}(x') = \f{\p x^\r}{\p x'{}^\m} \f{\p x^\s}{\p x'{}^\n} g_{\r\s}(x).
\ee
The transformation law of a scalar is such that its value at one point $P$ is the same after the diffeomorphism, since both $x$ and $x'$ identify the same point, just in different coordinates.\footnote{This is sometimes misstated by saying that scalars are invariant under coordinate transformations, which is not true. A quantity is invariant under coordinate transformations if it satisfies the stronger property that its value does not depend on the coordinates used, which for the scalar field would be the equation $\phi(x')=\phi(x)$. This is not true in general, but only for isometries --- more on this below. A typical example of coordinate invariance is the integral 
over the whole manifold of a scalar times the volume form, as experience from solving integrals via change of coordinates should show.} 
The vector and metric do the same, but furthermore their indices are mixed up using the Jacobian of the coordinate transformation, or its inverse. A vector field is said to transform as a contravariant tensor of order one, and the metric as a covariant tensor of order two.
Since coordinate transformations are typically restricted to be differentiable, namely continuous and connected to the identity, they are also invertible, and correspond to mathematical transformations called diffeomorphisms. In this language, \eqref{gdiffeo0} is a diffeomorphism of the metric.

If the coordinate transformation is infinitesimal, we can write it as $x'{}^\m=x^\m+\xi^\m(x)$, and approximate the transformation rules in \eqref{gdiffeo0} using the Taylor expansion (applied to both the field's argument and the Jacobian). This defines the infinitesimal transformations
\begin{align}\label{phidiffeo}
& \d_\xi \phi:= \phi(x)-\phi'(x)=\xi^\m\p_\m\phi \equiv \pounds_\xi \phi, \\
& \d_\xi v^\m:=v^\m(x)-v'{}^\m(x)=\xi^\n\p_\n v^\m-v^\n\p_\n\xi^\m \equiv \pounds_\xi v^\m, \\
& \d_\xi g_{\m\n}:=g_{\m\n}(x)-g'_{\m\n}(x)  \label{gdiffeo1}
=\xi^\r\p_\r g_{\m\n}+2g_{\r(\m}\p_{\n)}\xi^\r= 2\nabla_{(\m}\xi_{\n)}\equiv\pounds_\xi g_{\m\n},
\end{align}
where we introduced the use of round brackets for index symmetrization (and we will later on also use square brackets for index anti-symmetrization).
Notice that in all cases we recover as infinitesimal transformation the Lie derivative. This is a general result valid for any tensor.
The third equality in \eqref{gdiffeo1} is on the other hand special to the metric tensor, and follows from the expression of the connection in terms of the metric.

Being written in terms of tensors, 
Einstein's equations are automatically covariant under general coordinate transformations. This is the principle of general covariance, that played a key role in guiding Einstein to formulate his theory. One immediate implication is that locally we can always find coordinates such that the metric takes the Minkowski expression at a point, which is one version of the principle of equivalence. A more subtle implication is that coordinate transformations must be \emph{symmetries} of the theory, in other words a solution can be equivalently written in any coordinate system. To understand this point, let us consider the Lagrangian description of the dynamics. 
The field equations \eqref{EE1} are Euler-Lagrange equations of $\cL = \cL_{\sscr EH}+\cL_{\sscr M}$, 
where
\be\label{LEH}
\cL_{\sscr EH} = \f{c^3}{16\pi G}(R-2\L)\sqrt{-g}
\ee
is the Einstein-Hilbert Lagrangian density, and ${\cal L}_M$ the matter contribution, left arbitrary for the moment. Here $g=\rm{det}(g_\mn)$, and the word density has a double meaning: in the physical sense, since $c \cL$ has the dimensions {\tt J m}$^{-3}$ of an energy density, but also in the mathematical sense, since $\sqrt{-g}$ makes it transform not as a scalar but as a scalar density of weight 1. This has the following consequence. Recall that the general formula for the variation of a determinant is $\d g=gg^{\m\n}\d g_{\m\n}$. This implies that $\pounds_\xi \sqrt{-g} = \f12 \sqrt{-g}g^{\m\n}\pounds_\xi g_{\m\n} = \sqrt{-g} \na_\m\xi^\m$,
whence $\pounds_\xi (\sqrt{-g}\phi) = \p_\m(\sqrt{-g}\xi^\m\phi)$ for any scalar $\phi$. Thus, a Lagrangian density transforms as a total derivative under diffeomorphisms:
\be\label{LiecL}
\pounds_\xi \cL = \p_\m(\xi^\m\cL).
\ee
Since total derivative do not affect the field equations, the transformed solutions are still solutions. We conclude that {\it any diffeomorphism is a symmetry of a general covariant Lagrangian.}

We stress that this result relies crucially on the fact that in a general covariant theory the metric is a dynamical field, and not a fixed background.
To appreciate this point and the difference with non-general relativistic physics, let us consider the matter Lagrangian 
$ {\cL}_{\sscr M} (g,\psi)$, which depends on both the metric $g_{\m\n}$ and the matter fields, which we denote collectively as $\psi$. Applying the chain rule, we find
\begin{align}\label{dxiLM}
\d_\xi {\cL}_{\sscr M} &= \f{\d {\cL}_{\sscr M} }{\d\psi} \d_\xi\psi + \f{\d {\cL}_{\sscr M} }{\d g_{\m\n}} \d_\xi g_{\m\n} +\p_\m\tl\th^\m
=  \f{\d {\cL}_{\sscr M} }{\d \psi}\pounds_\xi\psi + \f{\d {\cL}_{\sscr M} }{\d g_{\m\n}} \pounds_\xi g_{\m\n} +\p_\m\tl\th^\m \\
& = \pounds_\xi {\cL}_{\sscr M}+\p_\m\tl\th^\m =\p_\m(\xi^\m{\cL}_{\sscr M}+\tl\th^\m),\nn
\end{align}
where the second equality follows from \eqref{phidiffeo} and \eqref{gdiffeo1}, and $\tl\th^\m$ is the boundary term that arises because the Lagrangian depends on derivatives of the fields as well. Since the result is a total derivative, the field equations are unchanged, and this means that diffeomorphisms are symmetries also of the matter sector. But this conclusion relies crucially on treating the metric as a dynamical variable!
In non-general relativistic physics the metric is a non-dynamical, `background' field. Accordingly, there is no variation with respect to the metric, and no second term after the first equality of \eqref{dxiLM}. Lacking this term the second equality breaks down, and $\d_\xi {\cL}_{\sscr M} $ is no longer a boundary term. We find instead
\begin{align}
\d_\xi {\cL}_{\sscr M} =\p_\m(\xi^\m{\cL}_{\sscr M}+\tl\th^\m)- \f{\d {\cL}_{\sscr M} }{\d g_{\m\n}} \pounds_\xi g_{\m\n}.
\end{align}
This means that only those diffeomorphisms which are isometries are symmetries of the non-general relativistic physics, e.g. the familiar Poincar\'e invariance of special relativity. Whereas an arbitrary diffeomorphism does not map a solution into a new solution in non-general relativistic physics, and there is no invariance under general coordinate transformations. 

The discussion highlights why general covariance is often referred to as background independence, namely the absence of any fixed background metric in the theory, or as diffeomorphism invariance, since every physical observable should be independent of the coordinate used to describe it. General covariance, background independence, or diffeomorphism invariance, are thus different terms used to capture the same underlying property of general relativity.

The fact that every solution can be equivalently described in any coordinate system has a useful analogy with electromagnetism, where every solution can be described in any choice of gauge for the Maxwell potential. It is actually much more than an analogy, there is in fact a precise mathematical sense in which gauge transformations in Maxwell and Yang-Mills theories have the same property of coordinate transformations in general relativity, which we discuss next. Making this analogy precise is also useful in the context of gravitational waves as it will allow us to understand the origin of their gauge dependence and distinction between physical and unphysical modes.

\subsection{Noether's theorem and diffeomorphisms as gauge symmetries}\label{SecNoether}
\label{ssec:noether}

Noether's theorem proves that every differentiable\footnote{Namely, continuous and including the identity transformation.} symmetry defines a current $j^\m$ which is conserved on solutions, namely $\na_\m j^\m\eqons 0$. Here the symbol $\eqons$ means an equality valid only for solutions, or `on-shell', in theoretical physics jargon. Conserved currents are extremely useful to study the properties of the dynamics of the system, and to extract general physical predictions.
We have seen above that diffeomorphisms are symmetries of a general covariant Lagrangian, continuous and connected to the identity. Therefore Noether's theorem guarantees that there will be a conserved current associated to any diffeomorphism. 
However, the application of Noether's theorem to general relativity is quite subtle. 
Let us first recall the difference between `proper symmetries' and `gauge symmetries'. Both map solutions of the field equations into new solutions. If the new solution is physically distinguishable, we say that it is a proper symmetry, or a physical symmetry. If the new solution is on the other hand physically indistinguishable, we say that it is a gauge symmetry: This typically occurs when there is a redundancy of the field equations, which leaves some quantities undetermined but irrelevant for the physics. Noether's theorem provides a simple test to distinguish the two cases: in the latter, the Noether current itself vanishes on shell, and not just its divergence. This is precisely the case with diffeomorphisms in general relativity. In fact, the conserved current associated with diffeomorphisms of the Einstein-Hilbert Lagrangian \eqref{LEH}  is given by
\be\label{jxi}
j^\m_\xi = \f{c^3}{8\pi G}\Big( (G^\m{}_\n+\L \d^{\m}_{\n})\xi^\n - \na_\n \nabla^{[\m} \xi^{\n]}\Big).
\ee
One can immediately verify using the Bianchi identities that $\nabla_\m j^\m_\xi \eqons 0$.
On the other hand, the first term above vanishes on-shell, and the second term is a total derivative.
 Therefore, the Noether current itself vanishes on-shell, as anticipated, and there are no conserved quantities (in the absence of boundaries). 
Trivial conserved quantities is a hallmark of gauge symmetries as opposed to physical symmetries, hence the result provides a precise mathematical sense in which coordinate transformations in general relativity have the same status as gauge transformations  in Maxwell and Yang-Mills theories.\footnote{ A more rigorous approach is to look at the symplectic 2-form, and show that it is degenerate along gauge transformations and diffeomorphisms. 
}
 For this reason, diffeomorphisms are also referred to as the gauge symmetry of general relativity, and fixing a coordinate choice as \emph{fixing the gauge} in general relativity.

Having said so, there is a special situation that stands out: when the diffeomorphism corresponds to an \emph{isometry}, namely a transformation that does not change the metric. This occurs when \eqref{gdiffeo1} vanishes, and the corresponding equation $\na_{(\m}\xi_{\n)}=0$ is called  Killing equation, and $\xi$ a Killing vector. One should keep in mind that for a generic metric, this equation does not admit any solutions:
Isometries occur only for very special metrics. These special metrics are, however, important for physical applications,\footnote{A cow is always a spherical object in the initial investigations of a theoretical physicist.} 
hence Killing vectors play an important role.
First of all, anyone who is familiar with the study of geodesics on spacetimes with isometries knows that there are conserved quantities associated with the Killing vectors, and which can be derived as Noether charges for the test particles' dynamics.  

More importantly for us, isometries play also an important role in the study of gravitational waves, because the dynamics of perturbations on a given background is such that the isometries of the background induce proper symmetries for the perturbations.
We will see this in details in Section~\ref{ssec:pertE} below.

\begin{figure}[ht]\centering
  \includegraphics[width=6cm]
  {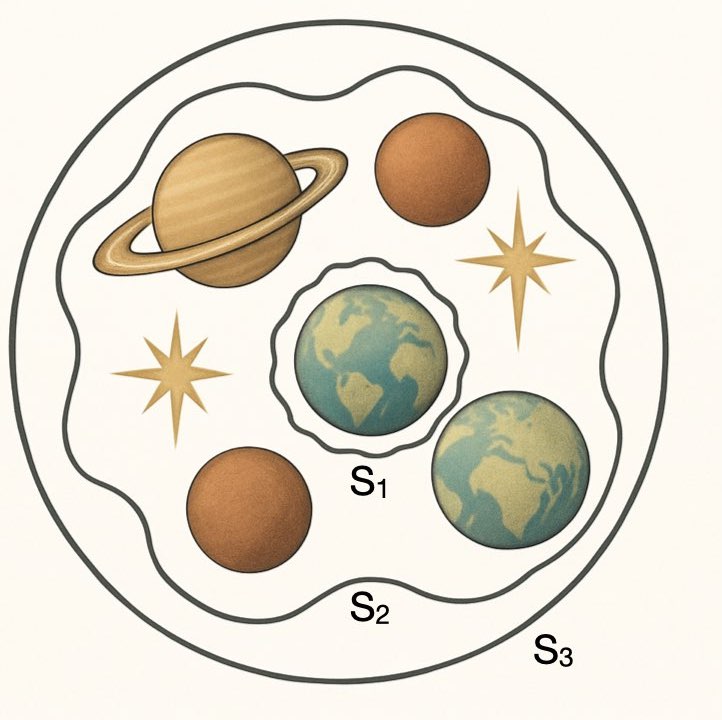}
  \caption{\small\emph{Conservation laws on stationary spacetimes. 
    Evaluating the surface integral \eqref{Komar} on the innermost surface $S_1$ gives a quantity proportional to the energy-momentum of the planet encompassed. Evaluating it on $S_2$ gives the total energy-momentum of all stars and planets. The difference between the two surface integral is proportional to the energy-momentum of the region between them. Finally since there is no source outside $S_2$, integrating on $S_2$ or $S_3$ gives the same result.}} 
   \label{FigStars}
\end{figure}

Before moving on, let us also mention another aspect in which isometries are important for the full theory. This is a more advanced topic, and will not be needed in the following, but it allows us to give a more complete picture, and also a first intuition of how boundaries introduce non-vanishing Noether charges for diffeomorphisms. If the spacetime has isometries, the Noether current \eqref{jxi} gives rise to a useful conservation law analogue to the Gauss law in electromagnism, which we recall states that the total charge in a region is equal to the flux of the electric field. To see this, we first observe that
\be
\na_\n \na^{[\m} \xi^{\n]} = \f12(R^\m{}_\n\xi^\n - \square \xi^\m+\na^\m \na_\n\xi^\n),
\ee
an identity which follows from the definition of the Riemann tensor as the commutator of two covariant derivatives.
If $\xi^\nu$ is a Killing vector, the second term gives $-R^\m{}_\n\xi^\n$ and the last term vanishes. Then integrating both sides of the equation over a 3d portion of space $V$ delimited by two boundaries $S_1$ and $S_2$, and using Stokes' theorem,  we find
\begin{align}\label{Komar}
& Q_\xi[S] =  \oint_S\na_\n \na^{[\m} \xi^{\n]} dS_\m, \\\label{Komarflux}
& Q_\xi[S_2]-Q_\xi[S_1] = \int_V R^\m{}_\n\xi^\n dV_\m \eqons  \f{8\pi G}{c^4} \int_V \left(T^{\m\n}\xi_\n-(\L+\f T2)\xi^{\m}\right) dV_\m.
\end{align}
The Noether charge \eqref{Komar} obtained in this way is known as Komar charge.
If the right-hand side of \eqref{Komarflux} vanishes, the Komar charge is conserved in the sense that it has the same value no matter which surface $S$ is used, and its value changes only when the deformations of $S$ include some source terms. If the right-hand side does not vanish, the Noether charge varies by an amount determined by the total quantity of energy-momentum in the enclosed region, see Fig~\ref{FigStars}. 
As an example, one can consider the Kerr solution, which possesses two Killing vectors corresponding to stationarity and axial symmetry. Evaluating \eqref{Komar} on an arbitrary 2-sphere $S$ encompassing the singularity gives respectively the mass and angular momentum (up to numerical coefficients to be fixed), independently of the coordinate used and independently of deformations of $S$.

While Komar charges are limited to isometries, it is possible to generalize the construction of Noether charges and canonical generators to arbitrary spacetimes, at least in so far as they admit boundaries with non-trivial residual diffeomorphisms.\footnote{There is also ongoing research on constructing charges on arbitrary regions in arbitrary spacetimes.} Let us mention three important examples. First, spacetimes that are asymptotically flat at spatial infinity. The residual diffeomorphisms compatible with the boundary conditions are the Poincar\'e transformations of the flat boundary metric. One can construct Noether charges and canonical generators for these boundary diffeomorphisms (see e.g. \cite{Iyer:1994ys}), and the result coincides with the Arnowit-Deser-Misner (ADM) charges that were previously derived with canonical methods. Second, spacetimes that are asymptotically flat a null infinity. This case is particularly relevant to understand gravitational waves at the non-perturbative level. The residual transformations are a generalization of Poincar\'e transformations in which translations are angle-dependent, an infinite-dimensional extension known as Bondi-Van der Burg-Metzner-Sachs (BMS) transformations.\footnote{Intuitively, the extension comes about because the induced metric on a null hypersurface is degenerate, hence any deformation along that direction leaves the system invariant.}  Noether charges for the BMS symmetry were constructed in \cite{Ashtekar:1981bq,Dray:1984rfa,Wald:1999wa}, and there is a large body of recent literature on the subject motivated by ongoing applications and developments. Among these, the application of the Noether approach to horizons and more generally null boundaries, see e.g. \cite{CFP,AKKL}.

\subsection{Gauge and observers}
\label{ssec:gauge}

Even though every physical phenomenon can be described in any coordinate system, some choices can stand out because they simplify the description, or because they are naturally associated with a class of observers of interest. For instance in flat spacetime, Cartesian coordinates make the Christoffel's symbols vanish, thus simplifying many calculations, and can be associated to inertial observers, in the sense that they label their rods and clocks.
In curved spacetimes the situation is more complicated because there are no Cartesian coordinates.
To take a simple example, consider the Schwarzschild black hole. The most common coordinates used to describe this solution are the so-called `static ones', which make time-independence manifest since the metric is independent of the time coordinate $t$. Or in better terms, since the time-translation Killing vector is simply $\p_t$. The vector field $\p_t$ describes a family of non-inertial observers static at a fixed distance outside the black hole, whose time delays are related by $\sqrt{-g_{tt}}$. The Schwarzschild static coordinates are naturally associated to these observers.
An alternative choice are Gullstrand-Painlev\'e coordinates, whose $t$ describes the proper time of observers radially free-falling into the black hole. 
Another choice is the `temporal gauge' defined by
\be\label{tempgauge}
g_{0\m}=(-1,0,0,0),
\ee 
in which the radially free-falling observes are all synchronized,\footnote{This requires giving them non-zero energy, as opposed to the Gullstrand-Painlev\'e observers that have zero energy.} whence the alternative name of `synchronous gauge'. 
For Schwarzschild, this gauge can be achieved using Lemaitre coordinates.
The synchronization may look like a nice feature, but in these coordinates the spatial part of the Schwarzschild metric is explicitly dependent on the coordinate $t$! Therefore its staticity is hidden, and has to be verified by the existence of a time-translational Killing vector.
In this example we can see the analogy between coordinate choices and gauge choices in electromagnetism very clearly. An electrostatic potential is more conveniently described in the Coulomb gauge because it makes the potential manifestly time-independent. But it can be described in any other gauge, and if we use the temporal gauge $A_0=0$, the potential acquires an inconvenient time dependence which is pure gauge and hides the staticity of the system.

The temporal gauge \eqref{tempgauge}  can be chosen for \emph{any} spacetime in a given coordinate chart. If it is done, it fixes completely the 4-dimensional diffeomorphism freedom in the chart. It is thus an example of complete gauge fixing,\footnote{Complete here refers to the 4-dimensional picture. There remains the freedom of time-independent 3-dimensional diffeomorphisms, namely of choosing the coordinates on one -- and one only -- given hypersurface.}  and the resulting coordinates describe free-falling observers with synchronised clocks. 
It means that in the temporal gauge the coordinates are attached to free-falling test bodies, and this can be a very useful choice in many situations.

One shortcoming of the temporal gauge is that it may `squeeze' physical information in field components one would not naturally expect, 
as the Schwarzschild example in Lemaitre coordinates shows. Another one is that it is non-symmetric among the components, in the sense that it relies on an initial choice of coordinate to be taken as the time.
An example of symmetric gauge is the harmonic gauge, which is satisfied by coordinates such that
\be\label{xharm}
\square x^\m =\G^\m_{\n\r}g^{\n\r} = \p_\n(\sqrt{-g}g^{\m\n})=0,
\ee
where $\square=\na_\n\na^\n$ is the curved spacetime d'Alembertian.
The symmetry makes this a convenient choice in many dynamical situations. For instance, it is the gauge in which it is easiest to see that the initial value problem is well-posed \cite{ChoquetBruhat1952}, and the one in which it is easiest to study gravitational waves, where it gives rise to a gauge preserving the Lorentz covariance of the Minkowski background. Notice that \eqref{xharm} is not a complete gauge fixing of the 4-dimensional diffeomorphism freedom, as there are infinitely many solution of the wave equation, hence infinitely many choices of harmonic coordinates for a given metric. To obtain a complete gauge fixing one has to specify a unique set of harmonic coordinates with additional conditions.

One version of the equivalence principle states that purely local experiments can not distinguish the presence of gravity. In the formalism of general relativity, this is embodied in the fact that at any given point in spacetime it is possible to find coordinates so that the metric is flat and its first derivative vanish. A coordinate system that achieves this is called a \emph{local inertial frame}. Only an experiment that can probe second-order variations in the metric would be able to see the effect of gravity in a coordinate-independent way, and these variations are the tidal forces that show up in the geodesic deviation equation. An example of local inertial frame is provided by Riemann normal coordinates, which are constructed around a given point so that the Taylor expansion of the metric components around that point taken as the origin gives
\be\label{gRNC}
g_{\m\n} = \eta_{\m\n} + c_{\m\n}R_{\m\r\n\s}x^\r x^\s + O(x^3),
\ee
where $c_{\m\n}=-1/3$ $\forall \m,\n$.\footnote{This formula is only valid in certain coordinate choices, hence the non-covariant notation with the $\m\n$ indices repeated but not summed over.} This gauge fixing specifies coordinates only in the neighbourhood of a point and not in a full coordinate chart, but it is still perfectly sufficient to describe the physics of a local inertial frame around that point. It is also possible to find coordinates such that the Christoffel's symbols vanish everywhere along a chosen time-like geodesic. Such coordinates are known as Fermi normal coordinates, and describe a free-falling local inertial frame. The metric takes the same form \eqref{gRNC}, but with $c_{00}=-1, c_{0a}=-2/3, c_{ab}=-1/3$. The difference between temporal gauge and Fermi normal coordinates is that in the first case every observer at constant spatial coordinates is free falling, whereas in the second case only the observer going through the origin along the $\p_t$ geodesic is free falling.

\section{Perturbative treatment of Einstein's equations}
\label{ssec:pertE}

\subsection{The idea: general background spacetime}

Perturbation theory can be set up choosing a background metric $\bar g_{\m\n}$ and writing 
\be\label{gexp1}
g_{\m\n}=\bar g_{\m\n} + h_{\m\n},
\ee
with the assumption that $|h_{\m\n}|\ll 1$ in some chosen coordinate system, so that it can be treated as a perturbation. One can then systematically Taylor-expand all metric functionals around the background, starting with the inverse metric $g^{\m\n}=\bar g^{\m\n}-h^{\m\n}+O(h^2)$ and Levi-Civita connection
\be
\G^\m_{\n\r}=\bar\G^\m_{\n\r} + \G^{\sscr (1)}{}^\m_{\n\r} +O(h^2), \qquad \G^{\sscr (1)}{}^\m_{\n\r}=\f12\bar g^{\m\s}(2\p_{(\n}g_{\r)\s}-\p_\s g_{\n\r}),
\ee
and attempt to solve the field equations order by order:
\be\label{EEgenpert}
\bar G_{\m\n} + G^{\sscr (1)}_{\m\n}+G^{\sscr (2)}_{\m\n}+\ldots = \f{8\pi G}{c^4} (\bar T_{\m\n} +T^{\sscr (1)}_{\m\n}+T^{\sscr (2)}_{\m\n})+\ldots
\ee
Here $\bar G_{\m\n} = G_{\m\n} (\bar g)$, $\bar T=T(\bar g)$, $G^{\sscr (1)}_{\m\n} = G_{\m\n} (\bar g; h)$, and so on. Explicit expressions will not be needed here, but are given in Appendix \ref{app:perturbed-action} for completeness.
The lowest order of the procedure is straightforward, one just has to be consistent with the treatment of the matter energy-momentum tensor: if we want to expand around a background which is a vacuum solution, we need $\bar G_{\m\n}=0$ hence we should treat matter as a first order perturbation. Then the first order equation to solve is
\be\label{EElin1}
G^{\sscr (1)}_{\m\n}  = \f{8\pi G}{c^4} \bar T_{\m\n}.
\ee
Solving this equation determines $h_{\m\n}$ in terms of its independent degrees of freedom, the background solution and the matter content.
To go to second order, we add a second perturbation, writing 
\be\label{gexpsecond}
g_{\m\n}=\bar g_{\m\n} + h_{\m\n} + h^{\sscr (2)}_{\m\n}. 
\ee
Then $G^{\sscr (2)}_{\m\n}=G^{\sscr (2)}_{\m\n}(h)+G^{\sscr (1)}_{\m\n}(h^{\sscr (2)})$ has two contributions, and we solve for $h^{\sscr (2)}$ using
\be\label{G2}
G^{\sscr (1)}_{\m\n}(h^{\sscr (2)})  = \f{8\pi G}{c^4} \left(T^{\sscr (1)}_{\m\n} +t^{\sscr G}_{\m\n}\right), 
\qquad t^{\sscr G}_{\m\n}:=- \f{c^4}{8\pi G}G^{\sscr (2)}_{\m\n}(h).
\ee
Notice that the first order solution feeds back as a source for the second order solution, a standard procedure from perturbatively solving non-linear equations. The same procedure applies also to the case when matter fields contribute to the background as well.
Although we wrote the perturbative expansion using dimensionless quantities only, it is always possible to rescale the metric perturbation by $\sqrt G$, so that the kinetic term of the free-field Lagrangian is canonically normalized. Then the expansion is a power series in $G$, and it is also referred to as {\it post-Minkowskian} (PM) expansion, when the background is flat.
In many cases the equations are still to hard to solve at each order, and one needs to look for additional approximations. A very common one is the non-relativistic approximation, also known as {\it post-Newtonian} (PN) expansion, in which one starts from a source that moves at small velocity  with respect to the background flat metric, namely with $v\ll c$, and expands in powers of $v^2/c^2$. 
Alternative approximation schemes include the Bondi asymptotic expansion which is perturbative in the inverse distance from the source but valid at all orders in $G$ and $v^2/c^2$ \cite{Bondi:1962px,Sachs:1962wk}; the extremal mass ratio inspiral (EMRI) and self-force expansion for a two-body system where the small parameter is the mass-ratio \cite{Barack:2018yvs}; the effective one-body approach based on resummed PN results \cite{Buonanno:1998gg}, and more recently the effective field theory approach based on tools from quantum field theory \cite{Goldberger:2004jt}.

On top of the technical difficulties, there is also a conceptual one: {\it truncating at any order beyond the lowest breaks covariance.} 
Expanding both sides of \eqref{gdiffeo1} with \eqref{gexp1}, we obtain
\be\label{totalVar}
\d_\xi \bar g_{\m\n} + \d_\xi h_{\m\n} = \pounds_\xi \bar g_{\m\n} + \pounds_\xi h_{\m\n}.
\ee
We assume that the background metric is fixed once and for all and unaffected by diffeomorphisms, namely $\d \bar g_{\m\n}=0$ and $\d_\xi\bar g_{\m\n}=0$,\footnote{It is also possible to interpret \eqref{totalVar} as 
$\d_\xi \bar g_{\m\n} = \pounds_\xi \bar g_{\m\n}$ and $\d_\xi h_{\m\n} =  \pounds_\xi h_{\m\n}$. The proof that this is a symmetry is identical. 
However this definition of perturbed transformations is less interesting physically, because it is more natural to compare perturbations when they are defined with respect to the same background.} 
 and define the effect of the linearized diffeomorphism on the perturbation to be 
\be\label{dxih2}
\d_\xi h_{\m\n} =  \pounds_\xi \bar g_{\m\n} + \pounds_\xi h_{\m\n}.
\ee
The key point is that the right-hand side contains terms of different order in $h_{\m\n}$. This is why different orders of the perturbative expansion must be included in order for diffeomorphisms to be a symmetry, and conversely truncating at a fixed order breaks covariance.  It is instructive to prove this in detail, because it highlights the special features that occur for the quadratic Lagrangian, namely the free theory, for which covariance is on the other hand possible.

To study the symmetries of the perturbative expansion, we write the perturbed Lagrangian as follows
\be
\cL(\bar g+h) = \bar \cL + \bar\cL^{\sscr (1)}{}^{\m\n} h_{\m\n} + \f12 h_{\m\n} \bar\cL^{\sscr (2)}{}^{\m\n\r\s} h_{\r\s}+\ldots,
\ee
where barred quantities only depend on $\bar g$ and not on the perturbation. 
Using \eqref{dxih2}, the infinitesimal variation gives
\begin{align}\label{dxiL}
\d_\xi \cL 
&= \bar\cL^{\sscr (1)}{}^{\m\n} \pounds_\xi \bar g_{\m\n} + \bar \cL^{\sscr (1)}{}^{\m\n} \pounds_\xi h_{\m\n} + h_{\m\n} \bar \cL^{\sscr (2)}{}^{\m\n\r\s} \pounds_\xi \bar g_{\r\s} 
\\\nn &\qquad + h_{\m\n} \bar\cL^{\sscr (2)}{}^{\m\n\r\s} \pounds_\xi h_{\r\s} + \f12 h_{\m\n} h_{\r\s}\bar \cL^{\sscr (3)}{}^{\m\n\r\s\t\l} \pounds_\xi \bar g_{\t\l} + \ldots .
\end{align}
All terms can be collected into Lie derivatives:
\begin{align}
& \bar \cL^{\sscr (1)}{}^{\m\n} \pounds_\xi \bar g_{\m\n}=\pounds_\xi \bar \cL, \\ 
& \label{symL2} \bar \cL^{\sscr (1)}{}^{\m\n} \pounds_\xi h_{\m\n} + h_{\m\n} \bar \cL^{\sscr (2)}{}^{\m\n\r\s} \pounds_\xi \bar g_{\r\s}
 =\pounds_\xi  (\bar \cL^{{\sscr (1)}\m\n}h_{\m\n}), \\ 
& \label{symL3} h_{\m\n} \bar\cL^{\sscr (2)}{}^{\m\n\r\s} \pounds_\xi h_{\r\s} + \f12 h_{\m\n} h_{\r\s}\bar \cL^{\sscr (3)}{}^{\m\n\r\s\t\l} \pounds_\xi \bar g_{\t\l}  =\f12\pounds_\xi (\bar \cL^{{\sscr (2)}\m\n\r\s}h_{\m\n}h_{\r\s}), 
\end{align}
and so on.  Each Lie derivative gives a boundary term through \eqref{LiecL}, hence \eqref{dxih2} is indeed a symmetry of the full Lagrangian. 
However, in every case except the lowest one, getting a boundary term for $\bar\cL^{\sscr (n)}$ requires both $\bar\cL^{\sscr (n)}$ and $\bar\cL^{\sscr (n+1)}$, hence if we truncate the series at a fixed order, we lose covariance. 

There are  two special features that occur at the quadratic order. First, if we take the background to be a solution, the first term in \eqref{symL2} vanishes, hence
\be\label{dxih1}
\d_\xi h_{\m\n}=\pounds_\xi\bar g_{\m\n}
\ee 
is a symmetry of the quadratic Lagrangian. 
Second, if the background has isometries, then the term proportional to $\bar\cL^{\sscr (3)}$ in \eqref{symL3} drops out, and then 
\be\label{dxihK}
\d_\xi h_{\m\n}=\pounds_\xi h_{\m\n}
\ee
is a symmetry of the quadratic Lagrangian.
This shows that isometries play a special role in perturbation theory. For a generic on-shell background the symmetry of the quadratic Lagrangian is \eqref{dxih1}.
But if the background has isometries, we have two different realization of the diffeomorphism symmetry in the quadratic Lagrangian: \eqref{dxih1} for a generic diffeomorphism, and \eqref{dxihK} for a Killing vector. To make this consistent with the perturbative expansion, we treat a Killing $\xi$ as zero-th order, and a non-Killing $\xi$ as first order. 

We stress that \eqref{dxih1} and \eqref{dxihK} are symmetries only for the quadratic Lagrangian. From the cubic Lagrangian onwards, there is no symmetry at fixed order in $h_{\m\n}$. The only symmetry is the combined \eqref{dxih2} and requires two different perturbative orders of the Lagrangian.
This fact has immediate consequences for the expanded Einstein tensor \eqref{EEgenpert}. Since the leading order $G^{\sscr (1)}$ comes from the quadratic Lagrangian, it is invariant under \eqref{dxih1}. 
But $G^{\sscr (2)}_{\m\n}$ which is derived from the cubic Lagrangian is not invariant.
{\it As a consequence, the quantity $t^{\sscr G}_{\m\n}$ appearing in \eqref{G2} is not gauge invariant.} This means that it does not provide a meaningful notion of gravitational energy-momentum tensor. We will come back to this important point below.

Even though both symmetries \eqref{dxih1} and \eqref{dxihK} descend from the same diffeomorphism invariance of the theory, which is a gauge symmetry, they have a different status at the perturbative level. 
Applying Noether's theorem to the quadratic Lagrangian, we find that the generic diffeomorphisms still have vanishing conserved current, and therefore maintain their status of gauge symmetries. However the diffeomorphisms corresponding to isometries of the background have non-vanishing Noether charges, indeed just like a standard theory on flat spacetime. A related difference is that the field equations are \emph{invariant} under \eqref{dxih1}, hence there are linear dependencies in the equations and some field components are left undetermined, and \emph{covariant} under \eqref{dxihK}, and there are no constraints associated with them.

\subsection{Weak-field approximation}
\label{ssec:weakfield}

The  weak-field approximation is a special case of the perturbative expansion in which the background spacetime is Minkowski,
\be\label{glin}
g_{\m\n}=\eta_{\m\n}+h_{\m\n}.
\ee
For instance in the solar system spacetime is approximately flat, hence it can be described using this approximation.\footnote{In first approximation, we can describe the solar system's metric using the Schwarzschild solution with $M$ the mass of the sun. Then the curvature scale can be estimated writing the Kretschmann scalar in static coordinates, and this gives $R_{\m\n\r\s}R^{\m\n\r\s}=48G^2M^2c^{-4}r^{-6}\sim 10^{-27}{\tt m}^{-4}(R_\odot/r)^6$ which is extremely small.}
We further use Cartesian coordinates, so all covariant derivatives become partial derivatives. 
In this case the linearized Einstein equations \eqref{EElin1} take the simple form
\be\label{linE}
\square h_{\m\n} -2 \p_{(\m} \p_\r h^\r{}_{\n)} + \p_\m \p_\n h + \eta_{\m\n}( \p_\r \p_\s h^{\r\s} - \square h) = -\f{16\pi G}{c^4} T_{\m\n},
\ee
where $\square$ is the  d'Alembertian in flat spacetime. 

Let us study the symmetries of \eqref{linE}. 
The Minkowski background has ten isometries, the Poincar\'e transformations. We parametrize them with vector fields 
\be\label{xiP4}
\xi^\m=a^\m{}_\n x^\n +b^\m, 
\ee
where $a^\m{}_\n$ and $b^\m$ are constants, and $a_{(\m\n)}=0$.  
According to the general discussion of the previous section, we expect two different types of symmetries in \eqref{linE}, both induced from diffeomorphism invariance of the full theory, and corresponding to those linearized diffeomorphisms that are isometries or not of the background. 
For diffeomorphisms corresponding to isometries,  \eqref{dxihK} gives
\be\label{hP4}
\d_\xi h_{\m\n} = \pounds_\xi h_{\m\n} = (a^\r{}_\s x^\s+b^\r)\p_\r h_{\m\n} + 2a^\r{}_{(\m}h_{\n)\r}.
\ee
This is the transformation of a rank-2 tensor in Minkowski under Poincar\'e transformations. 
For generic diffeomorphisms that are not isometries, \eqref{dxih1} 
gives
\be\label{diffeoeta}
\d_\xi h_{\m\n} = \pounds_\xi \eta_{\m\n} = 2\p_{(\m} \xi_{\n)}.
\ee
Both transformations are symmetries of \eqref{linE}, the difference being that the first changes the equations but in a covariant way, whereas the second leaves them invariant. This can be checked easily on the left-hand side. For the right-hand side, one has to pay attention to the behaviour of the energy-momentum tensor. 
Assuming the background $\eta$ to be a solution requires that we are treating the matter fields as first order in perturbation theory. Then it transforms as $\pounds_\xi T_{\m\n}$ for a Killing $\xi$, whereas its transformation under generic diffeomorphisms is second order, hence it does not affect the linearized equations.

The presence of the d'Alambertian in the linearized field equations suggests that wave solutions are indeed possible. However, there is an intricate tensorial structure that needs to be dealt with. 
Before doing so, let us discuss what it means for a wave to be `tensorial'.
The waves that we are most familiar with, such as water waves or sound waves, are \emph{scalar} waves, namely the quantity whose perturbation propagates following the wave equation is a scalar function, such as the height of the water or the pressure of the air. Electromagnetic waves on the other hand propagate changes in the electric and magnetic field which are described by vectors, and are thus `vectorial' waves. 
The difference between scalar, vectorial and tensorial waves can be described in terms of {\it spin.}
The reader already familiar with these concepts, or the reader interested in looking first at the explicit solutions without these details, can skip the next subsection.

\subsection{Spin and helicity}
\label{ssec:helicity}

Tensors in Minkowski spacetime belong to finite-dimensional representations of the Lorentz group.
This property can be used to decompose each tensor into irreducible parts, namely parts that are not mixed with one another by a Lorentz transformation. 
The irreducible parts can be labelled by a pair of half-integers $(j_1,j_2)$, and contain $(2j_1+1)(2j_2+1)$ components. For instance, a 4-vector $v^\m$ transforms under the irreducible Lorentz representation $\bf (\f12,\f12)$ with 4 components, and a symmetric tensor $h_{\m\n}$ transforms under the reducible Lorentz representation
${\bf (1,1) \oplus (0,0)}$ with $9+1=10$ components. 

The representations can be further subdivided if we pick a time direction $\t^\m$, and restrict attention to the rotation subgroup of the Lorentz group that preserves it. 
The subsets of the tensor which are irreducible with respect to the rotation subgroup are called  \emph{spin} representations, and their allowed values given by the Clebsch-Gordan addition rule $(j_1+j_2, j_1+j_2-1,\dots, |j_1-j_2|)$. For instance a vector contains the two spin representations $\bf 1\oplus 0$, and a symmetric tensor the 
four spin representations $\bf 2\oplus1\oplus 0\oplus 0$.
To be more explicit let us take $\t^\m=(1,0,0,0)$. The rotation subgroup that preserves it has the form 
\be\label{R4d}
R^\m{}_\n=\mat{1}{\vec 0}{\vec 0}{R^a{}_b}, \qquad R^a{}_b\in\SO(3).
\ee 
It is then immediate to see that  the spin-1 and spin-0 representation of the 4-vector are the spatial vector $v^a$ and the spatial scalar $v^0$ respectively.
Similarly for a 1-form $v_\m$, the spatial and time components $v_a$ and $v_0$. For a symmetric tensor, the four spin representations are
\be\label{spindec0}
h_{\la ab\ra} = h_{ab}-\f13\d_{ab}h^c_c, \quad h_{0a}, \quad h_{00}, \quad h_s:=h^c_c.
\ee

The spin 0 representation has one component, the spin 1 has three, and the spin 2 has five. These different components can be classified choosing a 
reference spatial axis and looking at the eigenmodes of the rotation generator along that axis. To fix ideas let us choose the $z$ axis. The rotation matrix that preserves it is
\be\label{Rz}
R^a{}_b = \left(\begin{array}{ccc}
\cos\th & -\sin\th & 0 \\ \sin\th & \cos\th & 0 \\ 0 & 0 & 1
\end{array}\right).
\ee
Inserting this in \eqref{R4d} and acting on a 4-vector $v^\m$ we obtain $v'{}^\m = R^\m{}_\n v^\n$, where
\ba
&& v'{}^0=v^0, \qquad v'^{x} = \cos\th v^x - \sin\th v^y,
\nn
\\
&& v'^{y} = \sin\th v^x + \cos\th v^y, \qquad v'^{z}=v^z.
\ea 
The temporal and the $z$ component, which is longitudinal along the rotation axis, are invariant.
The components $x$ and $y$, which are transverse to the rotation axis, transform among themselves, 
and we can diagonalize the action introducing
\be
v^{\pm}:=v^x\pm i v^y, \qquad v'{}^{\pm} = e^{\pm i\th}v^\pm.
\ee
In this way we have identified all 4 eigenmodes of the vector. We can equivalently describe the decomposition in terms of a basis of eigenvectors. If we start from the canonical basis $e^\m_I:=\d^\m_I$, where $I=0,\ldots 3$, then the eigenvectors are
\be\label{vecmodes1}
\eps_{0}^\m=e^\m_0, \qquad \eps_{\sscr L}^\m = e^\m_3, \qquad \eps^\m_{\pm}=\f1{\sqrt 2}(e^\m_1\mp i e^\m_2),
\ee
and their corresponding eigenvalues
\begin{align}\label{vecmodes}
R^\m{}_\n \eps_{0}^\n = \eps_{0}^\m,\qquad R^\m{}_\n \eps_{\sscr L}^\n = \eps_{\sscr L}^\m, \qquad R^\m{}_\n \eps_{\pm}^\n = e^{\pm i\th} \eps_{\pm}^\m.
\end{align}
The integers 0 and $\pm1$ that appear in front of the rotation angle $\th$ are the projection of the spin along the $z$ axis.
Since our convention for vector rotations is anti-clockwise, the $+1$ (resp. $-1$) component can be also referred to as right-handed (resp. left-handed).

Acting on a symmetric tensor $h_{\m\n}$ we obtain $h'_{\m\n}=R^\r{}_\m R^\s{}_\n h_{\r\s}$, where

\begin{subequations}\begin{align}
& h'_{00}= h_{00}, \qquad h'_{0z} =  h_{0z}, \qquad h'_{s} =  h_s,\qquad h'_{zz} =  h_{zz}\\
& h'_{0x} =  \cos\th h_{0x}+\sin\th h_{0y}, \qquad h'_{0y} =  -\sin\th h_{0x}+\cos\th h_{0y}, \\
& h'_{xz} =  \cos\th h_{xz}+\sin\th h_{yz}, \qquad h'_{yz} =  -\sin\th h_{xz}+\cos\th h_{yz},
\end{align}
and
\begin{align}
& h'_{xx}=\cos\th (h_{xx} \cos\th + h_{xy} \sin\th) + \sin\th  (h_{xy} \cos\th + h_{yy} \sin\th),\\
& h'_{xy}=-\sin\th (h_{xx} \cos\th + h_{xy} \sin\th ) + \cos\th (h_{xy} \cos\th + h_{yy} \sin\th), \\
& h'_{yy}=-\sin\th (h_{xy} \cos\th - h_{xx} \sin\th) + \cos\th (h_{yy} \cos\th - h_{xy} \sin\th).
\end{align}
The last three can be disentangled if we define
\be\label{defhpt}
h_+:=\f12(h_{xx}-h_{yy}), \qquad h_\times := h_{xy}, \qquad h_s:= h_{xx}+h_{yy}+h_{zz}.
\ee
Then,
\be
h'_+ =  \cos 2\th h_+ +  \sin 2\th h_\times, \qquad h'_\times =  -\sin 2\th h_+ +  \cos 2\th h_\times,
\ee
and
\be
h^{\pm2}:=\f1{\sqrt{2}}(h_+\mp ih_\times), \qquad h'{}^{\pm2} = e^{\pm 2i\th}h^{\pm2}.
\ee
\end{subequations}
We have thus identified all ten eigenmodes of a symmetric rank-2 tensor under the $z$ rotations.

To write the eigenvectors, we introduce a canonical basis in the space of symmetric $4\times 4$ matrices, 
\be
e_{\m\n}^{IJ}=\left\{\begin{array}{cc}\d_\m^{I}\d_\n^{J} &\quad I=J \\ \f1{\sqrt 2}(\d_\m^{I}\d_\n^{J}+\d_\m^{J}\d_\n^{I}) &\quad I\neq J\end{array}\right.
\ee
Then
\begin{subequations}\label{spin2modes}\begin{align}
& \eps^{\sscr 0}_{\m\n} = e^{00}_{\m\n}, \qquad \eps^{\sscr s}_{\m\n} =\f1{\sqrt 3} (e^{11}_{\m\n}+e^{22}_{\m\n}+e^{33}_{\m\n}),
\nn
\\
& w^{\pm}_{\m\n} =\f1{\sqrt 2}(e^{01}_{\m\n}\pm ie^{02}_{\m\n}),
\qquad  w^{\sscr L}_{\m\n} = e^{03}_{\m\n}, \\
& \eps^{{\sscr T}\pm}_{\m\n} =\f12(e^{11}_{\m\n} -e^{22}_{\m\n})\pm \f i{\sqrt 2}e^{12}_{\m\n},
\qquad
\eps^{{\sscr L}\pm}_{\m\n} =\f1{\sqrt 2}(e^{13}_{\m\n}\pm ie^{23}_{\m\n}),
\nn
\\
&\eps^{\sscr LL}_{\m\n} =\f1{\sqrt 6}  (e^{11}_{\m\n}+e^{22}_{\m\n} - 2e^{33}_{\m\n})
\end{align}\end{subequations}
with eigenvalues
\begin{align}
& R^\r{}_\m R^\r{}_\n \, \eps^{\sscr i}_{\r\s} = \eps^{\sscr i}_{\m\n},\quad {\rm for} \   {i= 0,s},{\text {LL}} 
\qquad R^\r{}_\m R^\r{}_\n \, w^{\sscr L}_{\r\s} = w^{\sscr L}_{\m\n}, \\ \label{spin2hel}
& R^\r{}_\m R^\r{}_\n \, \eps^{\pm}_{\r\s} = e^{\pm 2i\th}\eps^\pm_{\m\n},
\quad R^\r{}_\m R^\r{}_\n \, w^\pm_{\r\s} = e^{\pm i\th} w^\pm_{\m\n},
\quad R^\r{}_\m R^\r{}_\n \, \eps^{{\sscr L}\pm}_{\r\s} = \eps^{\pm i\th} w^{{\sscr L}\pm}_{\m\n}.
\end{align}
The spin-2 components $e^{{\sscr T}\pm}$, $e^{{\sscr L}\pm}$ and $e^{\sscr LL}$  carry respectively the $\pm2,\pm1$ and $0$ modes of the $z$-projection.
The spin-1 components $w^{\pm}$ and $w^{\sscr L}$ carry $\pm1$ and $0$ modes as in \eqref{vecmodes}, and $e^{\sscr 0,s}$ are the remaining two spin-0 modes from \eqref{spindec0}.

The discussion so far concerned global vectors and tensors.
In the case of electromagnetism and  gravity linearized around Minkowski we work with vector and tensor \emph{fields}, $A_\m(x)$ and $h_{\m\n}(x)$. 
The mode decompositions described above are then applied locally in Fourier space. Each Fourier mode 
 $\tl A_\m(k)$ and $\tl h_{\m\n}(k)$ is characterized by a wave vector $k$, and we can use its spatial direction to classify the spin components. 
 With some abuse of language, we will often refer to the wave vector as the momentum of the wave.
 The eigenvalues of  the projection of the spin along the momentum are called \emph{helicities}, hence when this basis is chosen the mode decomposition is called helicity decomposition. The notion of helicity is closely related to the notion of polarization. More precisely, modes of helicity $\pm1$ describe waves of right-handed and left-handed circular polarization respectively,
and different linear combinations can be taken to describe for instance linear or elliptic polarizations. 
The helicity-1 linear polarizations are simply the $x$ and $y$ components, and the helicity-2 linear polarizations are
 \be\label{pol2}
 \eps^{+}_{\m\n} =\f1{\sqrt 2} \left(\begin{array}{cccc} 0&0&0&0\\0&1 & 0 & 0 \\ 0&0 & -1 & 0 \\ 0 & 0 & 0&0\end{array}\right), 
 \qquad \eps^{\sscr\times}_{\m\n} =\f1{\sqrt 2} \left(\begin{array}{cccc} 0&0&0&0\\0&0 & 1 & 0 \\ 0&1 & 0 & 0 \\ 0 & 0 & 0&0\end{array}\right).
 \ee
 The labels stand for `plus' (not to be confused with the plus used in the circular polarizations, and which will not be used in the following) and `cross', and the reason will become clear below.
 
 The helicity decomposition with an arbitrary momentum $\vec k$ can be conveniently described introducing the transverse and longitudinal projectors
\be\label{defTL}
T^a_b:=\d^a_b -\f{k^ak_b}{\vec k^2}, \qquad L^a_b:=\f{k^ak_b}{\vec k^2}.
\ee
Using these, we can decompose a spin-1 vector as follows,
\begin{align}\label{vecTL}
A^a = A_{\sscr T}^a + A_{\sscr L}^a, \qquad A_{\sscr T}^a= P^{\sscr{\bf (1)}}_{\sscr T}{}^a_bA^b, \qquad  A_{\sscr L}^a=P^{\sscr{\bf (1)}}_{\sscr L}{}^a_bA^b,
\end{align}
where
\be
P^{\sscr{\bf (1)}}_{\sscr T}{}^a_b=T^a_b,
\qquad P^{\sscr{\bf (1)}}_{\sscr L}{}^a_b=L^a_b, \qquad P^{\sscr{\bf (1)}}_{\sscr T}+P^{\sscr{\bf (1)}}_{\sscr L}=  P^{\sscr{\bf (1)}};
\ee
and a spin-2 tensor as follows,
\begin{align}\label{tenTL}
& h_{\la ab\ra } = h_{ab}^{\sscr TT} + h_{ab}^{\sscr L} + h_{ab}^{\sscr LL},\qquad h_{ab}^{\sscr TT} =P^{\sscr{\bf (2)}}_{\sscr TT}{}^{cd}_{ab}h_{cd}
\nn
\\
& h_{ab}^{\sscr L} =P^{\sscr{\bf (2)}}_{\sscr L}{}^{cd}_{ab}h_{cd}, \qquad h_{ab}^{\sscr LL} =P^{\sscr{\bf (2)}}_{\sscr LL}{}^{cd}_{ab}h_{cd},
\end{align}
where
\begin{align}
\label{P2}
& P^{\sscr{\bf (2)}}_{\sscr TT}{}^{ab}_{cd}=T^a_{( c} T^b_{d)}-\f12 T^{ab}T_{cd}, \qquad
 P^{\sscr{\bf (2)}}_{\sscr L}{}^{ab}_{cd}=T^a_{( c} L^b_{d)}+T^b_{ (c}L^a_{d)}, \\ &
 P^{\sscr{\bf (2)}}_{\sscr LL}{}^{ab}_{cd}=\f13\left(\f12 T^{ab}T_{cd}+2L^{ab}L_{cd}- T^{ab}L_{cd}-L^{ab}T_{cd}\right),
 \nn
 \\
& P^{\sscr{\bf (2)}} = P^{\sscr{\bf (2)}}_{\sscr TT}+P^{\sscr{\bf (2)}}_{\sscr L}+P^{\sscr{\bf (2)}}_{\sscr LL}.
\nn
\end{align}
We can then identify the helicities of the different projectors studying how they transform under a rotation with axis $\vec k$. For simplicity let us consider the case when $\vec k$ is in the $z$ direction, so that we can use the formulas already derived for the eigenvectors. 
In this case,
\be\label{TLz}
T^a{}_b = \left(\begin{array}{ccc}
1 & 0 & 0 \\ 0 & 1 & 0 \\ 0 & 0 & 0
\end{array}\right), \qquad
L^a{}_b = \left(\begin{array}{ccc}
0 & 0 & 0 \\ 0 & 0 & 0 \\ 0 & 0 & 1
\end{array}\right).
\ee
Then \eqref{vecTL} reduces to $A_a^{\sscr T}=(A_x,A_y,0)$ and $A_a^{\sscr T}=(0,0,A_z)$.
Comparing these to the earlier decomposition \eqref{vecmodes}, we conclude that the transverse projector contains the $\pm1$ helicity modes of a spin-1 field, and the longitudinal projector the $0$ helicity mode. 
The spin-2 projectors based on \eqref{TLz} give
\begin{subequations}\begin{align}
P^{\sscr{\bf (2)}}_{\sscr TT}{}^{ab}_{cd} h^{cd}&=(T  h T)^{ab}-\f12 \tr(Th) T^{ab} 
\nn
\\
&= 
\left(\begin{array}{ccc}\f12(h_{xx}-h_{yy}) & h_{xy} & 0 \\ h_{xy} & -\f12(h_{xx}-h_{yy}) & 0 \\ 0 & 0 & 0 \end{array}\right)
=\left(\begin{array}{ccc}h_{+} & h_{\sscr \times} & 0 \\ h_{\sscr \times} & -h_+ & 0 \\ 0 & 0 & 0 \end{array}\right), 
\label{PTTA}
\\
P^{\sscr{\bf (2)}}_{\sscr L}{}^{ab}_{cd} h^{cd}&= \left(\begin{array}{ccc}0 & 0 & h_{xz} \\ 0 & 0 & h_{yz} \\0 & 0 & 0
\end{array}\right), \qquad
P^{\sscr{\bf (2)}}_{\sscr LL}{}^{ab}_{cd} h^{cd}=\f16(h_{xx}+h_{yy}-2h_{zz}) \left(\begin{array}{ccc}1 & 0 & 0 \\ 0 & 1 & 0 \\ 0 & 0 & -2
\end{array}\right).
\end{align}\end{subequations}
Comparing these to the earlier decomposition \eqref{spin2modes}, we conclude that the spin-2 part of the gravitational perturbation can be decomposed into 5 helicity modes $\pm2,\pm1,0$, which are carried respectively by the  TT,  L and LL components.

The spin-helicity interpretation of the components of $h_{\m\n}$ is a kinematical classification based on a choice of reference frame given by the time direction\footnote{Which we have chosen to coincide with $t$ of the background Cartesian coordinates. However the whole spin-helicity description is Lorentz covariant. The spin projectors for an arbitrary time-direction $\t^\m$, $\t^2=-1$, are
\be\nn
P^{\sscr{\bf (1)}}{}^\m_\n=q^\m_\n:=\d^\m_\n + \t^\m \t_\n, \qquad P^{\sscr{\bf (0)}}{}^\m_\n=- \t^\m \t_\n,\qquad \Id^{\bf (\f12,\f12)} = P^{\sscr{\bf (1)}}+P^{\sscr{\bf (0)}},
\ee
for a vector, and
\begin{align}\nn
& P^{\sscr{\bf (2)}}{}^{\m\n}_{\r\s} =\d^\m_{(\r}\d^\n_{\s)} + \t^\m \t_{(\r}\d^\n_{\s)}-\f12q^{\m\n}q_{\r\s},\qquad P^{\sscr{\bf (1)}}{}^{\m\n}_{\r\s}=-\t^{\m}\t_{(\r} q^{\n}_{\s)},
\\
& P^{\sscr{\bf (0)}}{}^{\m\n}_{\r\s}=\t^\m \t_\r \t^\n \t_\s, \qquad
P_s^{\sscr{\bf (0)}}{}^{\m\n}_{\r\s}=\f12 q^{\m\n}q_{\r\s},\\\nn
&\Id^{{\bf (1,1) \oplus (0,0)}} = P^{\sscr{\bf (2)}}+P^{\sscr{\bf (1)}}+P^{\sscr{\bf (0)}}+P^{\sscr{\bf (0)}}_s,
\nn
\end{align}
for a symmetric tensor. One can also write the helicity projectors as 4d covariant objects encoding $\t$ in a choice of null vector transverse to $k^\m$, see e.g. \cite{Weinberg}.} and the spatial direction $\vec k$.
The next question is which of these components are dynamical. If the field satisfied the simple wave equation $\square h_{\m\n}=0$, then all components would be dynamical, and the field would carry 10 independent degrees of freedom corresponding to all the helicity states described above. 
But this is not the case of \eqref{linE}, because of the additional derivative operators present, and the gauge redundancy. One way to identify the degrees of freedom is to fix the gauge and study the resulting solutions. This is what we do next.

\subsection{De Donder and TT gauges \label{SecDeD}}

The gauge symmetry can be exploited to simplify the linearized field equations.
This can be done using \eqref{diffeoeta} to put the metric perturbation in a form where it satisfies
\be\label{DeDg}
\p_\m h^{\m}{}_\n - \f12\p_\n h=0.
\ee
Doing so eliminates all terms containing divergences. The linearized field equations \eqref{linE} are thus equivalent to 
\be\label{boxbarh}
{\square \bar h_{\m\n} = -\f{16\pi G}{c^4}T_{\m\n}, \qquad \p_\m\bar h^{\m\n}=0, \qquad \bar h_{\m\n}:= h_{\m\n}-\f12\eta_{\m\n}h.}
\ee
The short-hand notation $\bar h_{\m\n}$ is introduced for convenience, and it is known as trace-reversed perturbation, since $\bar{h}=-h$.
The condition \eqref{DeDg} is called De Donder gauge  in the literature, 
but also Lorenz gauge, in analogy with electromagnetism, and harmonic gauge,
because it preserves harmonic coordinates, as can be seen linearizing \eqref{xharm}.
The compatibility of the coupled system is guaranteed by the fact that $\p_\m T^{\m\n}$ vanishes on solutions of the matter field equations, more on this in Section~\ref{SecMatter} below.

The condition \eqref{DeDg} does \emph{not} fix completely the 4-dimensional diffeomorphism symmetry. Intuitively, this occurs because we are not fixing metric coefficients, but only their derivatives. To make the residual freedom explicit, observe that the diffeomorphism required to put an arbitrary metric perturbation in De Donder form is given by a solution of the equation
\be\label{DDc}
\square \xi^\m =- \p_\n h^{\m\n} + \f12 \p^\m h.
\ee
But for a given initial metric, this equation admits infinitely many solutions, parametrized by the zero modes of the d'Alembertian operator. In other words, once the De Donder condition is satisfied, there remains a residual freedom of gauge transformations that satisfy $\square \xi^\m=0$. 
Thus the De Donder gauge contains in fact an infinite family of inequivalent gauge fixings.
For this reason, it may be more appropriate to refer to \eqref{DeDg} not as a gauge fixing condition, but rather a \emph{family} of gauge fixings. 
The simpler albeit vaguer term De Donder gauge is, however, the one in use in most of the literature.

If need be, a unique representative of the De Donder gauge can be specified fixing the solution of $\square \xi^\m=0$ in terms of initial data at a reference $t=0$ hypersurface. Such initial data can always be chosen such that any four components of $ h_{\m\n}$ and their time derivatives vanish there. For vacuum solutions, this means that the chosen components vanish everywhere. It is convenient to apply this procedure and set to zero the components $\bar h$, so that $\bar h_{\m\n}=h_{\m\n}$, and $h_{0a}$. Then the De Donder condition implies that also $\p_a h^{ab}$ and $\p_0 h_{00}$ vanish everywhere. The vacuum equations then imply $\vec\p^2h_{00}=0$, hence the only solution with vanishing boundary conditions is $h_{00}=0$.
We conclude that in this gauge, vacuum solutions satisfy
\be\label{TTvacuum}
h=h_{00}= h_{0a}=\p_a h^{ab}=0, \qquad h_{ab}=h_{ab}^{\sscr TT}.
\ee
This shows explicitly that there are only two independent degrees of freedom.
We call this choice the \emph{transverse-traceless gauge} (TT gauge in short), since in this gauge, vacuum solutions coincide with the 
transverse-traceless perturbations.
We stress that this property and the equations \eqref{TTvacuum} are only valid for vacuum solutions. 
In other words, \eqref{TTvacuum} is not the definition of a gauge condition, but rather the specific value that solutions take in a certain gauge.
The analogue for solutions with sources will be discussed below. Notice also that $h_{0\m}=0$ is the linearized approximation of \eqref{tempgauge}.
Therefore the TT gauge of vacuum solutions implies the temporal gauge, hence the TT coordinates describe free-falling observers.

In the presence of sources on the other hand, even if we set to zero 4 components and their time derivatives on the chosen initial hypersurface, they will no longer be zero in the future causal domain of the sources. In this case there are more convenient choices to specialize the De Donder gauge condition, as we will discuss below.

\subsection{Vacuum solutions}


Let us flesh out these considerations by looking at the explicit form of the vacuum solutions. This will make it clear that the identification of the degrees of freedom with the transverse-traceless modes is a gauge-invariant statement. The vacuum equations can be solved straightforwardly taking
linear combination of plane waves via the Fourier transform
\be
h_{\m\n}(x) = \re \int d^4k \,\tl h_{\m\n}(k) e^{ik \cdot x}.
\ee 
Imposing the vacuum equations and the De Donder condition  \eqref{boxbarh} requires
\be
k^2 = 0, \qquad k^\m \tl h_{\m\n}=\f12 k_\n \tl h.
\ee
The first equation is solved by $k^0=\pm|\vec k\, |$, and we denote $\om:=k^0c>0$ for a future-pointing 4-momentum, thus
\be
k\cdot x=-\om(t-z/c).
\ee
The second equation gives a linear system of 4 conditions on the 10 components of the matrix. 
In the frame where $\vec k$ is along the $z$ axis, we take as independent components 
\be
\tl h_{0\m}, \qquad 
 h_{\sscr +}=\tl h_{xx}, \qquad h_{\sscr\times}=\tl h_{xy},
\ee
and then the system is solved by 
\be\label{gtransf1}
\tl h_{xz} = -\tl h_{0x}, \qquad \tl h_{yz} = -\tl h_{0y}, \qquad \tl h_{zz} = -\tl h_{00}-2\tl h_{0z}, \qquad \tl h_{yy} = -\tl h_{\sscr +}.
\ee
Therefore the general solution for a real, monochromatic wave with frequency $\om$ and propagating along the $z$ axis is a 6-parameter family given by
\be\label{PWgauge}
h_{\m\n}(x) = \left(\begin{array}{cccc}
\tl h_{00} & \tl h_{0x} & \tl h_{0y} & \tl h_{0z} \\  \qquad & h_{\sscr +} & h_{\sscr\times} & -\tl h_{0x} \\ & \qquad & -h_{\sscr +} & -\tl h_{0y} \\ & &\qquad & -\tl h_{00}-2\tl h_{0z}
\end{array}\right) \cos(k\cdot x).
\ee

The solution, however, contains gauge redundancy. Recall in fact that there is residual gauge freedom in the form of diffeomorphisms $\xi^\m$ that satisfy the vacuum wave equation. In Fourier space the gauge transformation \eqref{diffeoeta} read $\d_\xi \tl h_{\m\n} = 2ik_{(\m}\tl \xi_{\n)}$, where $\tl\xi^\m(k)$ is the Fourier transform of $\xi^\m(x)$ with $k^2=0$ in order to satisfy the vacuum wave equation and be an admissible residual gauge. Under the residual gauge transformation,
\be
\tl h_{0\m} \to \tl h_{0\m} + i k_0 \tl\xi_\m + i k_\m \tl\xi_0, \qquad h_{\sscr +,\times} \to h_{\sscr +,\times}.
\ee 
The second property follows immediately from the fact that $k_\m$ has no transverse components.
We conclude that only the two components $h_{\sscr +,\times}$ are gauge-invariant and thus physically relevant. These can be recognized as the TT components. 
The four coefficients $\tl h_{0\m}$ can be changed arbitrarily, and have no physical meaning. 
We can in particular set them to zero simultaneously, choosing $i\tl \xi_\m= \f c\om(\f12\tl h_{00}, \tl h_{0x}, \tl h_{0y}, \tl h_{0z}-\f12\tl h_{00})$. Then
\be\label{PWsol}
h_{\m\n}(x) 
= \left(\begin{array}{cccc}
0 & 0 & 0 & 0 \\ 0 & h_{\sscr +} & h_{\sscr \times} & 0 \\ 0 & h_{\sscr \times} & -h_{\sscr +} & 0 \\ 0 & 0 & 0 & 0
\end{array}\right) \cos\big(\om(t-z/c)\big).
\ee
We have thus achieved the TT form \eqref{TTvacuum}. 
In this gauge, the only components of the vacuum solutions are the gauge-invariant transverse-traceless ones. Accordingly, the solution can also 
be written as a sum over the two TT polarization tensors \eqref{pol2}, namely
\be\label{hsumpol}
\tl h_{\m\n}(k) = \sum_{p}  A_p(k) \eps^p_{\m\n}(k),
\ee
where $p=+,\times$ and $A_p=h_p$.\footnote{The polarization decomposition can also be used to solve the wave equation. 
In this approach one writes \eqref{hsumpol} with the sum including all ten polarization modes  \eqref{spin2modes} and ten arbitrary coefficients $A_i$.
Then the relations \eqref{gtransf1} between Cartesian components are replaced by relations between the polarization coefficients. The algebra is slightly more involved but the end result is the same. We will see the use of the the polarization mode procedure on cosmological backgrounds in Sec~\ref{sec:cosmology}.}

The resulting perturbed metric is
\ba
ds^2 &=&-dt^2+(1+h_{\sscr +}\cos k\cdot x)dx^2 + (1-h_{\sscr +}\cos k\cdot x)dy^2 
\nn
\\
&& \qquad \qquad \qquad \qquad  + 2h_{\sscr \times}\cos k\cdot x \, dx dy+dz^2.
\label{linmetric}
\ea

Let us pause for a short historical digression, about some of the confusion that hindered historically the understanding of gravitational waves. 
Let us consider any member of the general solution \eqref{PWgauge} 
with $h_{\sscr +}=h_{\sscr \times}=0$. This looks like a genuine wave, and it is a solution of the linearized Einstein's equations. However, it 
is a pure gauge solution, and can be set to vanish identically without loss of physical information. In other words, there are wave solutions which are in the end only coordinate artefacts. For the same reason, it is also possible to change coordinates so that the argument $ct-z$ in its cosine is replaced by $vt-z$ for an arbitrary constant $v$. Hence the pure gauge modes don't really propagate, and if a gauge is chosen so that they look like they are propagating, well one can do this with an arbitrary speed, the speed is not constrained in any way by the dynamics. To use Eddington's words, the non-physical gauge modes propagate at the ``speed of thought". This initial confusion was clarified by the identification of gauge-invariant components, and the fact that for the physical modes, the propagation speed is fixed to be the speed of light by Einstein's equations. However, additional doubts persisted, because the metric \eqref{linmetric} is \emph{not} a solution of the exact Einstein's equations; only of the linearized theory. In other words, there are vacuum solutions that take approximately the form \eqref{linmetric} in some regions of spacetime, but none that has that exact form everywhere in spacetime. This raised the issue of whether gravitational waves existed in the full theory, or were only an artefact of the linearized approximation.
This more complicated issue was solved only much later, when a non-perturbative identification of the wave degrees of freedom and their energy was made possible by the work of Bondi, Sachs and many others \cite{Bondi:1960jsa,Bondi:1962px,Sachs:1962wk,Newman:1968uj,Ashtekar:1981bq,Dray:1984rfa,Wald:1999wa}. %
For a historical overview, see \cite{Kennefick:1997kb}.

 Coming back to the physical solution \eqref{PWsol}, we see that the resulting wave tensor is transverse to the direction of propagation.
Comparing with \eqref{spin2modes}, we conclude that the physical gauge-invariant modes have helicity $\pm2$. All modes of  helicity $\pm 1$ and $0$ have dropped out, either by gauge fixing, or by solving the vacuum field equations. 
Because the physical modes have helicity $\pm2$, it takes a rotation of an angle $\pi/4$ to turn one mode into the other. This can be compared with the electromagnetic case, whose modes have helicity $\pm1$, and it takes a rotation by $\pi/2$ to turn one mode into the other, see Fig.~\ref{Figpols}.\footnote{The general formula is that two modes of helicity $\pm j$ are related by $\pi/2j$.} For the electromagnetic case, this angle $\pi/2$ corresponds to the orthogonality of the oscillations of the electric and magnetic fields. For the gravitational case, it corresponds to the antipodal symmetry in tidal forces. We will see below in Section~\ref{sec:detectionofGWs} how this intuition can be made precise by studying the effect of a gravitational wave on test particles, see in particular Fig.~\ref{Figpols}.

The solution \eqref{PWsol} describes a monochromatic wave. The most general solution has
$\infty^3$ Fourier components (one per choice of $\vec k$), and two independent degrees of freedom per component (the values of $h_{\sscr +,\times}$), therefore it 
is described by $2\times \infty^3$ arbitrary numbers. These are the independent degrees of freedom of gravitational waves. It is the same number of the full theory, so the linearized approximation simplifies the dynamics but preserves the number of independent variations of the gravitational field that can occur in the full theory. 
Notice that the number of independent degrees of freedom is the same of electromagnetism, or of two scalar fields. Apart from the dynamical behaviour, what changes is also the behaviour of these degrees of freedom under Lorentz transformations, because of their different spins.

The analysis has also shown that only the transverse-traceless modes are gauge invariant. We did this using the partial gauge fixing provided by the De Donder condition, but it can be proved in full generality starting from the projector $P^{\sscr {\bf (2)}TT}$ on transverse-traceless modes defined in \eqref{P2}, and observing that it annihilates gauge transformations.  Since we are going to use this projector often, we drop the label ${\bf (2)}$ from now on. We also introduce the notation $\hat k=\vec k/k^0$, using which we can write the explicit form of the projector as
\ba
P^{\sscr TT}{}^{ab}_{cd}(\hat k) 
&=& \d^a_{(c}\d^b_{d)} - \f12\d^{ab}\d_{cd} - \d^a_{(c}\hat k^b\hat k_{d)}
- \hat k^a\hat k_{(c}\d^b_{d)} 
\nn
\\
&& \qquad \qquad \qquad + \f12 ( \d^{ab}\hat k_c\hat k_{d}+\hat k^a\hat k^b\d_{cd}+ \hat k^a \hat k^b\hat k_{c}\hat k_{d}).
\label{eq:projector}
\ea
Note that the symmetrization on the indices here and in \eqref{P2} is  omitted in some books \cite{poisson,Maggiore1}, under the premises that one is applying it to symmetric tensors only anyway. 
In conclusion, we do not need to use the TT gauge, nor even the De Donder gauge, in order to identify the independent degrees of freedom. Whatever gauge we are using, we can always extract them via
\be\label{PTT}
h^{\sscr TT}_{ab} = P^{\sscr TT}{}_{ab}^{cd}h_{cd},
\ee
and the result of this projection is gauge invariant. 
If we align the frame so that $\hat k$ coincides with the $z$ axis, it is given by \eqref{PTTA}.
To treat a general direction, we parametrize it using polar coordinates $(\th,\varphi)$ on the sphere, and write
\be\label{hatk}
\hat k(\th,\varphi) = (\sin\th\cos\varphi,\sin\th\sin\varphi,\cos\th)  = R_{\hat z}(\varphi)R_{\hat y}(\th)\hat e_z=: R(\th,\varphi)\hat e_z.
\ee
We then pick a basis in the plane orthogonal to $\hat k$, given by
\be\label{mbasis}
\hat m_1 = (\cos\th\cos\varphi,\cos\th\sin\varphi, -\sin\th), \qquad \hat m_2= (-\sin\varphi,\cos\varphi,0).
\ee
We use this basis to distinguish the two polarizations, and define
\be\label{hTTk}
h^{\sscr TT}_{ab} = (\hat m_{1a}\hat m_{1b}-\hat m_{2a}\hat m_{2b})h_+ + (\hat m_{1a}\hat m_{2b}+\hat m_{2a}\hat m_{1b}) h_{\times}.
\ee
We obtain in this way
\begin{subequations}\label{hgendir}\begin{align}
h_+ &= \f12\Big(h_{xx}(\cos^2\th\cos^2\varphi-\sin^2\varphi) + h_{yy}(\cos^2\th\sin^2\varphi-\cos^2\varphi) + h_{zz}\sin^2\th  \nn\\
& \hspace{2cm} + h_{xy}(1+\cos^2\th)\sin 2\varphi - h_{xz}\cos\varphi\sin 2\th - h_{yz}\sin\varphi\sin 2\th  \Big),
\\ 
h_\times &= h_{xy}\cos 2\varphi\cos\th + h_{xz}\sin\varphi\sin\th- h_{yz}\cos\varphi\sin\th - \f{h_{xx}-h_{yy}}2\sin 2\varphi\cos\th,
\end{align}\end{subequations}
which coincides with (11.46) of \cite{poisson}.
If the wave is travelling along the $x$ axis for example, then $(\th,\phi)=(\f\pi2,0)$, and
$h_+=\f12(h_{zz}-h_{yy})$, $h_\times=-h_{yz}$.

\begin{figure}[ht]\centering
  \includegraphics[width=9.5cm]{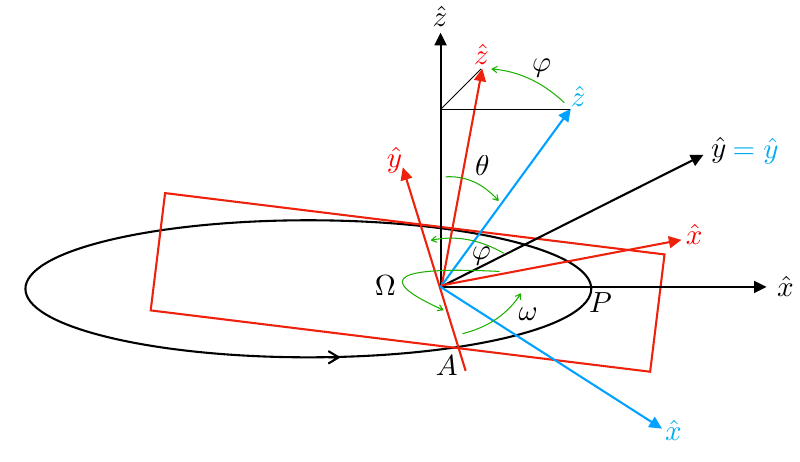}
  \caption{\small\emph{Rotation between the orbital plane of a binary system and the direction of the observer. The black reference frame is adapted to the orbit, with $\hat z$ aligned with the angular momentum, and $\hat x$ with the pericenter $P$. The first rotation along $\hat y$ by an angle $\theta$ rotates the frame to the blue one, and the second rotation along $\hat z$ by an angle $\varphi$ rotates it to the final red frame. The red frame has $\hat z$ pointing in the direction of the observer, and $\hat y$ axis along the line of nodes, pointing towards the descending node. In the red frame, the longitude of the ascending node $A$ is $\Om=3\pi/2$, and the longitude of pericenter is $\om$.}} 
   \label{FigNodes}
\end{figure}

We can also use this formula to relate the polarizations in an arbitrary direction $\hat k$ to the polarizations in the $\hat z$ direction. In this case
$h_{xx}=-h_{yy}=h_+(\hat z)$ and $h_{xy}=h_\times(\hat z)$, while all other components vanish. The expressions then simplify to 
\begin{subequations}\label{hgendir2}\begin{align}
h_+(\hat k) &= \f{1+\cos^2\th}2(h_+(\hat z)\cos 2\varphi + h_\times(\hat z)\sin 2\varphi ), \\
h_\times(\hat k) &= \cos\th(h_\times(\hat z)\cos 2\varphi - h_+(\hat z) \sin 2\varphi).
\end{align}\end{subequations}
In physical applications, the coordinate system may be set from characteristics of the system, for instance for binaries, adapted so that the orbital plane coincides with the $(x,y)$ plane, and $x$ is in the direction of the pericenter. Then our rotation identifies $\th$ with the inclination $\iota$ of the orbital plane, and the $y$ axis is along the `line of nodes', namely the intersection between the orbital plane and the plane of sight, pointing in the direction of the descendent node. On the other hand, astronomers often choose the convention to align the line of nodes along the $x$ axis, pointing in the direction of the ascendent node. This means that in our reference system, the `longitude of ascending node' $\Om$ is $-\pi/2$, and the `longitude of pericenter' $\om$ taken from the ascending node is $\varphi-\pi/2$.
That is,
\be
\iota=\th,\quad \om=\f\pi2-\varphi, \quad \Om=-\f\pi2.
\ee
This maps our basis (\ref{hatk}-\ref{mbasis}) to (3.44-3.45) of \cite{poisson}. See Fig.~\ref{FigNodes}.

\subsection{Gauge-invariant description: independent and constrained degrees of freedom}
\label{ssec:dofcount}

The  analysis of vacuum solutions has allowed us to identify the independent degrees of freedom of the gravitational field. There are also dependent degrees of freedom, namely components of the field that are gauge invariant, but uniquely determined from the sources. These can be exposed looking at the constraints, namely the $00$ and $0a$ components of \eqref{linE} which give rise to elliptic equations, as opposed to hyperbolic ones. The intricacies of the tensorial structure make it however convenient to do first a kinematical analysis of gauge invariance using the transverse and longitudinal projectors.
To that end, we first observe that the projectors can be described also in configuration space without doing the Fourier transform. In this case \eqref{defTL} is replaced by
\be\label{TLconf}
T^a_b=\d^a_b -\f{\p^a\p_b}{\vec \p^2}, \qquad L^a_b:=\f{\p^a\p_b}{\vec \p^2}.
\ee
These expressions are somewhat implicit because one needs to specify boundary conditions in order to have a well-defined inverse of the Laplace operator $\vec\p^2$. Requiring the fields to vanish at spatial infinity makes this definition equivalent to the one in momentum space.
Notice that the transverse-longitudinal decomposition in configuration space can be recognized as the Helmholtz decomposition of a 3d vector field into solenoidal and irrotational parts. This representation also makes it clear that the projectors are local operators in Fourier space, but non-local in spacetime. It has the important consequence that it is not possible to identify exactly the gauge-invariant modes with local observations only.

It is also convenient to dispose of the projectors for the components with non-maximal helicities, by introducing auxiliary fields with smaller spin. 
For instance, we denote $W_a:=h_{0a}$ the spin-1 part of the gravitational perturbation, and write its longitudinal part as the gradient of a scalar:
\be
W^{\sscr L}_a := P^{\sscr {\bf (1)}}_{\sscr L}{}^a_bW^b= \p_a W, \qquad W=\vec \p^{-2}\p_aW^a.
\ee
Similarly for the spin-2 part, we can write the mixed transverse-longitudinal and fully longitudinal modes introducing a transverse vector $B^a$ and a scalar $B$, 
\begin{align}
& P^{\sscr{\bf (2)}}_{\sscr L}{}^{ab}_{cd} h^{cd} = 2\p^{(a} T^{b)}_c  \f{\p_d h^{cd}}{\vec \p^2}
= 2\p^{(a}B^{b)}, 
\qquad B^b:=\f{2}{\vec\p^2} T^{b}_c\p_d h^{cd},
\nn
 \\
& P^{\sscr{\bf (2)}}_{\sscr LL}{}^{ab}_{cd} h^{cd} = \f32\left(\f{\p^a\p^b}{\vec \p^2} -\f13 \d^{ab}\right)\f{\p_{\la c}\p_{d\ra}}{\vec \p^2}h^{cd}
= (\p_a\p_b-\f13\d_{ab}\vec\p^2)B, 
\nn
\\
& B:=\f32\f{\p_{\la c}\p_{d\ra}}{\vec\p^4}h^{cd}.
\nn
\end{align}

Let us summarize. 
With this new notation at hand, the gravitational perturbation around Minkowski can be decomposed into
\begin{subequations}\label{hdec}\begin{align}
& {\rm spin\ } 0 \qquad h_s   \\
&{\rm spin\ } 0 \qquad h_{00} \\
&{\rm spin\ } 1\qquad h_{0a}=W_a=W_a^{\sscr T}+\p_a W \label{hspin1}\\
&\smash{ {\rm spin\ } 2 \qquad h_{ab} = h_{ab}^{\sscr TT}+2\p_{(a}B_{b)} +(\p_a\p_b-\f13\d_{ab}\vec\p\,^2)B+\f13\d_{ab}h_s,}\label{hspin2}
\end{align}\end{subequations}
with
\begin{subequations}\begin{align}
& W^{\sscr T}_a = T^b_a W_b \qquad {\rm helicity} \pm1\\
& 
W=\vec \p^{-2}\p_aW^a\qquad {\rm helicity\ } 0, 
\end{align}\end{subequations}
and
\begin{subequations}\begin{align}\label{hTTdef}
& \hspace{3cm} h_{ab}^{\sscr TT}:=P^{\sscr{\bf (2)}}_{\sscr TT}{}^{ab}_{cd} h^{cd}= \left(T^a_{ c} T^b_{d}-\f12 T^{ab}T_{ cd}\right)h^{cd} \qquad {\rm helicity} \pm2 \\
& \hspace{3cm} B^b:=\f{2}{\vec\p^2} T^{b}_c\p_d h^{cd}\qquad {\rm helicity} \pm1 \\ 
& \hspace{3cm} B:=\f32\f{\p_{\la c}\p_{d\ra}}{\vec\p^4}h^{cd}\qquad {\rm helicity}\ 0
\end{align}\end{subequations}

Next, we look at the behaviour of these different helicities under gauge transformations. 
We have used the Poincar\'e symmetry of the Minkowski background to organize the ten components of $h_{\m\n}$ in terms of spin and helicity. 
But Minkowski is not invariant under the general diffeomorphism symmetry \eqref{diffeoeta}, hence there is no reason to expect that this decomposition be gauge-invariant. Consider for instance the spin-0 part $h_{00}$: this is just a metric component, and manifestly not invariant under diffeomorphisms.
Only the TT component is gauge-invariant, as can be seen explicitly replacing $h_{\m\n}$ in \eqref{hTTdef} with its gauge transformation \eqref{diffeoeta} and observing that it vanishes thanks to the transverse projection on both indices. This is the gauge-invariance of the two TT components already observed in (the second equation in) \eqref{gtransf1}.
The remaining components transform as follows:
\begin{align}
& h_{00}'=h_{00}+2\p_0\xi_0, \qquad h_{s}'=h_{s}+2\vec\p^2\xi^{\sscr L},  \qquad \xi^{\sscr L}:=\vec\p^{-2}\p_a\xi^a,
\nn
\\
& W_a^{\sscr T}{}' = W_a^{\sscr T}+\p_0\xi_a^{\sscr T}, \qquad W'=W+\xi_0+\p_0\xi^{\sscr L},
\nn
\\
& B_a'=B_a +\xi^{\sscr T}_a, \qquad B'=B + 2 \xi^{\sscr L}.
 \nn
\end{align}
It is possible to combine them to find a maximal number of 6 gauge-invariant quantities, given by
\begin{align}
& h_{ab}^{\sscr TT}, \qquad
c^{-2} \Phi:=-\f12 h_{00} +\p_0 W - \f12\p_0^2 B, 
\nn
\\
& c^{-3} \Phi_a:=\f14(W^{\sscr T}_a - \p_0 B_a), \qquad 
c^{-2} \Psi:= \f16(\vec\p^2 B-h_s ). 
\label{gicomp}
\end{align}
The numerical factors and powers of $c$ have been chosen for later convenience.
We will also use a dot for the derivative with respect to the time coordinate $t=x^0/c$, e.g. 
$\p_0 W=c^{-1}\dot W$.
Our conventions for the physical dimensions are summarized as follows: 
\begin{align}\nn
& [x^\m]={\tt m}  \qquad [t=x^0/c]={\tt s}  \qquad [\p_\m]={\tt m^{-1}} \qquad [\p_t=c\p_0]={\tt s^{-1}} \\\nn
&  [g_{\m\n}]=[h_{\m\n}]=[h]=[h_s]=[W^{\sscr T}]=1 \\& [W]= [B_a]={\tt m} \quad [B]={\tt m^2} \qquad [\Phi] = [\Psi] = {\tt m^2s^{-2}} \qquad [\Phi_a]={\tt m^3s^{-3}}\nn
\end{align}

Inserting the parametrization \eqref{hdec} in the linearized Einstein's equation \eqref{linE} one finds that gauge dependent quantities drop out and only the gauge-independent ones remain. This allows us to decouple the tensorial equations into two sets, an hyperbolic one featuring the d'Alambertian operator alone, and an elliptic one featuring the Laplace operator alone:
\begin{subequations}\label{linEgi}\begin{align}
& \square h_{ab}^{\sscr TT} = -\f{16\pi G}{c^4}\s_{ab}, \\\label{Newt}
& \vec\p^2\Psi= {4\pi G}\r, \qquad \vec\p^2\Phi_a = {4\pi G}s_a, \qquad \vec\p^2(\Phi-\Psi)= \f{12\pi G}{c^2} \left(\dot s+\f13\t \right).
\end{align}\end{subequations}
The sources on the right-hand side of these equations are the components of $T_{\m\n}$ projected in the same way as \eqref{hdec}, namely 
\ba
&T_{00}=c^2 \r, \qquad T_{0a}=c(s_a+\p_a s), 
\nn
\\
 &T_{ab}=\s_{ab}+2\p_{(a}\s_{b)}+\left(\p_a\p_b-\f13\d_{ab}\vec \p^2\right)\s+\f13\d_{ab}\t.
 \nn
\ea

Rewriting the linearized Einstein equations \eqref{linE} in the equivalent form \eqref{linEgi} makes their three-sided structure, mentioned in subsection \ref{ssec:gencov}, 
manifest. Four equations were redundant and have dropped out, hence the field has four undetermined components, which can be assigned arbitrarily choosing a specific gauge. Four equations are elliptic, and describe constrained degrees of freedom, namely dynamical components of the gravitational field which are uniquely determined by the sources. Finally, two equations are hyperbolic, hence they contain free data, and describe how these independent degrees of freedom propagate and react to the sources. This analysis therefore establishes that  the gravitational field has two independent degrees of freedom, which are carried by the two $h_{ab}^{\sscr TT}$ components of the metric. These are gauge-invariant, and describe the $\pm2$ helicities of a spin-2 wave.

As for the remaining gauge-invariant components, we have seen that they satisfy Poisson equations, hence these are degrees of freedom that are entirely determined by the sources. For this reason they are sometimes called `Coulombic' degrees of freedom.
Their meaning can be elucidated looking at the post-Newtonian expansion, in which sources are moving slowly with respect to the speed of light.
To begin with, let us first consider perfectly static sources.

For static sources in a given frame, $T_{00}=c^2\r$ is the only non-vanishing component on the right-hand side of Einstein's equations. Then the second and third equations in \eqref{Newt} imply\footnote{Assuming trivial boundary conditions, see App~\ref{app:Green}.}  $\Phi_{a}=0$ and $\Psi=\Phi$, whilst the first gives Newton's equation, and we can identify $\Phi$ with Newton's potential. This is also the way in which one fixes the coupling constant of the full Einstein's equations in terms of $G$ and $c$.
How about the other potentials? They are sourced by moving bodies, and their existence is a consequence of relativistic invariance, akin to the electromagnetic occurrence of the vector potential next to the Coulomb potential. They describe effects that collectively go under the name of  `gravito-magnetism'.
These include additional contributions to the precessions of equinoxes, light bending, and frame dragging or Lense-Thirring effect.
Their effect can be studied looking at the Lagrangian for a massive test particle, which gives
\ba
{\cal L} & = &-mc\sqrt{-g_\mn u^\mu u^\nu} =  -mc\sqrt{-(\eta_\mn + h_\mn) u^\mu u^\nu} 
\nn
\\
&=&  -mc^2 +\f12mv^2-m\Phi +
\nn
\\ && \; +\f m{c^2}\left(\f18v^4 + \f12\Phi^2 - \f12v^2\Phi - \Psi v^2 + 4\Phi_av^a + c^2h^{\sscr TT}_{ab}v^av^b\right)+O(c^{-4})
\nn
\ea
where $v^a = dx^a/dt$.
The term in round bracket is the \emph{first post-Newtonian correction}. 
As we will see below, $h^{\sscr TT}\sim c^{-4}$, hence the last term there is higher order: dissipative effects for massive particles only appear at 2PN.

\subsection{Gauge-fixing with sources}

Solving the decoupled equations \eqref{linEgi} determines the gauge-invariant quantities \eqref{gicomp} in terms of the sources and the initial conditions.
These solutions \emph{do not determine a metric}. Doing so requires the additional step of  specifying the coordinates to be used. Only after the coordinates are given, or in other words only after a gauge is chosen, the physical degrees of freedom can be described in terms of a metric tensor.

The gauge-invariant approach is conceptually satisfying because it identifies the physical degrees of freedom and decouples the equations, making them easier to solve in principle. However, it is very limited in applicability. Firstly, the decoupling and simple identification of gauge-invariant quantities occur only for very special backgrounds, such as flat spacetime or homogeneous and isotropic.\footnote{On a general background, one can still build the analogue of the spin-helicity decomposition replacing the partial derivatives with covariant derivatives, although care is needed to invert the Laplacian and handle its non-commutativity with the covariant derivative. However, the metric now enters explicitly the decomposition of the energy-momentum tensor, hence the decoupling will be lost in general. The non-commutativity of covariant derivatives also hinders the identification of gauge-invariant quantities.} Secondly, even when the background is Minkowski, a general identification of gauge-invariant quantities at a fixed order in perturbation theory is only possible at the linear level,  as explained earlier. 

To go beyond these limitations, it is easier to put to the side the gauge-invariant description, and work instead in a fixed gauge.
In the gauge-fixed approach, one chooses coordinates that impose restrictions on the metric, and then solves for individual metric components in that gauge, like we did in Section~\ref{SecDeD}. 
Namely we do not solve \eqref{linEgi}, but the original system \eqref{linE}, coupled to additional equations fixing the gauge.
The additional equations remove 
the problem that the field equations are redundant and do not determine all metric components.

A simple example of gauge fixing is the temporal gauge
\be\label{htempgauge}
h_{00} = h_{0a} = 0.
\ee
This is the linearized version of the non-perturbative temporal gauge described in Section~\ref{ssec:gauge}, and provides a complete gauge fixing of the 4-dimensional diffeomorphism symmetry. In this gauge the only non-trivial components of the metric are the spatial ones, and \eqref{gicomp} reduces to
\be
c^{-2} \Phi= - \f12\p_0^2 B, \qquad c^{-3} \Phi_a=-\f14 \p_0 B_a, \qquad c^{-2} \Psi= \f16(\vec\p^2 B-h_s ). 
\ee
All gauge-invariant potentials are encoded in the components of the spin-2 mode \eqref{hspin2}.
A related choice would be to replace the condition on $h_{00}$ with the trace condition $h=0$. In this case
\ba
&&c^{-2} (\Phi+3\Psi)=-h_{00} + \f12\square B, \qquad c^{-3} \Phi_a=-\f14 \p_0 B_a, 
\nn
\\
&& c^{-2} (\Phi-3\Psi)= -\f12(\p_0^2+\vec\p^2) B.
\label{tracelesspotentials}
\ea
This time it is the four metric components $B$, $B_a$ and $h_{00}$ that are fixed in terms of the sources, while $h_{0a}$ and $h$ vanish.

Another simple gauge fixing is the gravitational equivalent of the Coulomb gauge, defined so that both spin-2 and spin-1 parts are purely transverse, namely
\be\label{CoulGF}
\p_a h^{a}{}_{0}=0, \qquad \p_a h^{a}{}_{b} = \f13\p_b h_s \quad \Leftrightarrow\quad B_a=B=W=0.
\ee
In this gauge the non-trivial components of the metric are
\begin{align}\label{gfCoulomb}
h_{00}, \qquad h^{\sscr T}_{0a}=W_a^{\sscr T}, \qquad h_{ab}=h_{ab}^{\sscr TT}+\f13\d_{ab}h_s,
\end{align}
and their relation to the gauge invariant potential is given by
\be
c^{-2} \Phi=-\f12 h_{00}, \qquad  c^{-3} \Phi_a=\f14W^{\sscr T}_a, \qquad c^{-2} \Psi= -\f16h_s. 
\ee

These examples have the advantage that the gauge-invariant potentials are identified with individual components of the metric. On the other hand, the radiative gauge-invariant modes $h^{\sscr TT}_{ab}$ are complicated functions of the metric components. 
When fixing the coordinate gauge, one should keep in mind that some choices may be better than others, as we discussed with the Schwarzschild example in Section~\ref{ssec:gauge}.\footnote{If $GM/c^2\ll r$ in the static spherical coordinates, we can treat it as a perturbative solution with $h_{\m\n}= 2GM/(c^2r)(\d^t_\m\d^t_\n+\d^r_\m\d^r_\n)$. We then see that these coordinates correspond to the Coulomb gauge fixing. Changing coordinates so to have temporal gauge with $h_{00}=0$ would 
push the Newton potential in the $B$ component, and introduce a dependence on the $t$ coordinate, as mentioned earlier.} 
The examples above are non-covariant with respect to the Lorentz symmetry of the background, because they make reference to a given time foliation, and treat time and space components differently. In the presence of radiation, it is best to use a covariant gauge, because it simplifies the analysis of the solutions. 
We can, in fact, remark that while the above examples simplify the relation between the potentials and individual metric components, the remaining field equations for the propagating degrees of freedom are complicated. Whereas with the covariant De Donder gauge, all field equations took the simpler form \eqref{boxbarh}.
However while the De Donder condition \eqref{DeDg} can be also imposed in the presence of sources, we can no longer select a unique representative satisfying the TT condition \eqref{TTvacuum} globally. If need be, a unique representative of the De Donder gauge in the presence of sources can be specified as follows.
The general solution of \eqref{boxbarh} is
\be
 \bar h_{\m\n} = -\f{16\pi G}{c^4}\int d^4x' G(x,x')T_{\m\n}(x') + \bar h_{\m\n}^\circ, 
\ee
where $G(x,x')$ is the d'Alembertian's Green function (see Appendix~\ref{app:Green}), $\bar h_{\m\n}^\circ$ any solution of the homogeneous equation, and the gauge condition is maintained via $\p_\m T^{\m\n}=\p_\m\bar h^{\circ \m\n} = 0$. Since both $\bar h_{\m\n}^\circ$ and the residual diffeomorphism parameters $\xi^\m$ satisfy the vacuum wave equation, we can use the residual freedom to set to zero 
any four components of $\bar h^\circ_{\m\n}$. For instance, we can choose $\bar h^\circ=0$, so that $\bar h^\circ_{\m\n}=\bar h^\circ_{\m\n}$, as well as $\bar h^\circ_{0a}=0$. Then the De Donder condition implies that four more components of the homogeneous solution also vanish everywhere, specifically 
$\bar h^\circ_{00}$ and $\p_b \bar h^{\circ ab}$. At this stage the gauge is completely fixed, and the only components left in the homogeneous solution are  the gauge-invariant ones $\bar h^{\circ\sscr TT}_{\m\n}$, which carry the independent degrees of freedom of gravitational waves. The general solution is
\be\label{barhgen}
 \bar h_{\m\n} = -\f{16\pi G}{c^4}\int d^4x' G(x,x')T_{\m\n}(x') + h^{\circ\sscr TT}_{\m\n}.
\ee
It is a complete gauge fixing that singles out a unique element of the De Donder family, and reduces to \eqref{TTvacuum} for  vacuum solutions.  For generic sources, all components of the metric perturbation are non-zero in this gauge.\footnote{More precisely, in the region causally connected to the sources, since the retarded Green function vanishes outside the light cone. So if the sources are present at all times, the metric perturbation is non-zero everywhere, whereas if the sources are `turned on' at some initial time, then the perturbation vanishes outside the causal domain of the sources from that initial time.} 

Alternatively, it is also possible to select a unique representative of the De Donder gauge requiring that the metric components $h_{00}=h_{0a}$ and their derivatives vanish on a given initial value surface, as considered in \cite{WaldBook}. However these metric components will remain zero only in the region outside the causal domain of the sources from the initial value surface. 
This prescription thus achieves a TT gauge valid also in the presence of sources, but only in a finite region of spacetime. 
A similar construction of TT gauge in a local region of finite time outside the sources is presented in \cite{Flanagan:2005yc}. While interesting in principle, the `local TT gauge' seems to be of minor practical use. The procedure described in the previous paragraph achieves instead a structure of the solutions valid globally and at all times, and it is the analogue of the complete gauge-fixing used for instance in electro-magnetism by the Lienard-Wiechert potentials of a moving charge.

A special situation occurs if the sources are static. In this case, it is possible to specialize the De Donder gauge to the Coulomb gauge \eqref{CoulGF}.
To prove this, we first observe that when rewritten in terms of the gauge-invariant quantities \eqref{gicomp}, 
the De Donder condition  \eqref{DeDg}  implies
\begin{align}\label{eldor}
\square W = -\f4{c^3}\dot\Psi, \qquad \square B = \f2{c^2}(\Phi - \Psi), \qquad  \square B_a = \f4{c^4}\dot\Phi_a.
\end{align}
For static sources $\Phi_a=\dot\Phi=0$ and $\Psi=\Phi$, hence all right-hand sides vanish. It is then possible to specialize the De Donder gauge choosing $W=B=B_a=0$ everywhere. Having done so, $h_{00}=-2c^{-2}\Phi=h_s/3$.
We have thus achieved the Coulomb gauge described earlier, as opposed to the TT gauge, and again this is a complete gauge fixing singling out a member of the De Donder family.
In other words, the De Donder gauge is compatible with the Coulomb gauge for static sources. Notice that this is what happens also in electromagnetism, where the Lorenz gauge is compatible with the Coulomb gauge for static sources.

We conclude with a word on another important gauge fixing that can be used to study radiation including sources, the Bondi, or Bondi-Sachs gauge.
In this case the background Minkowski metric is not in Cartesian coordinates, but rather spherical coordinates, and time is replaced with retarded time $u=t-r$.
In these coordinates the Minkowskian Christoffel symbols do not vanish. However a crucial advantage is that it is a gauge that can be defined also in the full theory, and allows a non-perturbative description of gravitational waves, at the price of introducing an asymptotic expansion away from the sources.

\section{Detection of GWs }
\label{sec:detectionofGWs}

\subsection{Coordinate displacements versus physical displacements}

The simplest way to study the effect of a gravitational wave, is to compute how it changes the physical distance between free-falling test masses.
These follow time-like geodesics, whose tangent vector field $u^\m$ satisfies  the geodesic equation
\be\label{geo}
u^\n\na_\n u^\m = \f {du^\m}{d\t} +\G^\m_{\n\r}u^\n u^\r = 0, \qquad u^\m \p_\m= \f d{d\t}.
\ee
Here 
$\t$ is the proper time $\t$, and 
$u^2=-c^2$.
If the masses are initially at rest, we have $u^a=0$ and
\be\label{geo2}
\f {d u^\m}{d\t} = - \G^\m_{00} c^2 = \left(\f12\p^\m h_{00}-\p_0 h_{0}^\m\right)c^2 +O(h^2).
\ee
For vacuum solutions in the TT gauge $h_{0\m}=0$. Hence $\dot u^\m$ vanishes at lowest order, and the coordinate distance (as well as the coordinate time delay) between two nearby time-like geodetics remains the same during the passage of the wave.
This result provides us with an interpretation of the TT gauge: it is a choice of coordinates which are labelled by the position of free-falling test masses. It is thus the linearized version of the temporal gauge in the full theory discussed earlier. Now, even though the \emph{coordinate distance} between two test masses remains the same in this gauge, their \emph{physical distance} does not. 
It is given by 
\be\label{Lphys}
L = \int_0^{L_0}d\l \sqrt{g_{ab}\hat e^a\hat e^b}  = \int_0^{L_0} d\l \left(1+\f12 h^{\sscr TT}_{ab}\hat e^a\hat e^b\right) +O(h^2).
\ee
Here $\hat e^a$ is the tangent to the curve connecting the test masses, and $\l$ an arbitrary parametrization thereof. 
Since we are working in perturbation theory, we can consider as curve the background geodesic connecting the test masses, and since we are working in Cartesian coordinates of the background, we can choose $\hat e^a$ to be constant along a coordinate axis, and take $\l$ as one of the coordinates. Then $L_0$ is the coordinate distance. It coincides with the physical distance in the background flat metric, but differs from it when the spacetime is perturbed.
If we further assume that the wavelengths of the wave are much bigger than $L_0$, we can ignore the space dependence of $h^{\sscr TT}_{ab}$ and write the result as
\be\label{Lwave}
L\simeq \left(1+\f12 h^{\sscr TT}_{ab}\hat e^a\hat e^b\right)L_0.
\ee
The approximation becomes of course exact if the direction of propagation of the wave is orthogonal to the axis connecting the masses. 

The discussion offers an example of one of the most important lessons of general relativity, namely the importance of distinguishing coordinate effects from physical results. General covariance guarantees that all calculations can be performed in any coordinate system. But one has to always make sure that the physical consequences derived from the calculations are coordinate independent. 
In this example, we found that the coordinate position of test particles are unaffected by the passage of the wave. This is true only in the TT gauge. 
The relevant coordinate-independent quantity is the physical distance, and we found that it changes. Furthermore, it would change in the same way in any coordinate system preserving the endpoints of the integral: the geodesic distance between two physical points is a gauge-invariant observable. 

As we have already discussed, even though any coordinate system can be chosen, choosing a good coordinate system is important to simplify calculations. 
Even thought TT coordinates hide the passage of the wave, they are convenient because we can write the physical distance using fixed extrema in the integral at all times.\footnote{The same effect is used when choosing the synchronous gauge in cosmology, it is often convenient to choose a coordinate system such that the values of the coordinate grid represent galaxies, so that their coordinate distance does not change, while the physical one does.}
Another convenient gauge is the one of Fermi normal coordinates, which works in the opposite way: the metric is trivial at lowest order near two geodesics, see \eqref{gRNC}, hence the physical distance is entirely captured by the coordinate distance. To see this, let us consider the geodesic deviation equation.
It is given by
\be\label{geodev1}
u^\r\na_\r (u^\n\na_\n \xi^\m)= R^\m{}_{\n\r\s}u^\n u^\r \xi^\s,
\ee
where $\xi^\m$ is a vector connecting neighbouring geodesics, chosen such that $\xi\cdot u=[\xi,u]=0$. At first order in $h$, and with the assumption of vanishing initial velocities, it reduces to
\be\label{geodev2}
\f {d^2\xi^a}{d t^2} = -2\G^a_{0\n}u^0\dot\xi^\n - c^2 \xi^\n\p_\n \G^a_{00}.
\ee
In the TT gauge, the last term vanishes and the first one too if the initial velocity was zero. The coordinate distance between the geodesics stays constant, in agreement with the result already derived using the geodesic equation above.

The description changes completely if we use a gauge corresponding to a local inertial frame, such as the 
Fermi normal coordinates, which can be used to set to zero the Christoffel symbols all along a chosen geodesic. Doing so, the LHS of \eqref{geodev1} becomes simply a second partial derivative, and the equation reads
\be\label{accRiem}
\f {d^2\xi^a}{dt^2} = c^2R^a{}_{00b} \xi^b = \f12 \ddot h^{ab}_{\sscr TT}\xi_b+O(h^2),
\ee
where in the second equality we have neglected any contribution from the potentials. This shows that in a local inertial frame, the deviation between nearby geodesics can be described directly in terms of their coordinate separation. This is consistent with what previously seen in \eqref{gRNC}, the metric only changes at quadratic order in the coordinate distance from the origin, hence coordinate distances coincide with physical distance at first order. 
The geodesic deviation equation \eqref{accRiem} shows that an apparatus that can detect tidal effects will be sourced only by the physical components of the GW, and not by the gauge ones. The equation can be easily solved at first order in $h$, with
\be\label{xit}
\xi^a(t) = \xi^a(0) +\f12h^{ab}_{\sscr TT}\xi_b(0).
\ee
Since in this gauge the coordinate distance coincides with the physical distance at first order, we recover the gauge-invariant result \eqref{Lwave} but where this time the metric is unchanged, and it is the extremum of the integral that has moved.

\begin{figure}[ht]
\centering 
\includegraphics[width=12cm]{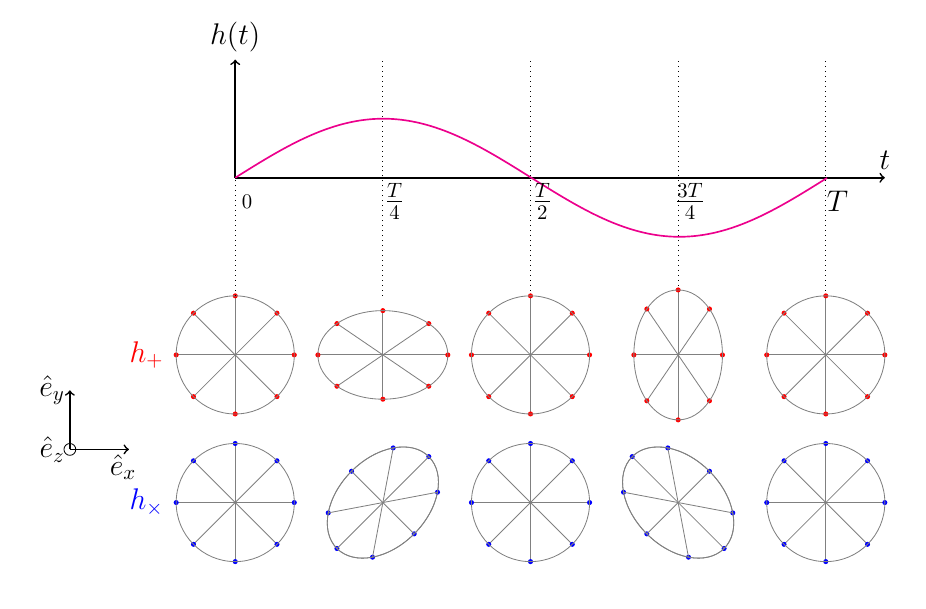}
\caption{\label{Figpols}
\small{{The effect of the two polarizations on a circular distribution of test masses. In the upper panel (red circles) $h_+\neq 0$ and $h_\times =0$. The lower panel has $h_\times \neq 0$ and $h_+ =0$.  } }}
\end{figure}

The change of proper distance \eqref{Lwave} also shows the meaning of the wave's polarizations.
Using the example \eqref{PWsol} of a monochromatic wave propagating along the $z$ axis, and setting the test masses at $z=0$, we can write the 
 relative change in physical distance as
\be
\f{\d L}{L_0}:=\f{L-L_0}{L_0}\simeq \left(\f12 h_{\sscr +}(\hat e^{x}\hat e^x-\hat e^{y}\hat e^y) +h_{\times}\hat e^{x}\hat e^y\right) \cos\om t.
\ee
An $h_{\sscr +}$ polarization would cause pairs of masses along the $x$ and $y$ axis to periodically approach and recede, hence drawing a $+$-like pulse in time, see Figure~\ref{Figpols}. An $h_{\times}$ polarization would cause the same effect but along the axis 
$\hat e^a = (1,1,0)/\sqrt 2$, namely rotated by 45 degrees.
This type of deformation is also called `shear' of the congruence of time-like geodesics followed by the test masses.
It can be visualized even more clearly if we use \eqref{xit}. We consider a circular distribution of test masses centered around the origin in the plane perpendicular to the direction of propagation of the wave, see Fig.~\ref{Figpols}. Then we can identify the displacement vector with the coordinate vector of each mass (labelled by $i$), and the effect of a monochromatic wave of frequency $\om$ is
\ba
x_i(t)&=&x_{i}(0) + h_{\sscr +}(t)x_i(0)+h_{\sscr \times}(t)y_i(0), 
\nn
\\
 y_i(t)&=&y_i(0) - h_{\sscr +}(t)y_i(0)+h_{\sscr \times}(t)x_i(0).
 \nn
\ea
The effect is shown in Fig.~\ref{Figpols}, where the period $T=2\pi/\omega$. The spin-2, quadrupolar nature of the gravitational force is evident from the shape of the shear deformation, and can be compared with the tidal deformation of earth's oceans. 

If external forces are present, on top of the gravitational one, then \eqref{geo} and \eqref{geodev1} acquire additional terms on the right hand side.
So in particular \eqref{accRiem} now reads
\be\label{accRiem2}
\f {d^2\xi^a}{dt^2} = c^2R^a{}_{00b} \xi^b +\f{F^a}{m}.
\ee
as an example, consider a material bar. While the effect of the gravitational wave is to stretch spacetime changing the physical distance between the molecules of the bar, there are also electromagnetic forces that hold the bar together, and which are intrinsically much stronger. For instance, the Coulomb interaction between two electrons one angstrom apart is  $\sim 10^{42}$ times stronger than its Newtonian counterpart. For this reason, one could in principle use simply a rigid ruler measuring the distance between two freely falling masses to detect gravitational waves. The problem with this idea is the weakness of the waves, and more sophisticated experiments are required to actually observe the waves.

A special case of external forces appearing on the right-hand side of \eqref{accRiem2} are the inertial forces. These, by the equivalence principle, can be reabsorbed in a coordinate transformation of the metric. For instance if the frame has both an acceleration $\vec a$ and an angular velocity $\vec\Om$ with respect to a local inertial frame, then \cite{Ni:1978zz}
\begin{align}
ds^2 &= - c^2dt^2\left((1+ \f1{c^2}\vec a\cdot \vec x)^2 - \f1{c^2}(\vec\Om\times\vec x)^2 + R_{0c0d}x^cx^d\right) \nn\\
&\qquad +2cdtdx^a\left(\f1c\eps_{abc}\Om^b x^c - \f23 R_{0cad}x^cx^d\right) +dx^adx^b\left(\d_{ab}-\f13R_{acbd}x^cx^d\right)\nn \\
& \qquad +O(x^2), 
\end{align}
and \eqref{accRiem2} becomes
\be\label{accRiem3}
\f {d^2\vec\xi}{dt^2} = -\vec a - 2\vec\Om\times \vec v +\f{\vec F}{m} +O(x^2).
\ee
All the gravitational effects as well as further non-inertial effects such as centrifugal acceleration are $O(x^2)$. So in order to be capable of detecting gravitational waves, a detector must first of all be freed from 
all the external forces that would otherwise drown the signal in noise.

In realistic physical systems, the emission will not be a plane wave, but rather a wave packet with finite temporal extension. The effect on the circular distribution will then be a superposition of different frequencies and different helicities, each with their own (time-dependent) amplitude. The temporal finiteness of the signal can also lead to a new type of effect: after the wave has passed, the distribution will stop oscillating, but its shape will in general not be the same as before the wave's arrival. This effect is called displacement  memory, and we will see below in Sec.\ref{subsec:hyper} an explicit example. The effect carries the memory of the wave, since 
it permits in principle to detect the passage of a gravitational wave even after the event. In practise though the detection is very difficult, because the external forces that make up the matter distribution will act and bring it back to its rest configuration. It is nonetheless one of the targets of future detectors \cite{Grant:2022bla}.

\subsection{Interferometers}

Let us briefly describe how the formulas above are used in the most common type of detectors, laser interferometers. 
Other chapters in this collection will cover more details as well as the types of detectors.
The basic idea of a laser interferometer is to detect physical changes like \eqref{Lphys} from the time-of-flight of monochromatic light signals. This can be done easily in the linear theory if we make the additional approximations that gravitational potentials can be neglected,\footnote{The potentials generated by the source can be naturally neglected because they fall off faster than the radiative modes, so this approximation concerns mostly the potentials of the local gravitational field of the observer.}
and that the wavelength $\l$ of the signal is much longer that the arms of the interferometer. The first approximation guarantees that the only source of curvature comes from the wave, hence the Riemann tensor scales like $\l^{-2}$. We can then set up a free falling frame say in Fermi normal coordinates centered on the beam splitter's geodesic. Thanks to the second approximation, the spatial projection of the null geodesics follows straight lines, hence the time of flight is directly related to the physical distance along the interferometer's arms. The latter is given by \eqref{Lwave} regardless of the direction of the wave, thanks again to the assumption that the wavelength is much larger than the arms' length. Denoting $\hat e^a_{1,2}$ the two axis, we have
\be
L_2-L_1 = \f{L_0}2 h^{\sscr TT}_{ab}(\hat e^a_1\hat e^b_1-\hat e^a_2\hat e^b_2).
\ee
For a typical signal $h\sim 10^{-21}$ (see overview Section \ref{ssec:binaries-characteristic}), hence the difference in arrival time would be $\D T=(L_2-L_1)/c\sim 10^{-26}s$ which is way too small to be measurable. Two ingenious ideas come to the rescue. The first is Michelson-Morley's idea to measure not time but phase interferences, and the second it to increase the effective path of light through Fabry-Perot cavities.
If we set the lasers so that the phases at the beam splitter are identical, the phase shift after the travel to and back from the mirrors will be
\be
\D\phi=\f{2\pi\n}cN_p(2L_1-2L_2),
\ee
where $N_p=1$ for a Michelson device and up to 300 for the Fabry-P\'erot one used by LIGO and Virgo, and the laser frequency $\n$ can go as high as $10{\tt kHz}=10^4{\tt s^{-1}}$, thus improving the strength of a typical signal to $10^{-20}$, which is actually -- and remarkably -- an observable phase difference.

\begin{figure}[ht]\centering
  \includegraphics[width=6cm]
  {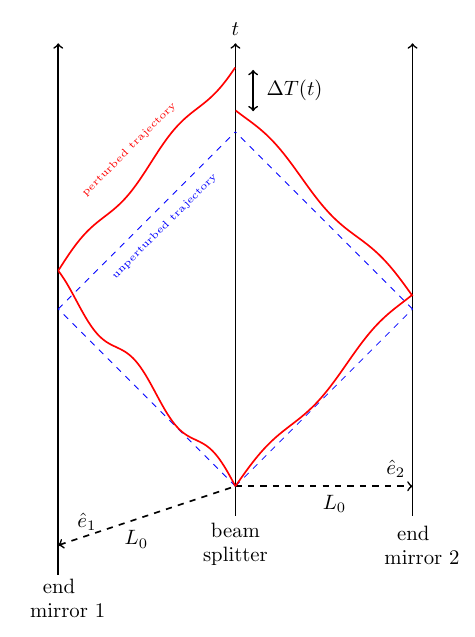}
  \caption{\small\emph{World lines of photon trajectories between the beam splitter and the end mirrors in an intertferometer with arms of equal length. Blue dashed lines: with no gravitational wave. Red lines: perturbed trajectories. In the long wavelength approximation $\l\gg L_0$ the red lines are straight (but still have different angles than the unperturbed blue lines), and $\D T=2\D L/c$.}} 
   \label{FigInter}
\end{figure}

Combining the last two formulas we have
\be
\D\phi=\f{2\pi\n}cN_p L_0 h_{ab}^{\sscr TT}(\hat{e}^a_1 \hat{e}^b_1-\hat{e}_2^a\hat{e}_2^b),
\ee
where now $\hat{e}^a$ are the unit vectors giving the direction of each arm.
It is possible to express the result in terms of the two wave polarizations by replacing $ h_{ab}^{\sscr TT}$ with \eqref{hTTk}. To compute the scalar products between the vectors corresponding to the detector's arms and the vectors defining the polarizations in the plane orthogonal to the propagation, we introduce a rotation 
from the detector's frame to the frame of propagation, plus a reflection to take into account the fact that the axis of propagation is opposite to the direction of the source acting also on $y$ to keep right-handed orientation of the frame. If we denote $(\th,\varphi)$ the angles identifying the direction of propagation, the result is
\be
\D\phi=\f{4\pi\n}cN_p L_0 (F_+h_++F_\times h_\times),
\ee
where the coefficients
\begin{align}
& F_+=\f12(1+\cos^2\th)\cos 2\varphi, \qquad
 F_\times=\cos\th\sin2\varphi
\end{align}
are called detector's pattern functions.\footnote{This formula assumes that the arms are perpendicular, and a fixed polarization basis. It can be generalized to include an additional rotation of the two polarization basis, as well as a non-perpendicular angle between the arms.}  This shows that while a single two-armed interferometer is sensitive to both polarizations, it cannot distinguish them.
It also shows that the sensitivity depends on the relative orientation with respect to the sources. The dependence is very strong, to the point that there are directions in which the detector is completely blind, like $(\th,\varphi)=(\f\pi2,\f\pi4)$. Hence the importance of multiple detectors in order to increase sensitivity in every direction and the possibility of distinguishing the polarizations. Multiple detectors also allow studying the localization of the source via triangulation.

If the approximation $\l\gg L_0$ is no longer valid, then one has to take into account the redshift changes during the time of flight, see for instance discussion in \cite{Cornish:2002rt,AnderssonBook}.

\section{Generation of GWs from sources}
\subsection{Introducing sources}\label{SecMatter}

Following the principle of general covariance, the matter Lagrangian should satisfy the property \eqref{dxiLM}, namely be written solely in terms of the dynamical matter fields and spacetime metric, and no additional background fields. The simplest way to obtain a viable matter Lagrangian is then to start from the one used in the absence of gravity, and `covariantize' it by the replacements
\be
\eta_{\m\n}\to g_{\m\n}, \qquad \p_\m\to \na_\m, \qquad d^4x\to\sqrt{-g}d^4x.
\ee
Doing so introduces a minimal coupling of matter to the gravitational field. Additional interactions can be included if phenomenologically or theoretically motivated, provided they respect the principle of general covariance, embodied for instance by condition \eqref{dxiLM}. Having done so, we define the matter energy-momentum tensor
\be
T_{\m\n} = - \f {2c}{\sqrt {-g}}\f{\d \cL_{\sscr M}}{\d g^{\m\n}}.
\ee
Inserting this definition in \eqref{dxiLM}, and using \eqref{gdiffeo1}, we obtain
\be
\xi_\n \na_\m T^{\m\n}= 
\f c{\sqrt{-g}}\left[\f{\d\cl_{\sscr M}}{\d\psi}\, \pounds_\xi \psi + \p_\m(\tl\th^\m_{\sscr M}-\xi^\m\cl_{\sscr M}+\f{\sqrt{-g}}c T^{\m\n}\xi_\n)\right].
\ee
The first term on the right-hand side is proportional to the matter's equations of motion, and vanishes on-shell. The rest is a total derivative and vanishes in the absence of boundaries or with appropriate boundary conditions. 
Since the identity holds for any $\xi$, we conclude that on-shell,
\be\label{naT0}
\na_\m T^{\m\n}\eqons 0.
\ee
This equation replaces the familiar conservation of the energy-momentum tensor guaranteed by Noether's theorem in flat spacetime. 
More precisely, the Noether current of the total Lagrangian $\cL_{\sscr EH}+\cL_{\sscr M}$ is
\be\label{jxitot}
j^\m_\xi = \f{c^3}{8\pi G}\Big( E^\m{}_\n\xi^\n - \na_\n \na^{[\m} \xi^{\n]}\Big),
\ee
where $E$ are Einstein's equations \eqref{EE1}, namely \eqref{jxi} with the vacuum equations replaced by the equations in the presence of matter), and whose conservation requires to be on-shell of both the Einstein's and matter's field equations:
\be
\na_\m j^\m_\xi = \f{c^3}{8\pi G} E^{\m\n}\na_\m\xi_\n-\f1c\na_\m T^{\m\n}\xi_\n \eqons 0.
\ee

\begin{figure}[ht]
\centering 
\includegraphics[width=4cm]{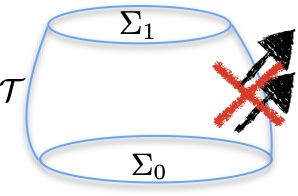}
\caption{\label{Figcons}
{{A region of spacetime bounded by two space-like hypersurfaces $\Si_{1,2}$ and a time-like one $\cal T$. With conservative boundary conditions on $\cal T$, \eqref{Qcons} establishes as many conservation laws as there are Killing vectors.}}}
\end{figure}

Even though \eqref{naT0} is often referred to as the general covariant version of energy-momentum conservation, it is important to remark that it is \emph{not} a  conservation equation in the usual sense. To understand this point, let us follow the usual procedure to obtain Noether charges from the current, and apply Stokes's theorem to a finite region $M$  with boundary $\p M$. To do so we need a scalar, which we obtain by contracting the left-hand side of \eqref{naT0} with a vector $\xi^\m$. After integrating by parts, we find
\ba
\int_M \na_\m T^{\m\n}\xi_\n \sqrt{-g}d^4x & =& \oint_{\p M} T^{\m\n}\xi_\n n_\m\sqrt q d^3y +\int_M T^{\m\n}\na_\m \xi_\n \sqrt{-g}d^4x,
\ea
for a boundary $\p M$ with induced metric $q$, normal $n_\m$ coordinates $y$. If $\na_{(\m}\xi_{\n)}$ vanishes, namely if the Killing equation is satisfied, then \eqref{naT0} can be turned into a conservation law. To do so, we consider the case in which $\p M$ consists of two space-like hypersurfaces $\Si_1$ and $\Si_2$
connected by a time-like boundary $\cal T$, see Fig.\ref{Figcons}. If the fields satisfy conservative boundary conditions at $\cal T$ (typically $\cal T$ asymptotically far away and fall-off conditions on the fields), then 
\be\label{Qcons}
Q_\xi:=\int_{\Si_1} T^{\m\n}\xi_\n n_\m\sqrt q d^3y = \int_{\Si_2} T^{\m\n}\xi_\n n_\m\sqrt q d^3y
\ee
for each Killing vector $\xi$. Therefore \eqref{naT0} gives as many conserved quantities as there are isometries in spacetime. For flat spacetime, these are the ten Poincar\'e charges. For a generic dynamical spacetime, there are none.

The validity of  \eqref{naT0} implies the matter equations of motion, as we have seen from its derivation. 
In particular, if matter consists of test particles, namely free motion without self-interaction and ignoring the back-reaction on the metric, this equation implies the geodesics equation in curved spacetime. This is for instance how one can derive the relativistic corrections to the Kepler problem, by evaluating \eqref{naT0} on the Schwarzschild background.
At lowest order in the weak-field expansion \eqref{glin}, \eqref{naT0} reduces to the energy-momentum conservation law in  flat spacetime,
\be\label{pT0}
\p_\m T^{\m\n}=0.
\ee
This means that at lowest order the matter can interact with itself, but not with the gravitational field: the sources follow geodesics in flat spacetime (that is, straight lines). To include the effect of gravity on the sources we must go beyond the lowest order. 
In other words, the linearized theory still describes gravity in the Newtonian way, namely as a force acting in flat spacetime. Of course, it already contains departures from Newton's theory, since it includes the special relativistic effects such as the gravito-magnetic interaction and radiation.

\subsection{Source multipoles}

Let us study the conserved quantities that arise on the Minkowski background. We choose $\Si$ to be a global hypersurface of constant time $t$, write its unit normal as $n_\m=-\p_\m t$, and $\xi$ is one of the ten Poincar\'e Killing vectors \eqref{xiP4}. We can then use  \eqref{Qcons} to identify ten conserved quantities. Four are the energy and momentum
\be\label{MandP}
c^2M:=\int d^3x \, T^{00}, \qquad cP^a := \int d^3x \, T^{0a},
\ee
corresponding to $a^{\m}{}_{\n}=0$ and unit values of $b^\m$. The remaining six are the relativistic angular momentum
\ba
cL^a &=&  \frac{c}{2}\eps^{a}{}_{bc} L^{bc} := \eps^{a}{}_{bc} \int d^3x \, x^bT^{0c}, \nn \\
 c^2K^a &:=& \int d^3x \, (T^{0a}ct-T^{00}x^a),
\label{LandK}
\ea
corresponding to $b^{\m}=0$ and unit values of $a^{a}{}_{\n}$ and $a^{0}{}_{\n}$ respectively. Here $\eps_{abc}$ is the completely anti-symmetric Levi-Civita symbol in flat spacetime.
Their conservation can be easily checked. 
We start by separating \eqref{pT0} in time and space components,
\be\label{pT0split}
c^{-1}\dot T^{00}+\p_a T^{a0} = 0, \qquad c^{-1}\dot T^{0a}+\p_b T^{ab} = 0.
\ee
Then using Stokes' theorem and vanishing boundary conditions we immediately see that
\be\label{Pcons}
\dot M= \dot { P}^a= \dot L^a = \dot K^a = 0.
\ee

The first of \eqref{MandP} is the total energy, but we followed the custom in the literature to denote it $M$ and refer to it as `mass', using Newtonian language. 
The first of \eqref{LandK} is the angular momentum with respect to the frame defined by $n_\m=-\p_\m t$. The second conserved quantity is the `boost charge', and can be rewritten as 
$K^a = I^a - t P^a$, where we introduce the center-of-mass position
\be
I^a=\f{1}{c^2} \int d^3x \, T^{00}x^a.
\ee
Conservation of $K^a$ is thus the statement that the center of mass moves following the total momentum.

The conserved quantities can be used to fix a reference frame as follows. First, we can choose the rest frame, where $P^a=0$. This removes the freedom of Lorentz boosts. Then, we can fix the origin to be in the center-of-mass, where $I^a=0$. This removes the freedom of spatial translations. The rotation freedom can be fixed choosing the axis so that $L^a$ has only one component (say $z$), and the remaining SO(2) freedom is fixed choosing an axis in the plane perpendicular to $L^a$. Finally the time translation symmetry is fixed setting the zero value of the clock.

The quantity $I^a$ is also called mass-dipole moment. The terminology comes about if we see $\r=c^{-2}T_{00}$ as a distribution, then $I^a$ is the first moment of that distribution. Following this logic, we introduce a multi-index notation for the higher multipole moments:
\ba 
I^{ab\ldots} &=& \f1{c^2}\int d^3x \, T^{00} x^a x^b \ldots,
\qquad P^{a,b\ldots} = \f1c\int d^3x \, T^{0a} x^b \ldots, \nn \\
S^{ab,c\ldots}&=& \int d^3x \, T^{ab} x^c \ldots.
\label{poledefs}
\ea
The conservation laws \eqref{pT0split} together with integration by parts in the absence of boundary terms provide relations between multipole moments and time variations of higher multipoles, such as
\ba
P^a&=&-\dot I^a, \qquad S^{ab} = \f12\ddot I^{ab}, \nn \\ \dot S^{ab,c} &=& \f16\dddot I^{abc}+\f13(\ddot P^{a,bc}+P^{b,ac}-P^{c,ab}), \qquad \dot{P}^{a,b} = S^{ab}
\label{polecons}
\ea
and so on. The first one above is the conservation of $K^a$ already seen, and relates the momentum monopole to the mass dipole time variation.
The second one allows one to determine the total effect of the stresses in the matter in terms of the second time derivative of the mass quadrupole.
These relations are useful because it is typically easier to measure and interpret the multipole moments of the mass and momentum distributions, rather than the spatial stresses.

When working with multipoles, it is typically convenient to organize them into irreducible representations of the rotation group, which are label by a an integer number $l$ and have $2l+1$ components each, as recalled earlier. This can be achieved expanding the distribution in spherical harmonics, e.g. $\r=\sum_{l,m}\r_{l,m}Y_{l,m}$, then the integrals of the modes $\r_{l,m}$ are the irreducible multipoles. It is possible although more cumbersome to do this composition directly in Cartesian coordinates without introducing spherical harmonics. One then gets
\begin{align}\label{Mmultipoles}
& M= \f{1}{c^2} \int d^3x \r, \qquad D^a=I^a= \f{1}{c^2} \int d^3x \r x^a, \\ & Q^{ab}= \f{1}{c^2} \int d^3x \r ( x^a x^b- \f{r{}^2}3\d^{ab}), 
\quad O^{abc} =\f{1}{c^2}  \int d^3x \r (15x^ax^bx^c - 9x^{(a}\d^{bc)}r^2),
\end{align}
and so on. Notice that while the multipole moments are useful at all orders in perturbation theory, the conservation laws \eqref{Pcons} are only valid at lowest order.
 
\subsection{Solving the wave equation with sources}

We are interested in the emission of gravitational waves from matter sources, without incoming radiation. This  can be imposed choosing the retarded Green function and setting to zero the independent degrees of freedom $h^{\circ\sscr TT}_{\m\n}$. The general solution is then
\ba
 \bar h_{\m\n} &=& -\f{16\pi G}{c^4}\int d^4x' G(x,x')T_{\m\n}(x')\nn \\
&=&  \f{4 G}{c^4}\int d^3x'\, \f{T_{\m\n}(t-\tfrac 1c|\vec x-\vec x'|, \vec x')}{|\vec x-\vec x'|}.
\label{hwithT}
\ea
using \eqref{Gbox} and the specialized De Donder gauge with no homogeneous solution. Even in the linearized approximation, the integral is in general very complicated and there is no analytic solution. So we resort to approximation schemes.
In particular, we introduce two independent approximations:
\begin{itemize}
\item[$(i)$] Wave-zone approximation: we assume to be very far away from the sources, that is $R:=|\vec x|\gg |\vec x\,'|$. 
This allows us to expand the integrand in powers of $1/R\ll 1$. For the numerator, we have
\be
|\vec x-\vec x\,'| = R - {\vec N\cdot\vec x\,'} +\dots
\ee
where $\vec N:=\vec x/R$, and 
\be\label{Texp}
T_{\m\n}(t-\tfrac 1c|\vec x-\vec x\,'|, \vec x\,')\simeq T_{\m\n}(t_{\sscr R},\vec x\,') + \frac{\vec N\cdot\vec x\,'}{c}\, \dot T_{\m\n}(t_{\sscr R},\vec x\,') + \ldots,
\ee
where we introduce the \emph{retarded time}\footnote{Namely the time at which a signal  travelling at the speed of light was sent in order to arrive at $t$.}
\be
t_{\sscr R}:=t-\f Rc.
\ee

For the denominator, we have
\be\label{rinvexp}
\f1{|\vec x-\vec x'|} = \f1R + \f{\vec N\cdot \vec x'}{R^2} + \f32 (x'_a x'_b- \f{r'{}^2}3\d_{ab})\f{N^aN^b}{R^3}+\ldots
\ee

Furthermore, the direction of propagation of the wave coincides with the direction from the source, namely $-\vec N$ if we take the origin of the coordinates inside the source. Hence the TT projector can be written in terms of $\vec N$ instead of the wave vector. 
 
\item[$(ii)$] Slow dynamics: We assume that the dynamics of the source is slow, so that time derivatives in \eqref{Texp} are small corrections.
To understand why, consider that the integration coordinate 
$\vec x'$ spans at most the size of the source, and if this has a typical frequency scale $\om_s$ (for instance in a binary, the frequency of the orbit), then $v_s:=|\vec x'|\om_s$ is the velocity scale of the source. It follows that
\be
\f{|\vec x'|}c\dot T_{\m\n} \sim \f{|\vec x'|\om_s}cT_{\m\n} \sim \f{v_s}cT_{\m\n}
\ee
is suppressed by $v/c$. The Taylor expansion \eqref{Texp} is therefore controlled by the parameter $v/c\ll1$, and it is called post-Newtonian expansion.

\end{itemize}

The approximated solution can thus be written as 
\begin{align}\label{hsources}
\bar h_{\m\n}(x) &= \f{4G}{c^4 R} \int d^3x' \, \bigg( T_{\m\n}(t_{\sscr R}, \vec x\,') +\f{N_a}c \dot T_{\m\n}(t_{\sscr R}, \vec x')x'{}^a+\f{N_a}R T_{\m\n}(t_{\sscr R}, \vec x')x'{}^a+\ldots \bigg)
\end{align}
The first term is the leading order; the second term is the first of the PN corrections; the third term is the first of the $1/R$ corrections.
Using the multipole definitions \eqref{poledefs} and their conservation laws \eqref{polecons}, we can rewrite the different components of the solution \eqref{hsources} as

\begin{subequations}\label{hsourcePN}\begin{align}
&\bar h_{00} =\f{4G}{c^2R}\left( M -\f {N^a}cP_a+\f{N^aN^b}{2c^2}\ddot I_{ab}+\f{N^a}R I_a+ \ldots \right)\Big|_{t_{\sscr R}}, \\
&\bar h_{0a} =-\f{4G}{c^3R}\left(P_a + \f{N^b}{2c}\ddot I_{ab}+\f{N^b}{2Rc} (L_{ab} +\dot I_{ab}) + \ldots \right)\Big|_{t_{\sscr R}}, \\\label{habsource}
&\bar h_{ab} =\f{4G}{c^4R}\left(\f1{2}\ddot I_{ab}+\f {N^c}{3c}\Big(\f12\dddot I_{abc} + \ddot P_{a,bc}+\ddot P_{b,ac}-\ddot P_{c,ab}\Big) + \ldots \right)\Big|_{t_{\sscr R}}.
\end{align}\end{subequations}

These are the first few terms of the double expansion in velocities and distance from the sources.  We are not giving all metric components to the same higher order, this partial result is sufficient for our purposes. In the PN expansion this is the metric parametrization in the wave zone, with the no-incoming radiation condition, and completely gauge fixed.
The lowest order of the time-time component reproduces the Newtonian result.\footnote{In particular the lowest order of the Schwarzschild metric is obtained as the special case with constant $M$ and all the rest vanishing. Notice in fact that $h_{00} = \bar h_{00}+\bar h/2=2GM/c^2R$.}
The first PN correction is the movement of the source, and can always be set to zero by going to the rest frame. Doing so eliminates the lowest order of the $\bar h_{0a}$ component. The first corrections in that component contain the gravito-magnetic effects relevant to the Lense-Thirring effect, for instance.
Notice also that the angular momentum is sub-leading in $R$, as one could have expected from a large distance expansion of Kerr's metric.

The radiative degrees of freedom are in the spatial components \eqref{habsource} and can be extracted acting with the projector \eqref{PTT}.
We have the mass quadrupole at leading order, and the first PN correction features the mass octupole and momentum quadrupole.
We can immediately remark the absence of monopole and dipole contributions to the emission of waves. This is  a direct consequence of the conservation laws, since they imply that the mass monopole and dipole have vanishing second time derivatives.
As a consequence, an oscillating spherical distribution would not emit gravitational waves, in agreement with Birkhoff theorem in the full theory, 
nor would a distribution with axial symmetry rotating at constant velocity, in agreement with Kerr's solution.

Applying the TT projector removes any trace, hence $h^{\sscr TT}=\bar h^{\sscr TT}$ and one can replace $I_{ab}$ with the irreducible quadrupole moment $Q_{ab}$, and obtain at lowest order
\be\label{Qformula}
{ h^{\sscr TT}_{ab}(t,\vec x) = \f{2G}{c^4R} \ddot Q^{\sscr TT}_{ab}(t_{\sscr R}). }
\ee
This is the celebrated \emph{first quadrupole formula}, derived by Einstein in 1918:
The dominant radiation in the slow-motion approximation arises from the acceleration of the quadrupole moment of the mass distribution.
From this we can also obtain the expressions for the two independent polarizations. If $\vec k=\hat z$, we can use  \eqref{PTTA} and
\be
h_+(t,r) = \f{G}{c^4R} (\ddot Q_{11}-\left. \ddot Q_{22})\right|_{t_{\sscr R}}, \qquad h_\times = \f{2G}{c^4R}\left. \ddot Q_{12}\right|_{t_{\sscr R}}.
\label{eq:inzdirn}
\ee
For a general $\vec k$ it is obtained replacing $h_{ab}\to (G/c^4R)\ddot Q_{ab}$ in \eqref{hgendir}. Notice also that $P^{\sscr TT}(Q)=P^{\sscr TT}(I)$ since the projector removes the trace, hence we can replace $Q_{ab}$ with $I_{ab}$ in these expressions.

Let us make some order-of-magnitude estimates. By dimensional analysis, the mass multipoles scale like $Mr^l$, where $r$ is the typical size of the source. If the dynamics of the system has a typical velocity scale $v$, then $Q\sim Mr^2$ and $\ddot Q\sim Mv^2$. This gives
\be
h\sim \f{GMv^2}{c^4R} = 5\times 10^{-19}\left(\f{M}{10M_\odot}\right)\left(\f{1{\rm Mpc}}R\right)\f{v^2}{c^2}.
\ee
For example, if we extrapolate this formula to relativistic speeds $v\sim c$ for the merger of two 10-solar-masses black holes, we get a $10^{-18}$  amplitude at galactic distances, and  $10^{-21}$ at 100 Mpc where the Virgo cluster is located.

This estimate is the lowest order of various approximations, which is useful to recap here: (1) weak-field, PM expansion; (2) long-distance, multipolar expansion; (3) small velocities, PN expansion. To obtain more accurate results, one has to include higher order corrections. Doing so is actually far from simple.
Not only do we have three different expansion parameters with non-trivial hierarchies among them, we also have to face both  technical and conceptual challenges. 
Let us list a few, and tools used to deal with them. The PN expansion is not a convergent series, but rather what is known as an asymptotic series. Its accuracy degrades as we increase $R$. Dealing with this mathematical problem requires techniques such as the matched asymptotic expansion. Related to this is also the more conceptual issue that the causal propagation determined by the Green's function at lowest order follows the null cones of the background Minkowski metric. But null cones are bent by the gravitational interaction, hence higher order corrections have to also modify the retarded time to the correct one. For instance for the Schwarzschild metric the correct retarded time is 
\ba
u&=&t-R/c - 2GM/c^2\ln(R-2GM/c^2)
\nn \\
&=& t_{\sscr R}+\f{2GM}{c^2}\ln R - \left(\f{2GM}{c^2}\right)^2\f1R +O(R^{-2}). 
\nn
\ea
Hence higher orders change the notion of retarded time.\footnote{This issue can be solved non-perturbatively using the Bondi coordinates mentioned earlier, but then one is switching from a PN expansion to an asymptotic expansion of the full theory, and the approach is both conceptually and technically different.} Another tricky effect comes in at higher orders: the waves backscatter and self-interact, causing a delay in part of the signal, which starts travelling inside the light-cone, similar to light slowing down in a medium due to interactions with the medium. Then the total signal includes a `tail' that comes after the main part of the signal. To take this into account one has to include effects that arise from integration over time.
Another problem is that divergences appear after the first iteration, because convolution of Poisson integrals diverge even if the initial source has compact support. To regularize this unphysical divergence one has to split the integrals into near-zone and far-zone integrations. 

These and other types of difficulties plagued the theory throughout most of the seventies, and were addressed thanks to the work of many brilliant researchers, including  pioneers like Thorne, Will, and Damour. On the phenomenological side, people thought for a while that the lowest quadrupole order would have been enough to match experiments, given the weakness of the waves. Later theoretical work, e.g. the seminal paper  \cite{Cutler:1992tc}, clarified the observational sensitivity to the PN corrections and justified the importance of the endeavour. The task is very challenging, and researchers have come up with different approaches. We refer to the specialized literature \cite{Thorne:1980ru,Blanchet:2006jqj,poisson,Goldberger:2004jt} for reviews of this more advanced topic. 
In the following we will content ourselves to stay at lowest order, which is enough to understand the basics of the physics, if not for a detailed match to observations.

\section{Dissipation by gravitational waves}

The first historical evidence of gravitational waves was the orbital decay of the Hulse-Taylor pulsar.
Accurate measurements showed that the orbital decay was consistent with the prediction of general relativity. Indeed, general relativity predicts that gravitational waves carry energy away from a system that produces them. In this Section we describe how this prediction arises at the lowest order in perturbation theory, and in the next Section we show how it can be applied to predict the orbital decay.

\subsection{Energy of gravitational waves}

Let us look at a gravitational wave as a spin-2 field propagating on the Minkowski background. Thanks to the Poincar\'e invariance of the background, we can apply Noether's theorem and derive a conserved energy-momentum tensor for $h_{\m\n}$.  
An explicit calculation starting from the linearized Lagrangian gives
\be\label{tNoether}
t^{\sscr N}_{\m\n} =  \f{c^4}{32\pi G}\left(\p_\m h^{\a\b} \p_\n h_{\a\b} - \f12 \eta_{\m\n} \p_\l h_{\r\s}\p^\l h^{\r\s} \right),
\ee
where the label N stands for Noether, and we assumed here the De Donder condition to simplify the expression.This tensor is conserved, namely $\p_\m t^{{\sscr N}\m\n}\eqons0$, but has no clear physical meaning, because it is \emph{not} gauge-invariant: It changes under a linearized diffeomorphism \eqref{diffeoeta}, including those compatible with the De Donder condition, and consequently assigns a non-zero value of energy-momentum to pure gauge modes.
Furthermore, we can make it vanish entirely at any point using Riemann normal coordinates, since in these coordinates the first derivatives of the metric vanish at that point. Since it is zero at one point in one coordinate system but not in others, it is not a tensor. It is usually referred to as `pseudo-tensor'.

The only gauge invariant quantities that can be extracted from \eqref{tNoether}, at least at lowest order, are global ones, such as the \emph{total} energy and momentum on a space-like hypersurface $\Si$ of constant time,
\begin{subequations}\label{Ptot}\begin{align}
& E_{\sscr GW} = \int_\Si t^{{\sscr N}0 0}d\Si = \f{c^2}{64\pi G}\int_\Si\dot h^{\a\b} \dot h_{\a\b} d\Si, \\
& P^a_{\sscr GW} = \int_\Si t^{{\sscr N}a 0}d\Si = -\f{c^3}{32\pi G}\int_\Si\p^a h^{\a\b} \dot h_{\a\b} d\Si.
\end{align}\end{subequations}
In fact, at lowest order in perturbation theory, a gauge transformation acts as \eqref{diffeoeta} with $\xi=O(h)$, under which \eqref{tNoether} transforms as a total derivative. Spatial total derivatives do not contribute to \eqref{Ptot} under the standard condition that the perturbation goes to zero at infinity, and temporal ones can be converted to spatial derivatives on shell of the wave equation, hence the same argument applies.
Similarly, one can define the total angular momentum as
\be\label{defJ}
L_{\sscr GW}^{a}:= -\f{c^4}{32\pi G}\eps^{abc}\int_\Si( \dot h_{\m\n} x_{b}\p_{c} h^{\m\n} + 2 \dot h_{b\m}h_{c}{}^\m)d\Si.
\ee

For our applications below, the total energy-momentum and angular momentum will be sufficient. They are not sufficient however for iterating the perturbation theory beyond lowest order, and we now briefly describe some of the conceptual aspects of what is done in the literature in this respect. First of all, the reader may recall that Noether currents are not unique, being defined only up to adding total derivatives whose conservation is trivial. In the case at hand, it means that any quantity of the type 
\be\label{tU}
t^{\sscr N}_{\m\n}+\p^{\r}\p^\s U_{\m\r\n\s},
\ee
where $U$ has the same index symmetries as the Riemann tensor, is an equally valid Noether current. One may hope that there exists a representative in the equivalence class \eqref{tU} that would be gauge invariant, but this is not the case: the lack of gauge-invariance is simply the linearized version of the fact that there cannot be any local tensor representing the energy of the gravitational field. 
Such quantity will have to be zero in a local free-falling frame where the effects of gravity are absent, and if it were a tensor, it would then be zero in any frame.\footnote{A tensorial quantity capturing \emph{some} aspects of gravitational energy can be constructed using the Bel-Robinson tensor, but it is fourth-order in derivatives, therefore does not have the right physical dimensions, and will capture only higher-order terms of the gravitational energy.}  
This is therefore a direct consequence of the equivalence principle, 
and its mathematical implementation via general covariance.\footnote{It is instructive to put this problem in perspective with what happens in the electromagnetic case. If one computes the canonical energy-momentum tensor of Maxwell's theory using the Noether formula, one also finds a meaningless gauge-dependent expression. However, the Noether construction only defines the tensor up to total divergences, and it is possible to find one that gives a gauge-invariant expression, and which is furthermore symmetric and coincides with the one derived from the variation with respect to the metric. In gravity there is an analogue problem, but even adding total divergences it is not possible to find a local gauge invariant quantity.}

The obstruction to identifying a local energy density for the gravitational field shows up very prominently in the full theory.
We have already discussed how Noether charges for generic diffeomorphisms are all trivial in the bulk of spacetime. 
Similarly, the bulk Hamiltonian one finds from the Legendre transform of the Lagrangian is a sum of constraints, and thus identically zero when evaluated on solutions. Any attempt to work around these facts and define quasi-local observables representing the gravitational energy unavoidably run into trouble with ambiguities and  dependence on coordinates or other unphysical background structures \cite{Szabados:2009eka}. The clearest well-defined resolution to this problem is to consider not the energy density, but only the total energy. This is useful when describing isolated systems, namely spacetimes that are fully dynamical in a certain region, but become well approximated by flat spacetime at large distances from this region. In this case, one can introduce a physically meaningful notion of boundary to the spacetime, and exploit the fact that the Hamiltonian picks up a boundary contribution which is non-vanishing on solutions. The resulting \emph{surface charges} can be used to characterise the total energy momentum and angular momentum of the system, and can be derived as Noether charges as well. Examples of this construction are the ADM charges at spatial infinity, and the BMS charges at future null infinity, as mentioned in Sec.~\ref{SecNoether}.

While looking at global quantities such as the total energy of a free GW \eqref{Ptot}, or the surface charges in asymptotically flat spacetimes mentioned at the beginning of the section, is the safest way to define energy in the full theory, the perturbative treatment offers an alternative, `quasi-local' possibility. 
Perturbatively in fact, it is possible to construct gauge-invariant quantities by introducing a spacetime averaging procedure based on the properties of the background.
We consider a region $L$ whose size is much larger than the typical wavelength $\l$ of the perturbation, but much smaller than the typical wavelength $\l_B$ of the background (which is infinite for a flat background), and we define the averaging of a functional $F$ as
$\mean F:=\f1L\int_L F$. If applied to an expression quadratic in the Fourier modes like \eqref{tNoether}, 
the procedure suppresses combinations with different frequencies or different phases, in a way completely similar to how the total energy in a standard background-dependent theory comes mainly from positive interference superposition of waves. The difference is that in background-dependent theories averaging the energy is a choice, since the local energy density is theoretically also well defined. In gravity it is not a choice but mandatory, since there is no meaningful local energy density, and furthermore care is needed to define correctly the procedure in a way to make it compatible with general covariance. Detailed analysis \cite{Isaacson:1968hbi,Isaacson:1968zza,Burnett} shows that the result of the procedure is that expressions under the averaging sign can be freely integrated by parts in space and, upon going on-shell, also in time derivatives since a wave propagates on the light-cone. For instance,
\be
\mean{\p_\m h_{\a\b}\p^\m h^{\a\b}}=-\mean{h_{\a\b}\square h^{\a\b}}=0
\ee
outside the sources. Under this procedure, we find
\be\label{deftNav}
\mean{t^{\sscr N}_{\m\n}} = \f{c^4}{32\pi G}\mean{ \p_\m h^{\a\b} \p_\n h_{\a\b}}
\ee
in a region outside the sources.
One can show that the averaging procedure makes the right-hand side gauge-invariant \cite{Isaacson:1968zza}.
This means that it can be expressed in terms of the TT projection and the gauge invariant potentials. The latter can be neglected if the sources variation (induced by the partial derivatives in the expression above) occurs over much longer time scales than the $h^{\sscr TT}$ wavelengths.\footnote{In \cite{poisson} this step is called short-wave approximation, and it is performed without the averaging, in the context of the Landau-Lifshitz approach described in Appendix \ref{appLL}. } 
This motivates the definition of
\be\label{deft}
t_{\m\n}:= \f{c^4}{32\pi G} \p_\m h_{\sscr TT}^{ab} \p_\n h^{\sscr TT}_{ab}, \qquad t_{\m\n}=\mean{t^{\sscr N}_{\m\n}}.
\ee
This quantity is actually gauge-invariant at lowest order, since the only non-invariant terms are the partial derivatives, and these transform linearly in  $\xi\sim O(h)$. It follows that in so far as only lowest order results are needed, we can use the simpler expression \eqref{deft} as a proxy for the averaging procedure. 
Notice also that it coincides with \eqref{tNoether} in the TT gauge.

The expression \eqref{tNoether} in the TT gauge is the one used by Einstein to determine  the energy carried away by gravitational waves.
It has, however, a limited applicability. We have already discussed its gauge dependence. Another issue is that it relies heavily on the specific background chosen, and had we worked with a non-isometric one, then there would be no Noether charge to begin with. Furthermore, it is not clear how to extend this construction to treat higher orders in perturbation theory.
These shortcomings can be addressed if we look at a different definition for the gravitational energy-momentum pseudo-tensor, based on the actual back-reaction on the metric caused by the gravitational waves. In fact the actual ``effective" source that determines the second-order metric perturbation is not \eqref{tNoether}, but rather the second order expansion of the Einstein tensor, that we denoted $t^{\sscr G}_{\m\n}$ in \eqref{G2}.
The candidate gravitational energy-momentum pseudo-tensor $t^{\sscr G}_{\m\n}$ obtained in this way is also conserved. In fact, an explicit calculation shows that it differs from \eqref{tNoether} precisely by a term like \eqref{tU}, with $U$ a certain quadratic expression in derivatives of $h_{\m\n}$.
It has the improved 
property that it depends on second derivatives of the metric, so it cannot be made to vanish at any given point. However, it is still not gauge-invariant.
Therefore one has to invoke again the averaging procedure in order to extract gauge-invariant information. Upon doing so, one finds that the two prescriptions give a consistent answer \cite{Isaacson:1968zza}:
\be
\mean{t^{\sscr G}_{\m\n}}=\mean{t^{\sscr N}_{\m\n}}=t_{\m\n}.
\ee

This matching supports averaging as a viable way to extract unambiguous and gauge-independent quantities. The prescription  $t^{\sscr G}$ overcomes some limitations of the Einstein-Noether construction. It can be used in perturbation theory around an arbitrary background, and can be  systematically extended to any order in perturbation theory, by computing higher order corrections $G^{\sscr (n)}_{\m\n}$ and evaluating them on the perturbed solution. 
This procedure is however not very handy, and a better scheme is the one proposed by the Landau-Lifshitz reformulation of Einstein's equations, see e.g.~\cite{Blanchet:2006jqj,poisson}. There one changes variables from the metric to a densitized inverse metric ${\mathfrak g}^{\m\n}:=\sqrt{-g}g^{\m\n}$. One advantage of this reformulation is that it provides a full non-perturbative expression for a candidate energy-momentum, known as Landau-Lifshitz pseudo-tensor.
This has the usual limitations (gauge-dependence and vanishing at any point in a local inertial frame) dictated by the equivalence principle, but has the 
merit of being set up in a way that makes it very natural to develop a systematic perturbative expansion, since the pseudo-tensor is defined already at non-perturbative level, and does not need to be determined order by order as in the previous approach. Furthermore, it provides a prescription for the energy, momentum and angular momentum as surface charges that, even though restricted in validity to Cartesian coordinates in the region far away from the sources, can be evaluated including higher orders, and bypasses the need for the spatial averaging of volume integrals.
For these reasons, the Landau-Lifshitz formulation is widely used by the community working in the post-Newtonian expansion. We review it briefly in Appendix~\ref{appLL}.
The lowest order in the weak-field approximation of the Landau-Lifshitz pseudo-tensor differs from the previous two options by a term like \eqref{tU}, and also matches the gauge-invariant result after averaging:
\be
\mean{t^{\sscr LL}_{\m\n}}=t_{\m\n}.
\ee

Similar considerations apply also to define the angular momentum of gravitational waves.
In this case, the result of the averaging procedure starting from any of the three prescriptions described above motivates the following definition \cite{dewitt2011gravitation,Thorne:1980ru,poisson}
\be\label{defj}
j^{a}:=\f12\eps^a{}_{bc}j^{bc} = -\f{c^4}{32\pi G}\eps^{abc}( \dot h^{\sscr TT}_{de} x_{b}\p_{c} h_{\sscr TT}^{de} + 2 \dot h^{{\sscr TT}}_{bd}h^{\sscr TT}_{cd}).
\ee
As before, one can drop the TT projector and still get a gauge-invariant quantity after global integration or averaging.

We conclude that in so far as one is interested only in lowest order results, any of these three choices are equally good. For a systematic perturbative expansion, the latest is the better one. In fact even if the construction of gauge invariant quasi-local quantities via averaging is conceptually useful to clarify how gauge-invariant information could be extracted in principle, it is not very practical. To set up a systematic perturbative expansion, it is easier to work with gauge-fixed quantities at all intermediate steps, and then extract only at the end the physical predictions in terms of gauge-invariant observables. For instance, there is no problem in working with the non-averaged notions of energy-momentum and angular momentum pseudo-tensors, as long as one does not attempt to give them a direct physical interpretation. The idea is to use them to perform calculations, and at the end read off the physical dynamics not from their evolution but from that of gauge-invariant quantities such as the amplitude and frequency of TT modes, or the evolution of relative distances such as the periastron of an orbit. This is the logic used in the PN expansion, and based on the Landau-Lifshitz reformulation \cite{Blanchet:2006jqj,poisson}.

\subsection{Dissipation equations}

The fact that \eqref{deft} is conserved means that we can derive identities between time and spatial derivatives like those that led to the conservation laws \eqref{Pcons} for the matter sources. The key difference however is that the matter sources had compact support, hence we could neglect boundary contributions when integrating by parts. This is no longer true for the gravitational contributions, since the waves have non-compact support. The non-vanishing of the boundary terms has the effect that the `charges' corresponding to energy, momentum and angular momentum are no longer conserved. This dissipation is precisely the statement that gravitational waves carry energy and have a physical impact on the system.

At lowest order, we do not even need to use the conservation equation in order to study the dissipation, because of a special property of the explicit solution
\eqref{hsourcePN}. Each metric component has functional dependence on coordinate of the form $f(t_{\sscr R},N^a)$. For such functions, it is easy to check that
\be\label{id}
\p_a f = -\f{N_a}c\dot f+O(R^{-1}).
\ee
The leading order of this approximation plays an important role in simplifying many formulas in the wave zone, where $R\gg 1$.

Let us begin our analysis from the flux of gravitational energy, namely the emitted power. This is given by  
\be\label{Edot1}
\dot E_{\sscr GW} = \int_\Si \dot t^{00} d^3x
= - c\oint_{\p\Si} t^{0a} N_a dS 
=\f{c^4}{32\pi G}\oint_{\p\Si} \dot h^{\sscr TT}_{cd}N^a\p_a h_{\sscr TT}^{cd}  dS,
\ee
Stokes theorem choosing as boundary a 2-sphere of radius $R$ in the asymptotic region 
(hence the outgoing unit normal is simply $N_a$, 
and $dS=R^2d^2\Om$ where $d^2\Om=\sin\th d\th d\phi$) and \eqref{deft} in the last equality.
The spatial derivative can be replaced at lowest order with a time derivative using again \eqref{id}, and we arrive at
\begin{align}\label{Energyloss}
\dot E_{\sscr GW} &= -\f{c^3}{32\pi G}\oint_{\p\Si}  {\dot h^{\sscr TT}_{ab}\dot h_{\sscr TT}^{ab} } dS =
- \f{G}{8\pi c^5R^2} \oint_{\p\Si} {\dddot Q^{\sscr TT}_{ab}\dddot Q_{\sscr TT}^{ab}} dS|_{t_{\sscr R}}, 
\end{align}
where in the last step we used the explicit form \eqref{hsourcePN} of the solution, in particular \eqref{Qformula}.
To evaluate the integral, we observe that the only angular dependence occurs in the TT projector \eqref{PTT}. Using the following formula,
\be\label{intPTT}
\oint_{S^2}P^{\sscr TT}{}_{ab}^{cd}\, d^2\Om = \f{8\pi}{5}\left(\d^{c}_{(a}\d^{d}_{b)}-\f13\d_{ab}\d^{cd}\right),
\ee
we find  
\begin{align}\label{Energyloss1}
\dot E_{\sscr GW} = -\f{G}{5c^5}{\dddot Q_{ab}}\dddot Q^{ab}|_{t_{\sscr R}}.
\end{align}
This is the second famous quadrupole formula of Einstein \cite{Einstein1918}. 
It gives the instantaneous power radiated at a distance $R$ from the source and a time $t$, as a function of the quadrupole time variation
at the retarded time $t-R/c$. 
Notice that the index contraction occurs over \emph{all} indices of the (traceless) quadrupole moment. The effect of the $TT$ projection goes into a numerical factor,  after the integration \eqref{intPTT}.

For the linear momentum loss,
\ba
\dot P^a_{\sscr GW} &=& \f1c\int_\Si \dot t^{0a} d^3x = - \int_\Si \p_b t^{ab} d^3x  = - \oint_{\p\Si} t^{ab} N_b dS 
=-\f{c^2}{32\pi G}\oint_{\p\Si} {N^a \dot h^{\sscr TT}_{cd}\dot h_{\sscr TT}^{cd}}  dS,
\label{Pdot1}
\ea
where we used twice \eqref{id} in the last equality. Since $N^a$ is an odd function on the sphere, the integral vanishes: \emph{there is no loss of momentum at lowest order} (namely at order $G/c^6$, once we use the first quadrupole formula). 
A change in the total momentum of the system caused by the emission of GWs (`kick') occurs only at the next order $G/c^7$, when mixing of multipoles of different parity occurs.

For the angular  momentum loss,
\begin{align}
& \dot L^a_{\sscr GW} = \int_\Si d^3x \p_t j^a = -\f1c\oint_{\p \Si}j^adS 
\f{c^3}{16\pi G}\eps^{abc}\oint_{\p\Si} ({ \dot h^{\sscr TT}_{de} x_{b}\p_{c} h_{\sscr TT}^{de} + 2 \dot h^{{\sscr TT}}_{bd}h^{{\sscr TT}d}_{c} } ) dS.
\end{align}
Using the quadrupole formula \eqref{Qformula} and performing the integrals using identities similar to \eqref{intPTT}, one arrives at
\be\label{Jloss}
\dot L^a_{\sscr GW} = - \f{2G}{5c^5}\eps^{abc}{\ddot Q_{bd}\dddot Q_{c}{}^{d}}|_{t_{\sscr R}}.
\ee
Angular momentum loss occurs at the same order as energy loss, and involves one lesser time derivative.

\subsection{On the validity of the quadrupole formula}\label{SecVal}

The quadrupole formula \eqref{Qformula}, and its application leading to the second quadrupole formula \eqref{Energyloss1}, were derived in the linear approximation.
In this approximation, the energy-momentum tensor satisfies the flat spacetime conservation law \eqref{pT0}, and the geodesics of matter are straight lines. It is thus valid only for systems whose gravitational interaction is negligible. 
It is not valid, in particular, for a binary system held together by gravity, even at the non-relativistic, Newtonian level. To treat a gravitational binary, one has to go beyond the linear approximation, using the iterative scheme described around \eqref{gexpsecond}. In the first step, one should include only the kinetic and non-gravitational pieces of the dynamics in the moment of inertia sourcing the quadrupole formula, consistently with the motion being along straight lines. The result is then used to source the effective energy-momentum tensor $t_{\m\n}$  needed for the second iteration of the field equations, which determines $h^{\sscr (2)}_{\m\n}$ of \eqref{gexpsecond}. When this approach is applied to a gravitational binary, one finds that the second-order correction restores precisely the Newtonian contribution to the quadrupole moment. This is quite remarkable, and means that \emph{the same quadrupole formula is also valid if one includes the Newtonian interaction}. See for instance \cite{poisson,DeruelleUzan2018} for details. 

We will make use of this fact below, and deduce the right results for gravitational binary systems using the quadrupole formula with the Newtonian potential included in the moment of inertia, without going into the technical details required to solve the second iteration, for which we refer to the cited literature. However, we should keep in mind that this formula can only be trusted \emph{because it has been derived including the second iteration}, and not from the linearized approximation alone. We should also keep in mind that restoring the validity of the quadrupole formula after the second iteration in gravitational binaries is a special and remarkable fact, and not a general result. For instance, the linearized quadrupole formula gives the wrong answer for gravitational binaries in modified theories of gravity, see e.g. \cite{Taherasghari:2025mlf}.

The fact that the linearized approximation provides potentially correct answers beyond its regime of validity contributed to the controversy that heated the debate around the quadrupole formula, and that were ultimately solved only with the systematic and rigorous developments of the early 80's, see \cite{Damour:1982wm,Kennefick:1997kb,Will:2011nz} for discussions.

\subsection{Back-reaction}

Observable consequences of the emission of gravitational waves can be studied looking at how they impact the  dynamics of the source. This can be done for instance evaluating the first PM correction to the source trajectories, by solving
\be\label{naTexp}
0= \na_\m T^{\m\n} = \p_\m \bar T^{\m\n} + \G^{\sscr (1)}{}^\m_{\m\r} \bar T^{\r\n}+ \G^{\sscr (1)}{}^\n_{\m\r} \bar T^{\m\r} +\p_\m \bar T^{{\sscr(1)}\m\n}+O(h^2)
\ee
at various orders in the PN expansion. The details of this calculation can be found in \cite{poisson}. 
In some cases however, it is possible to consider the  following shortcut.
At zeroth order, we have the Newtonian dynamics, and this comes with a clear identification of conserved quantities such as energy $E$ and angular momentum $J$.
We then \emph{assume} that the first order correction is obtained allowing these quantities to be not conserved, and equating their change 
to the dissipation caused by gravitational waves. That is, we posit
\be
\dot E = - \dot  E_{\sscr GW}, \qquad \dot L^a  = - \dot L^a_{\sscr GW},
\label{Eloss}
\ee 
insert the expressions \eqref{Ptot} and \eqref{defJ} on the right-hand side, and solve the resulting equations. From the solutions we deduce how the source should change in time in order for its dynamics to be consistent with the dissipation caused by gravitational waves.
This is what we do in the next Section, and for which the total expressions \eqref{Ptot} and \eqref{defJ} are enough. Detailed calculations using the proper method \eqref{naTexp}, see e.g.  \cite{poisson}, confirm the validity of this shortcut, at least in so far as the lowest order in the PN expansion is concerned.

\section{GWs from binary systems: elliptical, circular and hyperbolic orbits}
\label{sec:CBCs}

We now apply the results of the previous section to determine the GW signal from binary systems. We will first consider the case of a bound system, with circular or elliptical orbits. These provide a simple yet realistic model of astrophysical sources that corresponds to the signals observed by LVK. We will see how one can express the two quadrupole formulas (and more generally the dissipation equations) in terms of the dynamics of the sources, compute the backreaction leading to orbital decay and increased wave emission, and produce analytic waveforms. We will also see explicitly the importance of the averaging procedure, which in the case of bound binary systems neatly separates the effects related to the two time-scales involved: the period of each orbit, and the `secular' effects that cumulate over many orbits. We will then consider the case of unbounded, hyperbolic orbits, produce their waveforms. These orbits are interesting because they provide the simplest  examples of displacement memory and gravitational capture. Throughout this Section, we will approximate the gravitational bodies with non-spinning point particles. This provides a good approximation at lowest order: detailed PN analysis shows that spinning and finite-size effect only enter at higher orders.

\subsection{Newtonian equations}

\begin{figure}[ht]\centering
  \includegraphics[width=7cm]{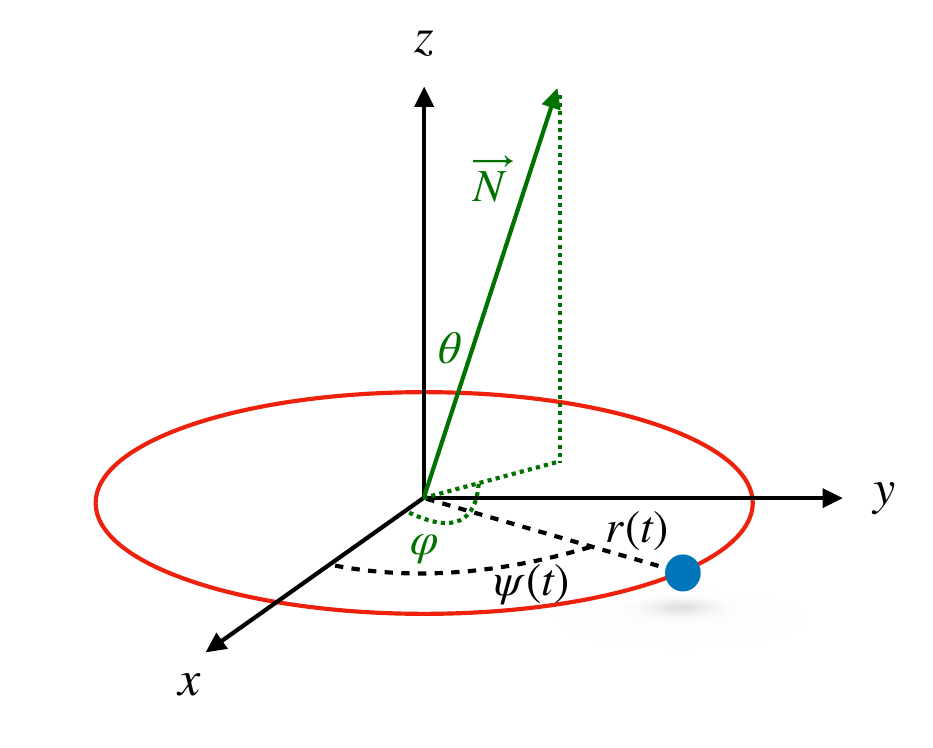}\hspace{1cm}\includegraphics[width=7cm]{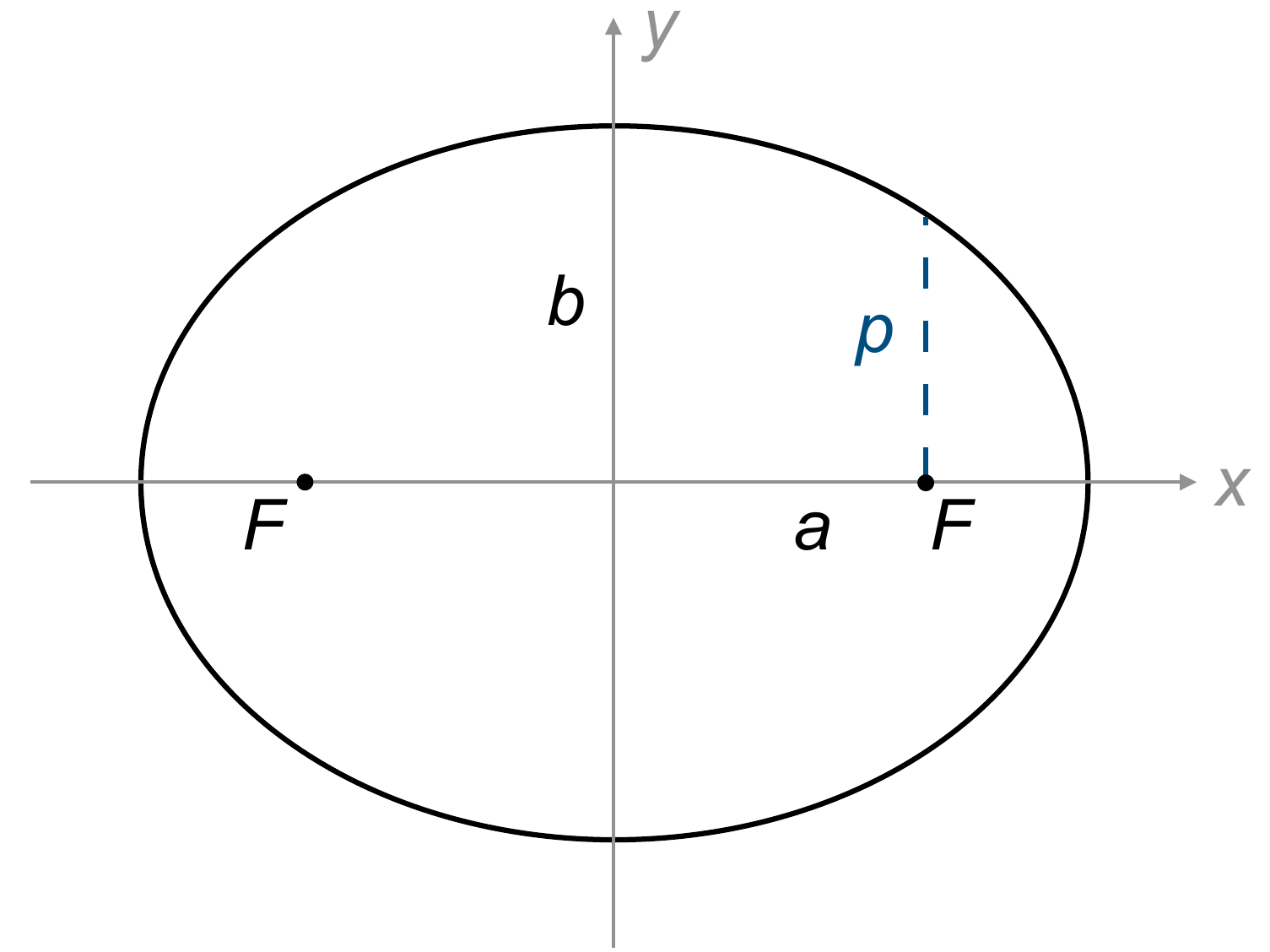}
  \caption{Left panel: {\it Bound binary system in the centre-of-mass frame: basic quantities and angles.} Right panel: {\it Geometric quantities of an ellipse.}} 
    \label{fig:setup}
\end{figure}

We first recall the Newtonian equations of motion for  two non-spinning point-particles of masses $m_{1,2}$, with relative position $\vec{r} = \vec{x}_1 - \vec{x}_2$ and relative velocity $\vec{v} = \vec{v}_1 - \vec{v}_2$. In the center-of-mass (CM) frame, the dynamics can be described by a single particle with position $\vec r$ and 
reduced mass 
\be
\m:=\frac{m_1 m_2}{m}, \qquad 
\nu:= \frac{m_1 m_2}{m^2}, \qquad m=m_1+m_2.
\ee
The dimensionless quantity $\n$ is introduced here for later convenience. The total energy $E=\f12\m v^2-G\m m/r$ and angular momentum $\vec L=\m\vec r\times\vec v$ are conserved. The latter implies that the motion is confined to a plane, and we choose coordinates so that this is the $(x,y)$ plane. We then parametrize
\be
     \vec{r} = r \vec{n}, \qquad \vec{n} = (\cos\psi,\sin\psi,0),
    \label{eq:vecr}
\ee
and introduce a second vector to form an orthogonal basis in the plane of the dynamics:
\be    \vec{v} = \dot{r}\vec{n} + r \dot{\psi} \vec{\lambda}, \qquad \vec{\lambda} = (-\sin\psi,\cos\psi,0).
    \label{eq:vecv}
\ee
The observer's detector is at position $\vec{R}=R\vec{N}$ where in spherical coordinates the unit vector is $\vec{N}=(\sin\varphi \cos\theta, \sin \varphi \sin\theta,\cos\varphi)$, see figure \ref{fig:setup}.
Newton's equations in the CM frame can be conveniently rewritten as
\begin{eqnarray}
    r(\psi) &=& \frac{p}{1+e\cos(\psi)},
    \label{eq:r}
    \\
    \dot{\psi} &=& \sqrt{\frac{G m}{p^3}} (1+e\cos(\psi))^2
    \label{eq:dotp}
\end{eqnarray}
where $e=\sqrt{1-(b/a)^2}$ is the \emph{eccentricity}, and $p=a(1-e^2)$ the \emph{semi-latus rectum}, of an ellipse with semi-major axis $a$ and semi-minor axis $b$,
see right panel of Fig.~\ref{fig:setup}.
These geometric quantities are related to the physical conserved quantities by 
\be
E=\f{G\m m}{2a}=\frac{G\m m}{2p}(e^2-1)=\nu \frac{Gm^2}{2p}(e^2-1).
\label{eq:energyKepler}
\ee
and
\be
\vec{L} = L \vec{e}_z, \qquad L = \m \sqrt{Gmp} =\n  \sqrt{Gm^3p}.
\label{eq:Lorb}
\ee
It follows from Eqs.~\eqref{eq:vecr}-\eqref{eq:dotp} that 
\be
\vec{v} = \sqrt{\frac{Gm}{p}}\left(-\sin\psi, e+\cos\psi,0\right),
\label{eq:v}
\ee
where we have chosen the origin $\psi=0$ at periastron, 
and
\be
\frac{|\vec{v}|^2}{c^2} = \left(\frac{Gm}{c^2 p} \right) \left( 1 + e^2 + 2e\cos \psi\right).
\label{eq:vello}
\ee
The Newtonian approximation requires $\vec{v}^2 \ll {c^2}$, and thus the dimensionless ratio $ {Gm}/{c^2 p}  \ll 1$.

Bound systems have eccentricity $e<1$, while unbound ones have $e>1$.  The border case $e=1$ corresponds to parabolic orbits.
\begin{itemize}
\item  {\bf Elliptical Orbits} have $0<e<1$ with $-\pi \leq \psi < \pi$, $r_{\rm min}=p/(1+e)$ and $r_{\rm max}=p/(1-e)$.  The orbital angular frequency $\omega_0$ and period $T$ satisfy Kepler's laws
\be
  \omega_0 = \sqrt{\frac{Gm(1-e^2)^3}{p^3}}  
   \, \qquad \text{and} \qquad T = \frac{2\pi}{\omega_0}.
 \label{eq:omega0}
\ee
\item {\bf Circular orbits} have $e=0$ and radius $r=p$, and  orbital frequency 
\be
  \omega_0 = \sqrt{\frac{Gm}{p^3}}.
 \label{omcirc}
\ee
\item {\bf Hyperbolic orbits} have $e>1$.  Now $\psi_-(e)\leq  \psi < \psi_+(e)$ where 
 \be
 \psi_\pm = \pm \arccos(1/e),
 \label{eq:psipm}
 \ee
and correspondingly $\sin\psi_{\pm} = 
 \pm e^{-1}{\sqrt{e^2-1}}$.
These orbits are not periodic, but have a characteristic time-scale (a burst time scale) related to the characteristic frequency scale
\be
\omega_c =   \sqrt{\frac{Gm(e^2-1)^3}{p^3} }.
\label{omegac}
\ee
  The closest distance of approach $r_{\rm min} =p/(1+e)$ at $\psi=0$.  As $\psi \rightarrow \psi_\pm$, $v \rightarrow v_\infty$ with 
\begin{equation}
    \frac{v_\infty^2}{c^2} = \frac{G m}{c^2 p}(e^2-1).
    \label{eq:vinfty}
\end{equation}
\end{itemize}

The GR modifications of the Newtonian equations of motion can be separated in two classes. First, the contributions of the potentials, of which the most famous is of course Einstein's initial precession calculation: over an orbital period the perihelion of elliptical orbits advances by $\Delta_h=2\pi(3Gm/c^2p)$, while for hyperbolic orbits $\Delta_h = (\Delta_r/3)\left\{ 6 \arccos(-1/e) + {e^{-2}}  {\sqrt{e^2-1}}  \left[  2(2+e^2) + 5\nu(e^2-1)     \right] \right\}$, see e.g.~\cite{Damour1985}. 
Second, non-conservative or dissipative effects caused by the emission of gravitational waves. In the following we will ignore the first, and focus on the second.
We will also ignore the precession introduced by dissipative effects, since this is higher order in the PN expansion.

\subsection{Energy and angular momentum fluxes}
 
In the quadrupole approximation, the TT component of the waveform is given by Eq.~\eqref{Qformula}, in terms of the traceless quadrupole tensor $Q_{ij} = I_{ij} - \frac{1}{3} I \delta_{ij}$. The mass quadrupole moment of the two point particles is
\be
{I}_{ab} = \m r_a r_b= \nu m r_a r_b.
\label{eq:I}
\ee
This is where the discussion of Section~\ref{SecVal} becomes crucial. If we are working in the linear approximation, we are only allowed to include in $I_{ab}$ the kinematical contribution from the velocities, and not the Newtonian gravitational potential, which would curve the geodesics away from the straight lines.
However going through the second iteration has ultimately the neat effect of restoring the quadrupole formula with the Newtonian potential included. We will make use of this fact, and assume in the following that the quadrupole formula was derived already going through the second iteration of the field equations,
and thus include the Newtonian potential in $I_{ab}$. Doing so, and using the equations of motion $d\vec{v}/dt = -Gm \vec{n}/r^{2}$, it follows that
\be
\ddot{I}_{ab} = 2\nu m \left( v_a v_b  - \frac{Gm}{r}n_a n_b\right).
\label{eq:ddI}
\ee
Thus from Eqs.~\eqref{eq:vecr} and \eqref{eq:v} the non-zero components of $\ddot{I}_{ab}$ are
\begin{subequations}\label{eq:ddotI22}
\ba
\ddot{I}_{11} &=& -2\n m c^2 \left(\frac{ Gm}{c^2 p}\right) \left[\cos(2\psi) + e\cos^3\psi \right],
\\
\ddot{I}_{12} &=& -2\nu m c^2 \left(\frac{ Gm}{c^2 p}\right) \left[\sin(2\psi) + e\sin\psi(1+\cos^2\psi) \right],
\\
\ddot{I}_{22} &=&2\nu m c^2 \left(\frac{ Gm}{c^2 p}\right) \left[\cos(2\psi) + e\cos\psi(1+\cos^2\psi) + e^2 \right],
\ea\end{subequations}
and $\ddot I=2\nu m c^2 \left(\frac{ Gm}{c^2 p}\right)e(e+\cos\psi)$.
We have written the coupling constants as
\be
\f{Gm_1m_2}p = m c^2 \left(\frac{ Gm}{c^2 p}\right) \nu,
\ee
in order to highlight that $\ddot{I}_{ab}$ has the dimensions of energy. The third derivatives of the quadrupole tensor are straightforwardly obtained from Eq.~\eqref{eq:ddotI22} and \eqref{eq:dotp} and read
\begin{subequations}\label{eq:trippleI}
\ba
\dddot{I}_{11} &=&2\nu (mc^2) \frac{c}{p} \left(\frac{ Gm}{c^2 p}\right)^{3/2} (1+e\cos\psi)^2  \left[2\sin(2\psi) + 3e\cos^2\psi \sin \psi \right]
\\
\dddot{I}_{12} &=&2\nu (mc^2) \frac{c}{p} \left(\frac{ Gm}{c^2 p}\right)^{3/2} (1+e\cos\psi)^2  \left[-2 \cos(2\psi) + e\cos\psi(1-3\cos^2\psi) \right]
\\
\dddot{I}_{22} &=& -2\nu (mc^2) \frac{c}{p} \left(\frac{ Gm}{c^2 p}\right)^{3/2}(1+e\cos\psi)^2  \left[2\sin(2\psi)+ e\sin\psi(1+3 \cos^2\psi) \right]
\ea\end{subequations}

\begin{figure*}[t]\centering
  \includegraphics[width=16cm]
  {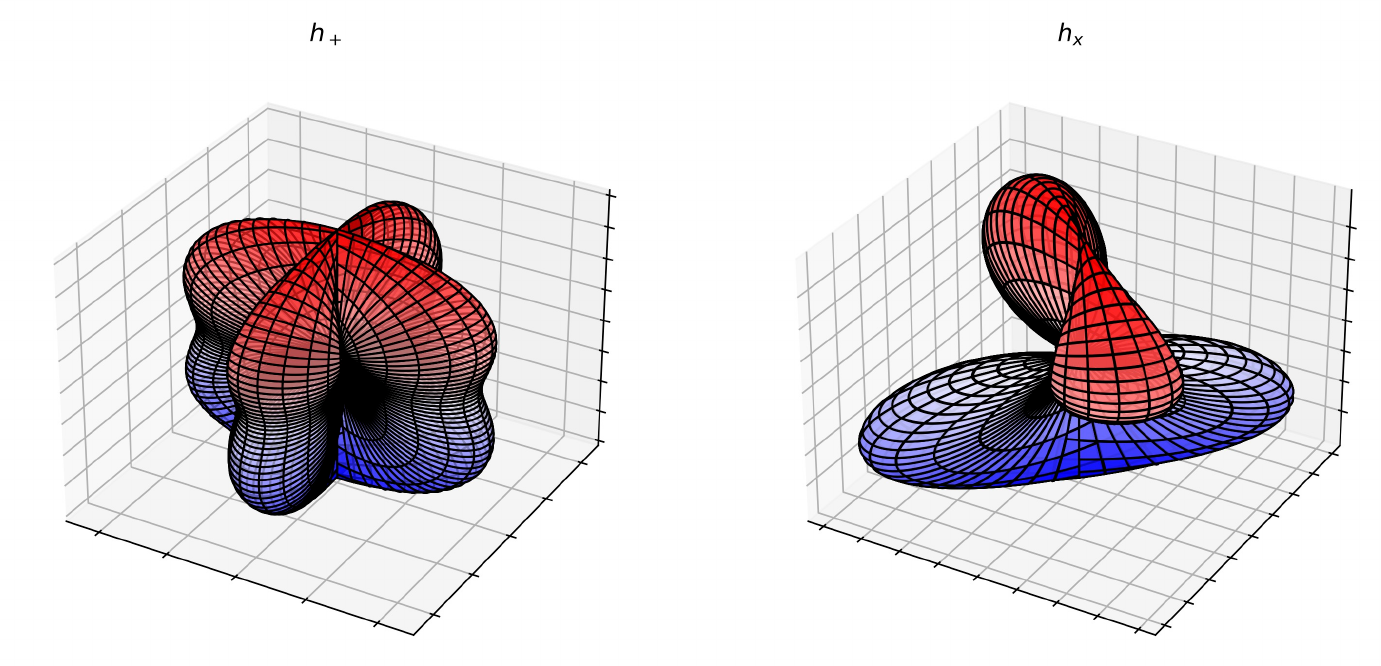}
  \caption{\small\emph{\it Surfaces of constant $h_+$ and $h_\times$ polarisations for circular orbits $e=0$. Here have chosen $\psi=\pi$. }}
    \label{fig:polarisns}
\end{figure*}

The GW perturbation is given by substituting these expressions into Eq.~\eqref{Qformula}. In the direction $\vec{N}=\hat{z}$ the plus and cross polarisations are given by (see Eq.~\eqref{eq:inzdirn}),
\ba
h_+(t) &=& \frac{G}{c^4 R} (\ddot{I}_{11} - \ddot{I}_{22} ) =-h_0 \left. \left[2\cos(2\psi) + e\cos\psi +2e\cos^3\psi  +e^2 \right] \right|_{t_{\sscr R}}
\label{eq:hpelliptical}
\\
h_\times(t) &=& \frac{2G}{c^4 R} \ddot{I}_{12} = -2 h_0 \left.  \left[\sin(2\psi) + e\sin\psi(1+\cos^2\psi) \right] \right|_{t_{\sscr R}}.
\label{eq:hxelliptical}
\ea
where the dimensionless amplitude is
\be
h_0 = \f{2G^2m_1m_2}{c^4Rp}= 2 \nu \left( \frac{G m }{c^2 R} \right) \left(\frac{ Gm}{c^2 p}\right) .
\label{eq:h0ampl}
\ee
The time-dependence is determined from $\psi(t)$ which is a solution of Eq.~\eqref{eq:dotp}. The polarizations $h_{+,\times}(t,\theta,\varphi)$ in an arbitrary direction $\vec{N}=(\sin\th \cos\varphi, \sin \th\sin\varphi,\cos\th)$ can be obtained plugging \eqref{eq:ddotI22} in \eqref{hgendir}, or written directly in terms of (\ref{eq:hpelliptical}-\ref{eq:hxelliptical}) using \eqref{hgendir2}. 
Figure \ref{fig:polarisns} shows surfaces of constant $h_{+,\times}$ as a function of $(\theta,\varphi)$, for a fixed value of $\psi=\pi$ and $e=0$. The quadrupolar nature is clearly visible.  

\begin{figure*}[t]\centering
  \includegraphics[width=13cm]
  {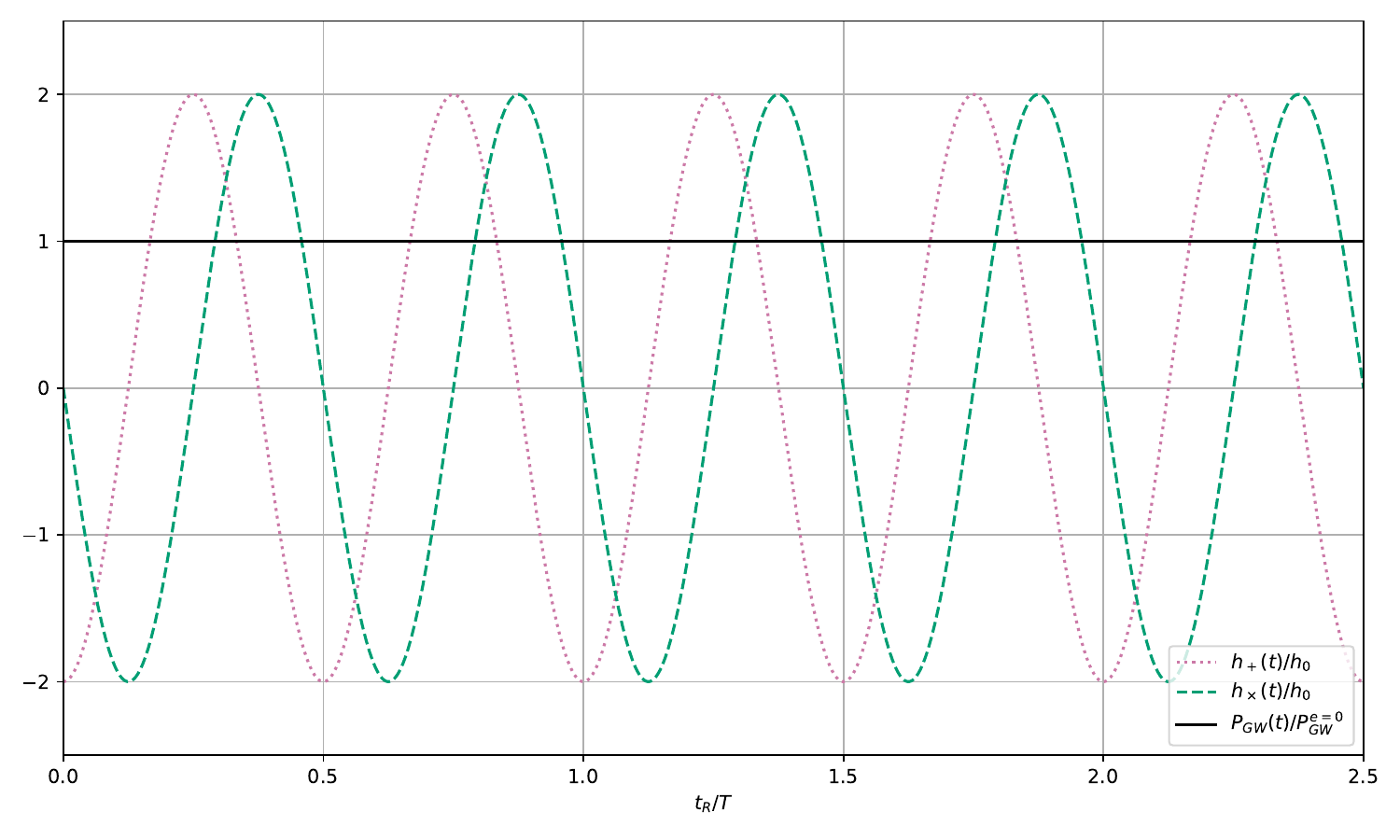}
  \caption{\small\emph{\it Circular orbits. The plot shows $h_+$ and $h_\times$ polarisations, and emitted GW power (solid line), as a function of retarded time in units of $T$ for 2.5 orbital periods. The emitted power is constant and given by Eq.~\eqref{Powero}. The GW wavelength is $cT/2$.}}
    \label{fig:hpx0}
\end{figure*}

\begin{figure*}[t]\centering
  \includegraphics[width=13cm]
  {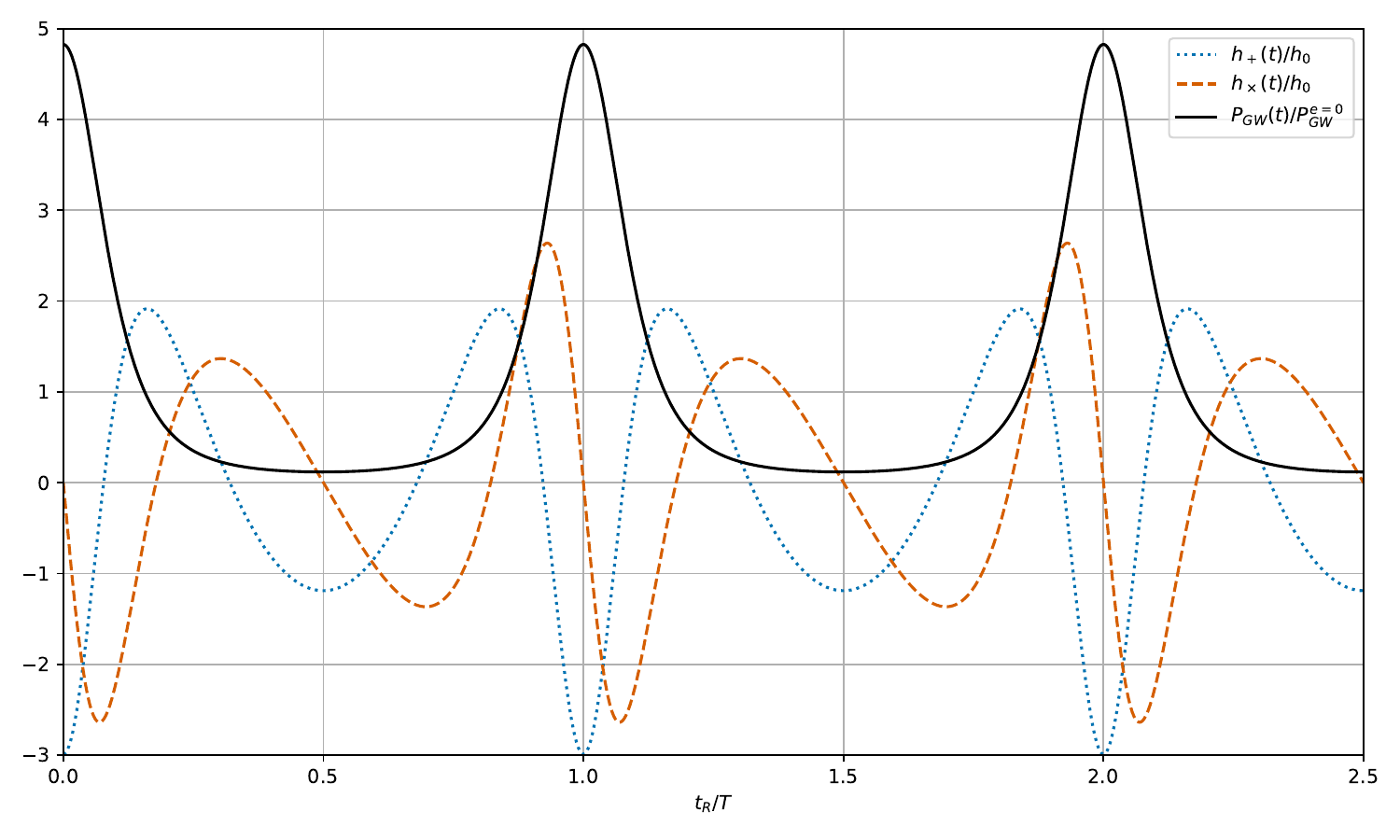}
  \caption{\small\emph{\it Elliptical orbit with $e=0.3$. The plot shows  $h_+$ and $h_\times$ polarisations, and emitted GW power, as a function of retarded time in units of $T$ for 2.5 orbital periods. The emitted power is largest at periastron $\psi=0$ (mod $2\pi$) where the orbital velocity is the largest.}}
    \label{fig:hpx03}
\end{figure*}

For circular orbits ($e=0$), only the terms in $\cos(2\psi)$ and $\sin(2\psi)$ remain, and furthermore the angular velocity is constant, in particular $\psi = \omega_0 t$ from \eqref{eq:dotp}. Thus {\it for circular orbits, the GW angular frequency is twice the orbital frequency}: 
\be\label{om2om0}
\om = 2 \omega_0.
\ee   Figure \ref{fig:hpx0} shows the waveforms as a function of retarded time, in units of $T$, for 2.5 orbital periods.  For elliptical orbits additional frequencies are present, both larger and smaller than $\om_0$. In fact since the angular velocity is not constant, infinitely many harmonics are emitted.
These are shown in Fig.~\ref{fig:hpx03}.

\begin{figure*}[t]\centering
  \includegraphics[width=13cm]
  {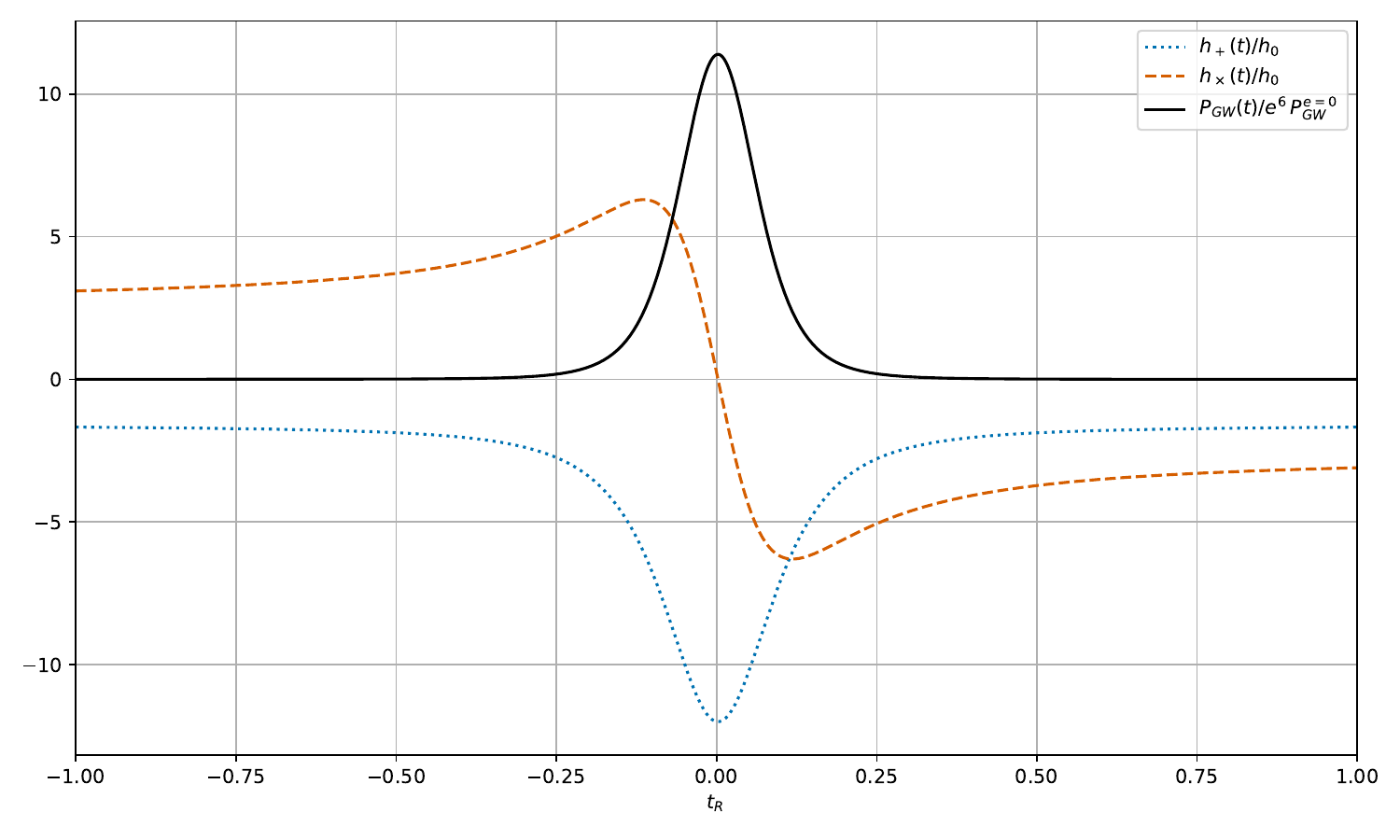}
  \caption{\small\emph{\it Hyperbolic orbits with $e=2$. The $h_+$ and $h_\times$ polarisations and emitted GW power as a function of retarded time, for an initial value $\psi = \psi_-$ see Eq.~\eqref{eq:psipm}.  The motion is no-longer periodic and a burst of GW energy emitted when $\psi=0$ at $t=0$.  Note that the emitted power scales as $e^6$ for large $e$, see Eq.~\eqref{eq:PGWelliptical}, and for that reason in the plot the power is normalised by an extra factor of $e^6$.  See section \ref{subsec:hyper} }}
    \label{fig:hpx-hyperbolic}
\end{figure*}

In the quadrupole approximation, the energy and angular momentum fluxes \eqref{Energyloss1} and \eqref{Jloss} read
\ba
P_{\rm{GW}}(\psi)  &=& \frac{G}{5c^5}  \dddot{Q}_{ab}  \dddot{Q}^{ab} =
 \frac{2G}{15c^5} \left[ \dddot{I}_{11}^2 + \dddot{I}_{22} ^2 + 3  \dddot{I}_{12}  \dddot{I}_{12} -  \dddot{I}_{11}  \dddot{I}_{22}\right]
 \label{energyflux}
\\
\dot{L}^z_{\rm{GW}}(\psi) &=&\frac{2G}{5c^5}\epsilon^{zab} \ddot{I}_{ac} \dddot{I}_{b}{}^c =\frac{2G}{5c^5}\left[ (\ddot{I}_{11}-\ddot{I}_{12}) \dddot{I}_{12}  + \ddot{I}_{12}(\dddot{I}_{22}-\dddot{I}_{11}) \right]
\label{angularMflux}
\ea
(only the $z$-component of angular moment is relevant since the binary is in the $xy$-plane ).  Substituting Eq.~\eqref{eq:trippleI} gives
\ba
P_{\rm{GW}}(\psi(t)) &=&P_{\rm{GW}}^{e=0} (1+e\cos\psi)^4\left. \left[ 1+2e\cos\psi + \frac{e^2}{12}(1+11\cos^2\psi) \right]\right|_{t_{\sscr R}}
\label{eq:PGWelliptical}
\\
\dot{L}^z_{\rm{GW}}(\psi(t))  &=& \dot{L}_{\rm{GW}}^{e=0} (1+e\cos\psi)^3 \left. \left[ 1 + \frac{3}{2}e\cos \psi - \frac{e^2}{4}(1-3\cos^2\psi)\right]\right|_{t_{\sscr R}}
\label{eq:JGWelliptical}
\ea
where for circular orbits the constant rates of emission are given by
\ba
P_{\rm{GW}}^{e=0}   &=&  \frac{32}5\f{G^4}{c^5}\f{\m^2m^3 }{p^5}  = \frac{32}{5}  \left( \frac{ c^5}{G}\right) \left(\frac{ G m}{c^2 p}\right)^{5}
\label{Powero}
\\
\dot{L}_{\rm{GW}}^{e=0} &=& \frac{32}5\f{G^{7/2}}{c^5}\f{\m^2m^{5/2} }{p^{7/2}}  = \frac{32}{5} \nu^2 (m c^2) \left(\frac{Gm}{c^2p}\right)^{7/2} 
\ea
in terms of the dimensionless coefficient $G m/c^2 p$.
The above expressions are valid for all $e\geq 0$ provided $\vec{v}^2 \ll c^2$.  
The power emitted is constant for circular orbits, and maximal at periastron for elliptical orbits.
Figure \ref{fig:hpx0} shows the waveforms Eq.~\eqref{eq:hpelliptical}-\eqref{eq:hxelliptical} and power emitted Eq.~\eqref{eq:PGWelliptical} over 2.5 periods of a circular orbit with $e=0$, in unit of $t_R/T$ where $T$ is the orbital period.  Figure \ref{fig:hpx03}  shows the same for an elliptical orbit with $e=0.3$.  In both cases the periodic motion is clear. Over longer time-scales $t \gg T$, however, the emission of energy and angular momentum backreact on the orbital trajectories and must be considered.  For a hyperbolic orbit, the corresponding plots are given in figure \ref{fig:hpx-hyperbolic}. The motion is obviously no-longer periodic and simply amounts to a fly-by: thus back-reaction effects do not accumulate over time and will be less significant (see subsection \ref{subsec:hyper}).

We now evaluate the effect of energy and angular momentum dissipation on the system's dynamics in order to study the back-reaction.

\subsection{Back-reaction and waveform: circular orbits}
\label{subsec:circular}
The effect of the gravitational wave emission produces an effect that modifies the dynamics of the system, known as \emph{radiation-reaction force}.
As explained at the end of the previous Section, this effect can be deduced equating the non-conservation of the energy and angular momentum with the corresponding gravitational wave fluxes, see \eqref{Eloss}, and solving to find how the orbital element change in time. In order to do so we have to replace the constant orbital elements with functions of time, a method already used to compute perturbations to Keplerian orbits and known as `osculating orbits'. For circular orbits, there is only one independent quantity, it is thus sufficient to look at the energy equation.
On the source side, we have
\eqref{eq:energyKepler}
\be
\dot E = - \f Ep \dot p.
\ee
On the gravitational wave side, we have \eqref{Powero}. Equating the two gives
\be
\dot p = \frac{64}5\f{G^3}{c^5}\f{\m m^2 }{p^3},
\ee
and this tells us how the orbit evolves. Using Kepler's law \eqref{omcirc}, we can obtain the evolution of the angular frequency as
\be
\dot \om_0 = A\om_0^{11/3}, \qquad A=\f{96}5\f{G^{5/3}\m m^{2/3}}{c^5} = \f{96}5 \left( \frac{G {\cal{M}}}{c^3} \right)^{5/3},
\ee
where ${\cal{M}}$ is the chirp mass \eqref{eq:chirpmass}. 
This in turns tells us the change in the frequency of the emitted waves, which as seen in \eqref{om2om0} is twice the orbital frequency:
\be
\dot{\omega}  = 2^{-8/3}A\omega^{11/3} = \frac{12}{5} 2^{1/3} \left( \frac{G {\cal{M}}}{c^3} \right)^{5/3} \omega^{11/3}, 
\label{eq:omegaevoln}
\ee
or equivalently in terms of $f=\omega/2\pi$,
\be
\dot{f} = \frac{96}{5} \pi^{8/3} \left( \frac{G {\cal{M}}}{c^3} \right)^{5/3} f^{11/3}. 
\label{eq:glo}
\ee
This shows that as gravitational waves are emitted, the orbit decays, the angular frequency increases, and so does the GW's frequencies as well, leading to an even greater orbital decay. This run-away effect stops when the orbit decays completely and the two body coalesce. Clearly before that happens, higher order effects become important and one should improve the calculation. It is however instructive to get the lowest-order approximation and a qualitative overall picture to assume that the  quadrupole approximation is valid throughout the evolution. We can then integrate between an initial time $t$ and the coalescing time $t_c$, where the frequency formally diverges. This gives
\be
f(t) = \frac{1}{\pi}\left(\frac{5}{256 (t_c-t)} \right)^{3/8} \left( \frac{G {\cal{M}}}{c^3} \right)^{-5/8}.
\label{eq:chirpf}
\ee
Thus for a binary inspiral on a circular orbit $f^{-8/3}$ is linear in time, with a slope which determines directly the chirp mass.
From this solution we can also immediately deduce the evolution of the orbital frequency $\om_0$, the orbital radius $p$, and the GW amplitude \eqref{eq:h0ampl}.

To get the complete waveform, we only need the time dependence of $\psi$. This can be computed observing from \eqref{eq:dotp} that $\dot \psi = \om_0$, 
therefore
\ba
\Phi(t) :=2\psi= 2\pi \int_{t}^{t_c} dt' f(t')+\Phi_c = -2 \left( \frac{5G{\cal M}}{c^3} \right)^{-5/8} (t_c-t)^{5/8}+\Phi_c,
\label{eq:Phiev}
\ea
where $\Phi_c$ is the phase at coalescing time.
Combining these results, we get for the polarizations (\ref{eq:hpelliptical}-\ref{eq:hxelliptical}) in the direction perpendicular to the orbital plane
\ba
h_+(t) &=&- \frac{4}{R}  \left( \frac{G{\cal M}}{c^2} \right)^{5/3} \left(\frac{\pi f(t_{\sscr R})}{c} \right)^{2/3} \cos \Phi(t_{\sscr R}),
\label{hplus1}
\\
h_\times(t) &=&- \frac{4}{R}  \left( \frac{G{\cal M}}{c^2} \right)^{5/3} \left(\frac{\pi f(t_{\sscr R})}{c} \right)^{2/3} \sin \Phi(t_{\sscr R}).
\label{htimes1}
\ea
For circular orbits, it is also easy to give the polarizations in an arbitrary direction $\vec{N}=(\sin\th \cos\varphi, \sin \th\sin\varphi,\cos\th)$, using Eq.~\eqref{hgendir2}: 
\ba
h_+(t) &=& \frac{4}{R}  \left( \frac{G{\cal M}}{c^2} \right)^{5/3} \left(\frac{\pi f(t_{\sscr R})}{c} \right)^{2/3} \cos [\Phi(t_{\sscr R}) -2 \varphi+\pi] \left(\frac{1 + \cos^2 \theta}{2} \right),
\label{hplus}
\\
h_\times(t) &=& \frac{4}{R}  \left( \frac{G{\cal M}}{c^2} \right)^{5/3} \left(\frac{\pi f(t_{\sscr R})}{c} \right)^{2/3} \sin [\Phi(t_{\sscr R}) -2 \varphi+\pi] \cos \theta.
\label{htimes}
\ea
As described after \eqref{hgendir2}, the angle $\theta$ can be identified with the inclination $\iota$ of the source relative to the detector, see Fig.~\ref{fig:characteristic-GW}, whereas $2\om=-2\varphi+\pi$ is the longitude of pericenter. Notice also that $\om$ can be absorbed into a redefinition of the initial time.
This is the analytic expression of the curve plotted in blue in figure \ref{fig:h}.

\subsection{Back-reaction: Elliptical orbits}

The case of elliptic orbits is more intricate, and offers a few interesting insight into the dynamics: there is a full spectrum of emission, and not a monochromatic one, which shows up in a non-symmetric waveform; there is loss of both energy and angular momentum, and at different rates, which shows up in the orbital back-reaction losing eccentricity faster than it decays; GW emission is not constant during the orbit, but stronger at periastron. 

To analyse the system, it is convenient to distinguish the effects on two different time scales: short-time effects, namely the variations within a single orbit; and long-time, or \emph{secular}, effects, namely the cumulative changes over many orbits. For instance, the dependence of the power emitted on the position $\psi$ illustrated in Fig.\ref{fig:hpx03} is a short-time effect. The orbital decay on the other hand is a secular effect. To consider secular effects, we introduce the average over one orbital period:
\be
\langle X \rangle = \frac{1}{T} \int_0^T dt X(t) = \frac{1}{T} \int_{-\pi}^{\pi} d\psi \frac{1}{\dot{\psi}} X(\psi),
\ee
where $\dot{\psi}$ is given in Eq.~\eqref{eq:dotp}.  We then replace \eqref{Eloss} with their time-averages,
\be
\dot E  = - \langle P_{\sscr GW} \rangle \qquad \dot L  = - \langle \dot{L}_{\sscr{GW}} \rangle.
\label{eqs:averaged}
\ee 
The aim of this subsection is to solve these equations to determine the secular evolution of $e(t)$, $p(t)$, and thus $h_{+,\times}(t)$ with backreaction included.

Substituting \eqref{eq:PGWelliptical} and integrating gives the {\it{Peter-Mathews}} formula \cite{Peters1963}:
 \be
\langle P_{\rm{GW}} \rangle = P_{\rm{GW}}^{e=0} 
  \; (1-e^2)^{3/2}
   \left[ 1 + \frac{73}{24}e^2 + \frac{37}{96}e^4 \right] .
 \label{eq:PM}
\ee
(This expression is only valid for $e< 1$ as we are dealing with elliptical orbits.) Keeping $p$ constant, the radiation increases from $e=0$, to a maximum at $e\sim 0.5$ before decreasing and vanishing at $e=1$.  The averaged angular momentum radiation is similarly determined using \eqref{eq:JGWelliptical} and gives
\be
\langle \dot{L}_{\rm{GW}} \rangle =\dot{L}_{\rm{GW}}^{e=0} \;
  (1-e^2)^{3/2} \left[ 1+\frac{7}{8} e^2 \right].
\label{eq:Jrad}
\ee

We now return to Eqs.~\eqref{eqs:averaged}, where on the left hand side the time-dependence is in $e(t)$ and $p(t)$.  By definition, see \eqref{eq:Lorb}, $L=\nu \sqrt{Gmp}$ from which
\be
\frac{dL}{dt} = \frac{\nu c}{2} \sqrt{\frac{Gm}{c^2 p}}  \frac{dp}{dt}.
\ee
This combined with Eqs.~\eqref{eqs:averaged} and Eq.~\eqref{eq:Jrad} gives
\be\label{pdot}
\frac{dp}{dt} = - \frac{64}{5} \nu c \left( \frac{Gm}{c^2 p}\right)^3 (1-e^2)^{3/2} \left[ 1 + \frac{7}{8}e^2 \right].
\ee
The energy of the orbit is given in Eq.~\eqref{eq:energyKepler}, from which
\be
\dot e = \f{1}{\nu G m^2 }\frac{p}{e}\dot E - \f{\dot{p}}{2pe}(1-e^2).
\ee
Then plugging in \eqref{eq:PM} and \eqref{pdot} gives
\be
\frac{de}{dt} = - \frac{304}{15} \nu c \left( \frac{e}{p}\right) \left( \frac{Gm}{c^2 p}\right)^3 (1-e^2)^{3/2} \left[ 1 + \frac{121}{304}e^2 \right].
\label{eq:dedt}
\ee
These coupled equations Eq.~\eqref{pdot} and \eqref{eq:dedt} can be solved  using hypergeometric function to get $e(t)$ and $p(t)$, or alternatively combined to determine $p(e)$.  

Observe that both $p(t)$ and $e(t)$ decrease with time.  An elliptical orbit with initial eccentricity $e \neq 0$ will thus become more circular due to GW radiation. Whereas an initially circular orbit with $e=0$ remains circular for all times.  {This is the reason why often it is a good approximation to consider circular orbits, particularly when studying the last moments before the merger of the binary system. (This is the case of the events observed by LVK.) 
At first order in $e$, we can approximate \eqref{eq:dedt} with
\be
\frac{de}{dt} \sim -\frac{e}{\tau_R},
\qquad \tau_R := \frac{1}{\nu} \left(\frac{Gm}{c^2 p}\right)^{-5/2} \frac{T}{2\pi},
\ee
where $T$ is the orbital period \eqref{eq:omega0}. This approximation provides a time-scale, $\tau_R$, of the radiative decay of $e$ and $p$. From Eq.~\eqref{eq:vello}, $ \left(\frac{Gm}{c^2 p}\right) \sim |\vec{v}|^2/c^2 \ll 1$, and thus $\tau_R \sim (c/v)^{-5} T \gg T$.

The decrease of $p$ and $e$ also implies that $T$ decreases with time.  Indeed from Eqs.~\eqref{eq:omega0}, \eqref{eq:dedt} and \eqref{pdot}
\be
\frac{dT}{dt} = -\frac{192}{5}\pi \left( \frac{G {\cal{M}}}{c^3} \frac{2\pi}{T}\right)^{5/3} \left[  \frac{1+ \frac{73}{24}e^2 + \frac{37}{96} e^4}{(1-e^2)^{7/2}}\right]
\ee
where ${\cal{M}}$ is the chirp mass \eqref{eq:chirpmass}.  Thus the orbital frequency $\omega_0$ increases, and GWs are emitted with increasing frequencies. The amplitude also increases, since has we can see from \eqref{eq:h0ampl} it is inversely proportional to $p$.

To have the quantitative behaviour of the waveforms $h_{+,\times}$, we also need the time evolution of $\psi$. This is obtained solving \eqref{eq:dotp} with $e(t)$ and $p(t)$ the solutions of Eqs.\eqref{eq:dedt} and \eqref{pdot}, and can be done numerically. The result is plotted in Fig.~\ref{fig:h}, for 4 different initial values of the eccentricity $e=0, 0.3, 0.5$ and $0.7$, for the plus polarization  in the $z$ direction  \eqref{eq:hpelliptical}.
The upper waveform in Fig.~\ref{fig:h}  is for circular orbits. The increasing amplitude and frequency of the GWs is clearly visibile and will be quantified in the discussion below.  The waveform diverges when $p$ reaches zero, though clearly this is beyond the regime of applicability of the quadrupole approximation which assumes $|\vec{v}|/c \ll 1$. Since $|\vec{v}| \sim  1/\sqrt{p}$ this is clearly violated as $p \rightarrow 0$. That is the reason why, in Section \ref{sss:mergerfreq}, we invoked the ISCO as a possible minimum distance, which then defined a merger frequency through Eq.~\eqref{eq:fmerger}.
The elliptic orbits are not symmetric, and the amplitude is maximal at the periastron, where we also have the maximum emitted power. The different plot also show that the waveform diverges at earlier and earlier times as $e$ increases, in agreement with the fact that the emitted powers increase with $e$. 

To get also a geometric intuition about the dynamics of the system during evolution, we plot in figure \ref{fig:horbits} the orbits for $e=0$ and $e=0.3$, showing the decrease in orbital radius and eccentricity.

\begin{figure*}[t]\centering
    \includegraphics[width=12cm] {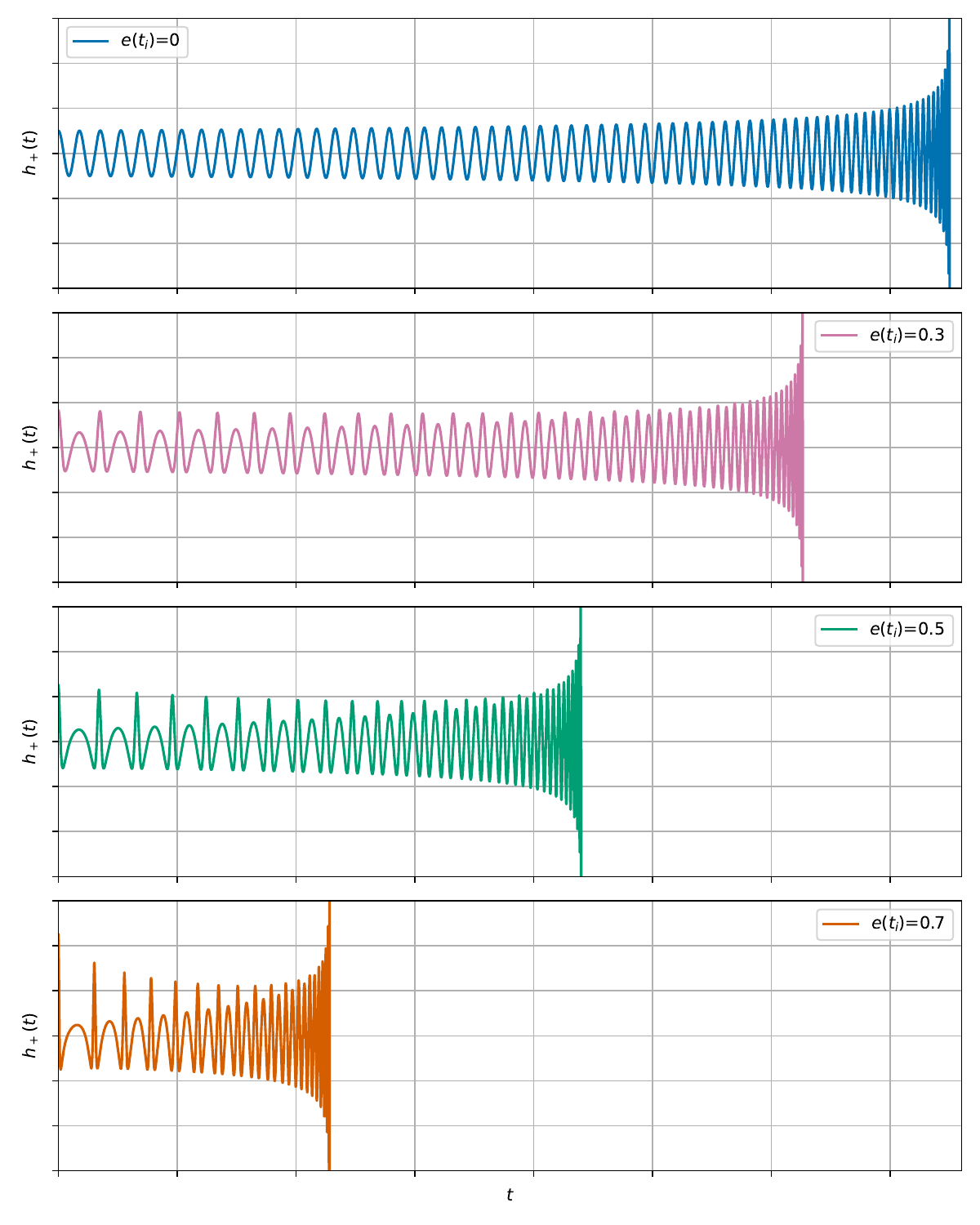}
  \caption{\small\emph{\it Four waveforms, in the lowest order PN expansion, with initial values of eccentricity given by $e=0, 0.3, 0.5$ and $0.7$. Most GW power is emitted near the pericenter where the orbital velocity is the largest. Also since more GW radiation is emitted as $e$ increases, the merger occurs earlier.} }
    \label{fig:h}
\end{figure*}

\begin{figure*}[t]\centering
  \includegraphics[width=14cm]
  {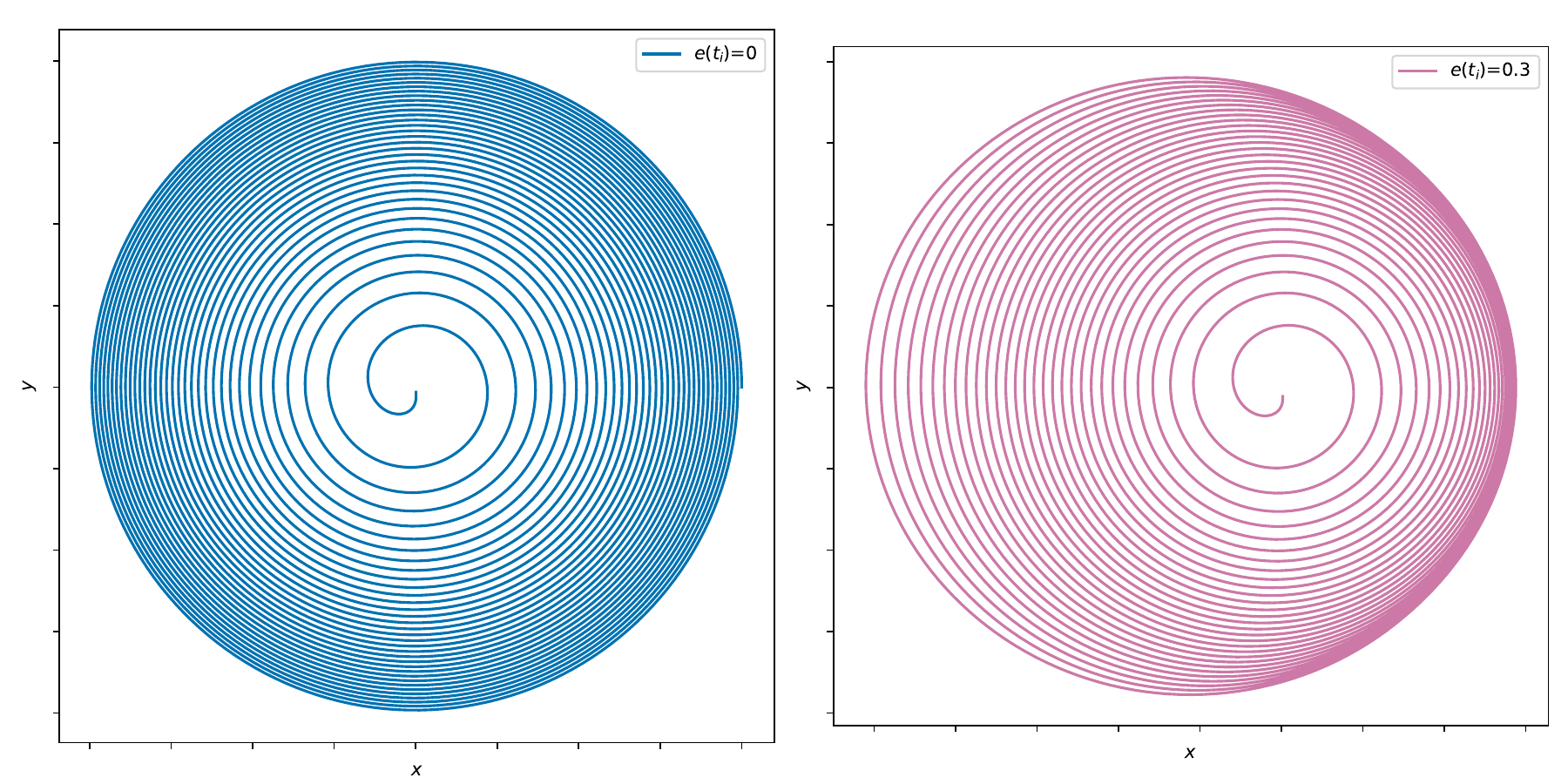}
  \caption{\small\emph{\it Corresponding to Fig.~\ref{fig:h}, the orbits $r(t)$ given by Eq.~\eqref{eq:vecr} for $e=0$ and $e=0.3$ } }
    \label{fig:horbits}
\end{figure*}

\subsection{Frequency content of elliptical orbits}
\label{sss:fk}\label{ss:powerelliptical}

As a side result, let us come back to elliptic orbits, without dissipation, and discuss how one can study the frequency content of the waves emitted
by computing the contributions at each frequency to the total power \eqref{energyflux}.
To that end, we expand the dynamics into Fourier modes multiples of the fundamental harmonics given by the orbital frequency $\omega_0 = \sqrt{Gm/p^3}$. So for instance we write
\be
I_{ab}(t) = \nu m p^2 \left\{ \tilde{A}_{ab}^{(0)} + \sum_{n=1}^{\infty}\left[ \tilde{A}_{ab}^{(n)} \cos(n \omega_0 t)  + \tilde{B}_{ab}^{(n)} \sin(n \omega_0 t) \right] \right\},
\label{eq:heree}
\ee
with the dimension-full factor in front so to have dimension-less Fourier components
\ba
&& \tilde{A}_{ab}^{(0)} = \frac{1}{\nu m p^2 T} \int_0^{T} dt I_{ab}(t), \\
&& \tilde{A}_{ab}^{(n)} = \frac{2}{\nu m p^2 T} \int_0^{T} dt I_{ab}(t) \cos(\omega_0 n t),
\quad \tilde{B}_{ab}^{(n)} = \frac{2}{\nu m p^2 T} \int_0^{T} dt I_{ab}(t) \sin(\omega_0 n t). 
\ea
Recall that $I_{ab}$ is given by \eqref{eq:I} with $\vec{r}$ parametrized as in \eqref{eq:vecr}. To solve this integral one then needs the explicit time dependence of $r$ and $\psi$, namely the solution of the (unperturbed) Keplerian orbit of ellipticity $e$. 
Details can be found in e.g.~\cite{Maggiore1}, and one finds
\be
0= \tilde{B}^{(n)}_{11}=  \tilde{A}_{12}^{(0)} = \tilde{A}_{12}^{(n)}=\tilde{B}^{(n)}_{22}, \qquad  \tilde{A}_{11}^{(0)} = \frac{1+4e^2}{2}, \qquad    \tilde{A}_{22}^{(0)} = \frac{1}{2}
\ee
and
\ba
\tilde{A}_{11}^{(n)}&=&\frac{1}{(1-e^2)^2}\cdot \frac{1}{n} \left[ J_{n-2}(ne) - J_{n+2}(ne) - 2e(J_{n-1}(ne) - J_{n+1}(ne)) \right],
 \\
 \tilde{B}_{12}^{(n)}&=&\frac{1}{(1-e^2)^{3/2}}  \cdot \frac{1}{n} \left[ J_{n+2}(ne) + J_{n-2}(ne) - e(J_{n+1}(ne) + J_{n-1}(ne)) \right],
  \\
\tilde{A}_{22}^{(n)}&=&\frac{1}{(1-e^2)}\cdot \frac{1}{n} \left[ J_{n+2}(ne) - J_{n-2}(ne) \right],
     \ea
     in terms of the Bessel's functions $J$.
Taking three time derivatives leads to factors of $n^3\omega_0^3$. Substituting into \eqref{energyflux} and averaging over one period to calculate $\langle P_{\rm GW}\rangle$ leads to terms such as $\langle \sin n\omega_0 t  \sin m\omega_0 t \rangle \sim \delta_{mn}$ meaning that the different harmonics do not interfere. In conclusion one finds
\be
\langle P_{\rm{GW}} \rangle = \sum_{n=1}^{\infty} \langle P_n \rangle
\ee
where 
\be
\langle P_n \rangle = P_{\rm{GW}}^{e=0}\cdot \frac{n^6}{96} (1-e^2)^4 \left[\left(\tilde{A}_{11}^{(n)}\right)^2+\left(\tilde{B}_{12}^{(n)}\right)^2 +3 \left(\tilde{A}^{(n)}_{22}\right)^2  - \tilde{A}_{11}^{(n)} \tilde{B}_{12}^{(n)} \right],
\ee
where $P_{\rm{GW}}^{e=0}$ is given in Eq.~\eqref{Powero}. 
The $\langle P_n \rangle$ are plotted in figure \ref{fig:GWpower-elliptical} for different values of $e$. 
For circular orbits all power is emitted via the $n=2$ mode. The power is the more and more distributed among other harmonics as $e$ increases, and the frequency at which the maximum power is radiated also increases with $e$. 
\begin{figure*}[t]\centering
  \includegraphics[width=12cm]
  {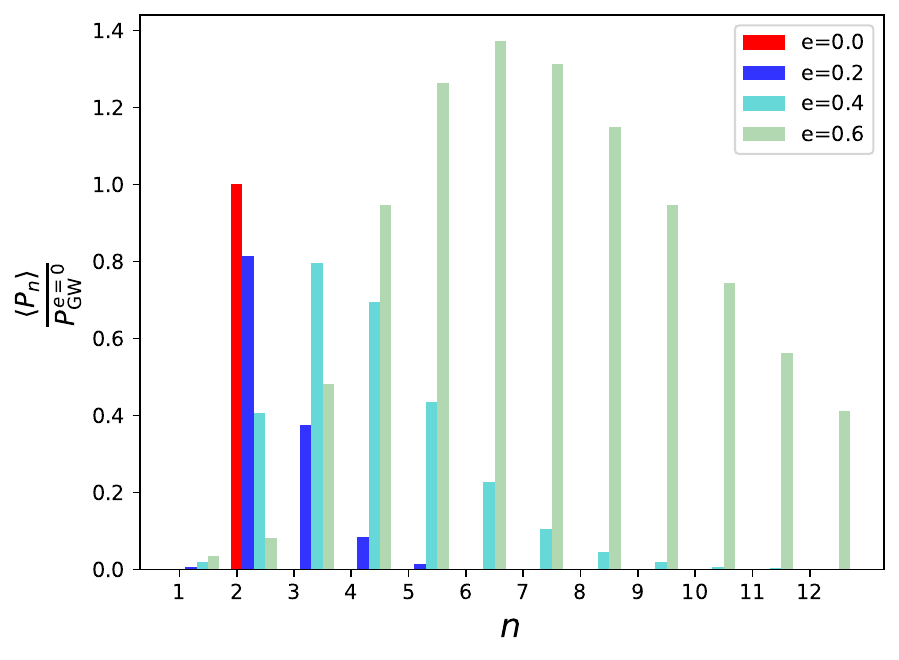}
  \caption{\small\emph{\it Plot of $\langle P_n \rangle/P_{\rm{GW}}^{e=0}$ for different values of $e=0$ (red), $e=0.2$ (blue), $e=0.4$ (cyan) $e=0.6$ (green). } }
    \label{fig:GWpower-elliptical}
\end{figure*}

\subsection{Hyperbolic orbits}
\label{subsec:hyper}
While all GW detections to date are from {\it bound} elliptical/circular CBCs with $e<1$, many other potential GW sources exist, for instance non-spherical spinning NSs and asymmetric core collapse Supernovae. In this brief subsection we discuss another possible source, namely {\it unbound binary systems} on {\it hyperbolic orbits}. That is, we consider cases in which the eccentricity $e>1$ see Eq.~\eqref{eq:energyKepler} and Fig.~\ref{fig:cfig-hyperbolic-orbit}.

Hyperbolic orbits are interesting not only because unbound orbits are expected to exist in nature (and hence the waveform for such events is and will be searched for by GW detectors \cite{NANOGrav:2023vfo,Gasparotto:2023fcg,Goncharov:2023woe,Inchauspe:2024ibs}), but also because this simple system provides a first example of a {\it gravitational wave memory effect}. There are many different kinds of memory effects (see e.g.~\cite{Favata:2010zu}), the simplest of which is the linear memory effect which occurs already at the lowest order in the PN expansion and which is illustrated by hyperbolic orbits.
 Memory effects occur when there is a permanent change $\Delta h_{ab}^{\rm{TT}}$ in the gravitational  waveform, and thus leads to permanent displacement $\Delta L$ of the arms of GW detector for example, see Eq.~\eqref{Lwave}.

\begin{figure*}[t]\centering
  \includegraphics[width=6cm]
  {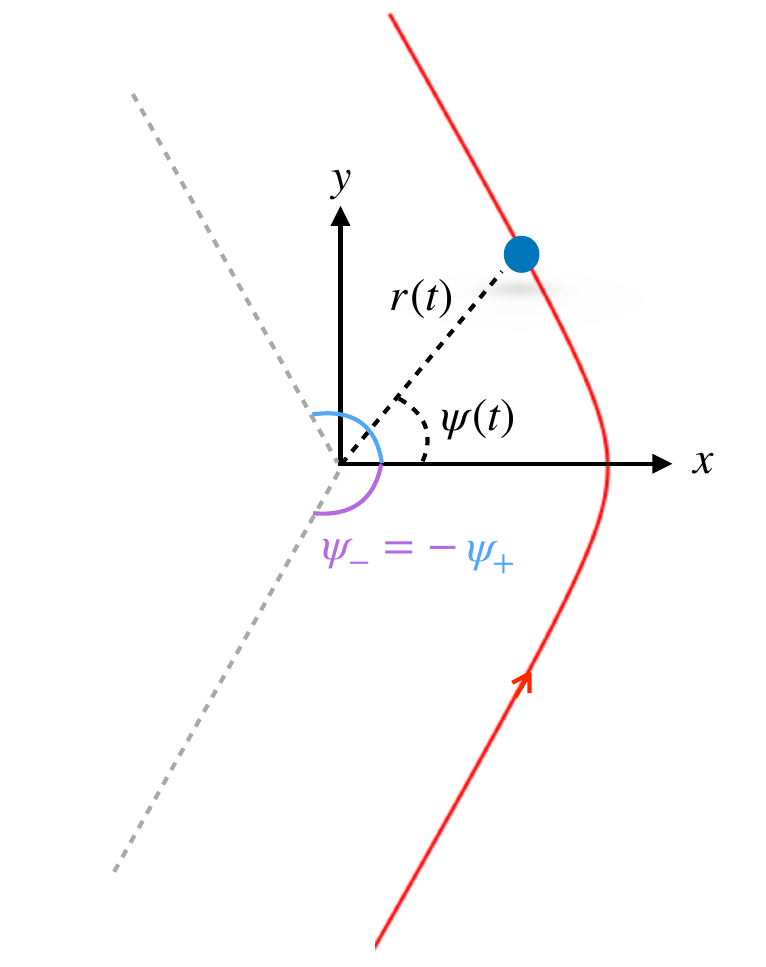}
  \caption{\small\emph{\it Sketch of a hyperbolic orbit in the CM, showing the angles $\psi_{\pm} =\pm \arccos(-1/e)$. The scattering angle is $\psi_+-\psi_--\pi=2\arccos(-1/e)-\pi=2\arcsin(1/e).$}} 
    \label{fig:cfig-hyperbolic-orbit}
\end{figure*}

\subsubsection{Linear memory effect and low-frequency GWs}

The two GW polarisations for hyperbolic orbits are shown in Fig.~\ref{fig:hpx-hyperbolic} for $\vec{N} = (0,0,1)$.  The waveform is not periodic but rather burst-like, and    $h_\times$ has a non-zero variation between $t=
\pm \infty$: this is the linear memory effect.
More generally, the variation of the metric perturbation between $t=
\pm \infty$ is given by
\be
\Delta h^{\text{TT}}_{cd}= \int_{-\infty}^{\infty} {\rm d}t \, \dot h^{\text{TT}}_{cd}(t) = h_{cd}^{\text{TT},+} - h_{cd}^{\text{TT},-},
\label{eq:thisoneidiot}
\ee
where we introduced the shorthand notation $ h_{cd}^{\text{TT},\pm}$ for the evaluation at $t=\pm\infty$.
It follows from \eqref{Qformula} that there will be a linear memory effect $\Delta h^{\text{TT}}_{ij} \neq 0$ when there is a {\it net change in the second time
derivatives of quadrupole moments} of the system. This is precisely the case for hyperbolic orbits, since at initial and final times the accelerations are zero, but the velocity vectors are different. 
At lowest PN order,
\begin{eqnarray}
    \Delta h^{\text{TT}}_{cd} &=& \frac{2G}{c^4 R} P^{\sscr TT}{}^{ab}_{cd}(\vec{N}) (\ddot{I}_{ab}^+ - \ddot{I}_{ab}^-)
= \frac{4G }{c^4 R} P^{\sscr TT}{}^{ab}_{cd}(\vec{N}) \sum_{i=1}^2 m_i v^i_a v^i_b\Big|^\infty_{-\infty}
= \frac{4G\m }{c^4 R}  v_\infty^2  P^{\sscr TT}{}^{ab}_{cd}(\vec{N}) n_a n_b\Big|^\infty_{-\infty},   \label{eq:realsp}
\end{eqnarray}
where $v_\infty$ is given in \eqref{eq:vinfty} and the unit-norm $n_a$ refers to incoming and outgoing directions.
From this one can compute the polarizations, for instance in the $\hat z$ direction,
\be
\D h_+(\hat z)= \frac{16G\m v_\infty^2}{c^4 R}   \frac{1-e^2}{e^4},
\qquad \D h_\times(\hat z)= \frac{8G\m v_\infty^2}{c^4 R}   \frac{(e^2-2)\sqrt{e^2-1}}{e^4}.
\ee
See  \cite{Braginsky:1987kwo} for more details.

The time-scale over which the GW signal varies in Fig \ref{fig:hpx-hyperbolic}, namely the burst time-scale, is determined by the inverse of the characteristic frequency $\nb$ given in Eq.~\eqref{omegac}.  
Up to factors of eccentricity, this also determines the characteristic frequency scale of the emitted GWs on hyperbolic orbits. Indeed, contrary to the case of periodic elliptical orbits discussed in section \ref{sss:fk}, GWs of all continous frequencies are emitted, and one can determine the GW power as a function of frequency by now Fourier transforming the power emitted.  Using the convention $\tilde{I}_{ab} (\omega)=\int dt I_{ab}(t) e^{-i\omega t}$, as well as the quadrupole approximation, the total energy emitted in GWs is
\ba
E_{\rm GW} &=& \frac{G }{5c^5} \int_{-\infty}^{\infty} dt (\dddot I_{ab})^2
=\frac{G  }{5 \pi c^5} \int_0^{\infty} {d} \omega \, \omega^6 \vert I_{ab}(\omega)\vert^2
\equiv \int_0^{\infty} {d} \omega \, \pgw(\omega) .
\nn
\ea
The emitted power in GWs is thus
\be
\pgw(\omega) = \frac{G}{5 \pi c^5} \left[ \omega^3 \tilde{I}_{ab}(\omega)\right]\left[ \omega^3 \tilde{I}_{ab}^*(\omega) \right].
\label{eq:Pomega}
\ee
Direct calculation analogous to that of section \ref{ss:powerelliptical} (see e.g.~\cite{Brax:2024myc}) shows that $\pgw(\omega)$ is peaked at a value fixed by $\omega_c$ but which increases with $e$, see figure \ref{fig:energy-hyperbolic-new}.    Notice that $\pgw(0)\neq 0$. This is due to the linear memory effect: indeed Eq.~\eqref{eq:thisoneidiot}, written in Fourier space reads
\be
\Delta h_{cd} ^{\rm TT}= -i \frac{2G }{c^4 R}P^{\sscr TT}{}^{ab}_{cd}(\vec{N})  \left. \left[\omega^3 \tilde{I}_{ab}(\omega) \right] \right|_{\omega=0},
\label{eq:nlm}
\ee
thus a non-vanishing linear memory effect implies $\pgw(0)\neq 0$.

\begin{figure*}[t]\centering
  \includegraphics[width=8cm]
  {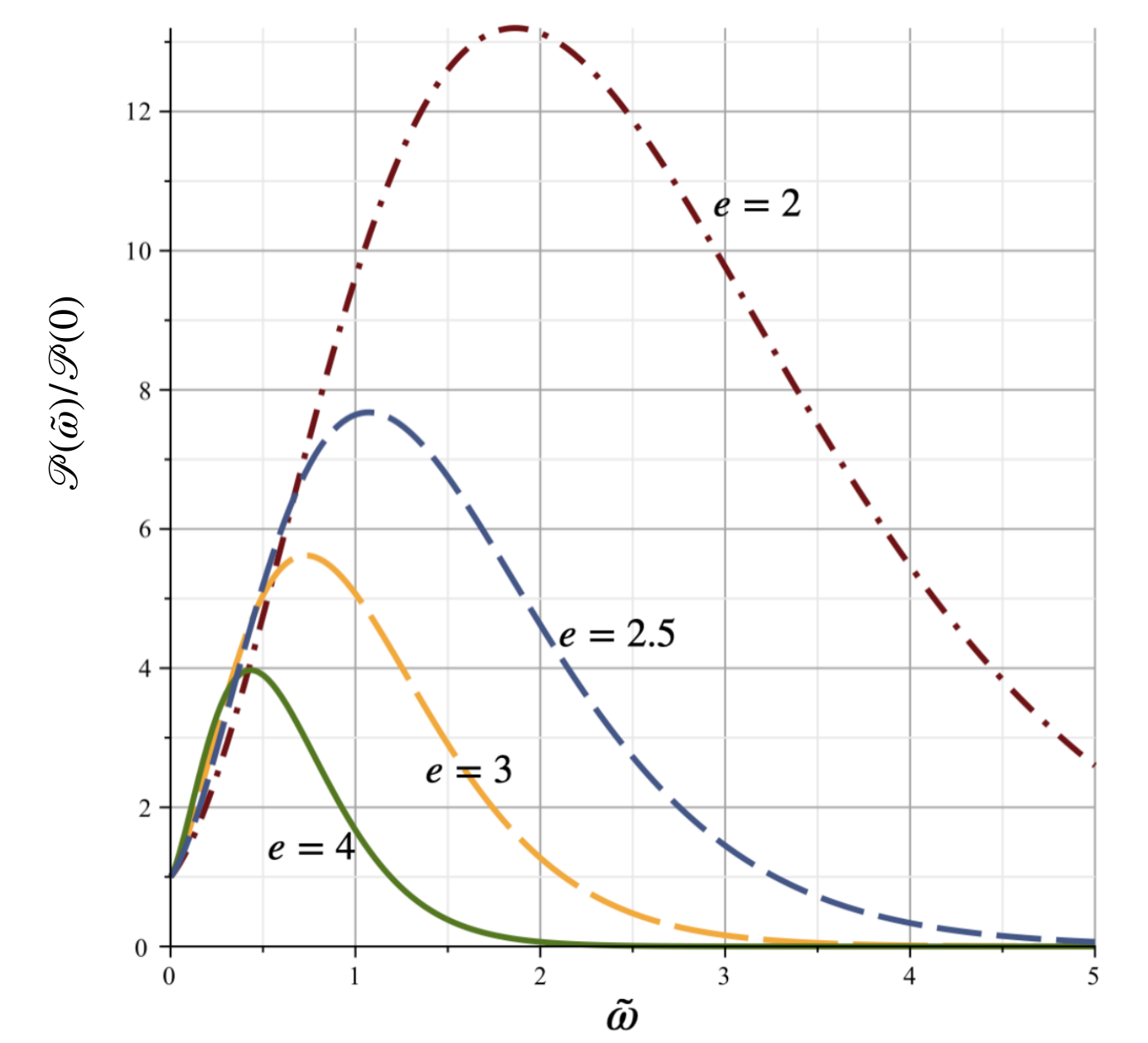}
  \caption{\small\emph{\it The energy spectrum $\pgw(\tilde{\omega})/\pgw(0)$ where $\tilde{\omega} = \omega / \omega_c$ for different eccentricities. Figure from \cite{Brax:2024myc}.} }
    \label{fig:energy-hyperbolic-new}
\end{figure*}

\subsubsection{Capture due to GW emission}

One can estimate the energy emitted in GWs between $\psi_- < \psi < 0$ by calculating
\be
\Delta E_{\rm{GW}} = 
\int_{\psi_-}^{0} d\psi \frac{1}{\dot{\psi}} P_{\rm{GW}}(\psi)
\label{eq:cap}
\ee
where $P_{\rm{GW}}(\psi)$ is given in Eq.~\eqref{eq:PGWelliptical}.  Let us consider an orbit which is only slightly unbound, thus $e=1+\epsilon$ with $0<\epsilon \ll 1$ so that the orbital energy is $E \simeq \nu Gm^2 \epsilon/p$, see Eq.~\eqref{eq:energyKepler}.  Then it is straightforward to determine $\Delta E_{\rm{GW}}$ to lowest order in $\epsilon$, leading to
\be
\Delta E_{\rm{GW}}  = \frac{85}{3} mc^2 \nu^2 \left( \frac{Gm}{c^2 p} \right)^{7/2} \pi + {\cal{O}}(1).
\ee
Note that here we assumed $p$ constant, that is we have neglected the backreaction of the emitted GR on $p$: this is a reasonable approximation for this individual burst process since $p$ changes on a time-scale $\sim (c/v)^{-5}\omega_c^{-1}$ as discussed previously.  The orbital energy $E \sim \nu Gm^2 \epsilon/p$ will thus be reduced by $\Delta E_{\rm{GW}}$, and if
\be
E - \Delta E_{\rm{GW}} < 0
\ee
then the GW energy loss will convert the hyperbolic orbit into a bound orbit before the pericenter.   This can be rewritten as the condition
\be
\epsilon < \frac{85\nu}{3} \left( \frac{Gm}{c^2 p} \right)^{5/2} \pi.
\label{eq:eps}
\ee

Instead of parametrising the orbit in terms of $(p,\epsilon)$, for such scattering trajectories it is more convenient to work with $(b,v_{\infty})$ where $b$ is the impact parameter and $v_{\infty}$ is the orbital velocity at infinite separation. From Eq.~\eqref{eq:vinfty}, $v^2_{\infty} \simeq 2 (Gm/p)\epsilon$ and substituting $\epsilon$ from Eq.~\eqref{eq:eps} gives
\be
 \left( \frac{Gm}{c^2 p} \right)^{7/2}> \frac{3}{170 \pi \nu }\left( \frac{v_{\infty}}{c}\right)^2.
\label{eqq:eq}
\ee
Furthermore, basic trigonometry gives
\be
b = \frac{p}{e\sin \psi_-} = \frac{p}{\sqrt{e^2-1}} \simeq \frac{p}{\sqrt{2\epsilon}} = \frac{\sqrt{Gmp}}{v_{\infty}}.
\ee
Saturating the bound in \eqref{eqq:eq} identifies the impact parameter $b_{\rm capture}$ for which capture occurs. The corresponding capture cross-section $\sigma_{{\rm GW}} := \pi b_{\rm capture}^2$ is thus given by
\be
\sigma_{{\rm GW}} = \pi \left(\frac{170\pi \nu }{3}\right)^{2/7} \left(\frac{Gm}{c^2}\right)^2 \left(\frac{c}{v_{\infty}}\right)^{18/7}.
\ee
Such a process could play an important role for instance in dense star clusters and galactic nuclei, see e.g.~\cite{Capozziello:2008ra,OLeary:2008myb,Hoang:2020gsi}.

\section{GWs in curved space-time, cosmology}
\label{sec:cosmology}

In the previous sections we discussed GWs in Minkowski space; our aim is now to generalise the results presented there to a cosmological space-time.
There is now a further scale of interest other than the characteristic size $d$ of the source and the GW wavelength $\lambda_{\rm{GW}}$, namely the {\it cosmological horizon}.  The results of the previous sections are valid in the so-called the {\it local wavezone} of the source, namely at distances scales $R$ which are large compared to the GW wavelength but {\it small} compared to the cosmological horizon, $d\ll \lambda_{\rm GW} \ll R \ll {\rm horizon}$.  We now aim to extend them to cosmological scales: we will see that the expansion of the universe dampen the GW amplitude, and redshift frequencies and masses.  

\subsection{General background metric}

\subsubsection{Linearised equations}

We now consider the linearized equations around an arbitrary curved background $\bar{g}_\mn$ satisfiying $\bar{G}_{\mn} = (8\pi G/c^4)\bar{T}_{\mn}$.  Thus as in \eqref{EElin1} $g_{\m\n}=\bar g_{\m\n} + h_{\m\n}$, and similarly the source is perturbed as ${T}_{\mn} = \bar{T}_{\mn} + T^{\sscr (1)}_{\m\n}$. The linearised equations are derived explicitly in Appendix~\ref{app:perturbed-action}, and they can simplified using the de Donder gauge \eqref{DeDg}, generalised to curved spacetime as\footnote{We are using a bar to indicate both the background metric, and the trace-reversed perturbation. We believe that no confusion should arise, because of the different places where these quantities enter.}
\be
\bar{\nabla}^\mu \bar{h}_\mn = 0, \qquad \bar{h}_{\m\n}:= h_\mn - \frac{1}{2}\bar{g}_\mn h.
\label{eq:LorenzC}
\ee
In this gauge, we have (see Eq.~\eqref{eq:bloodyfinalSO} for details)
\be
\bar{\square} \bar{h}_{\alpha \beta}  + 2 \bar{R}^{\mu}{}_{\a \nu \b} \bar{h}^\n{}_{\mu} + 2 \bar{G}_{\mu (\alpha}\bar h_{\b)}{}^{\nu}  - \bar g_{\a\b} \bar{R}_{\mu \nu} \bar{h}^{\mu\nu} = -\f{16\pi G}{c^4} T^{\sscr (1)}_{\ab}.
\label{eq:g}
\ee
Here $\bar{R}_{\mu \alpha \nu \beta}$ is the Riemann curvature tensor of the background space-time, and $\bar{R}_\mn$ the Ricci-tensor. The terms  containing the background curvature depend on the ratio of $\lambda_{\rm{GW}}$ relative to the scale of variation of the background metric.   In the following we will assume that they are negligible. The justification for this is that below we will restrict attention to the homogeneous and isotropic FLRW metric, where the scale of variation of the background is the cosmological horizon, which is much greater than $\lambda_{\rm{GW}}$.\footnote{In a perturbed FLRW metric, there could be local bumps in the curvature of scale similar to $\lambda_{\rm{GW}}$: we do not consider this case here.
See also \cite{Isaacson:1968hbi} for more details.} 
The right-hand side has generically two contributions, one from the metric perturbations, and one from the matter perturbation. 
In cosmology, it is often useful to approximate matter using a perfect fluid description, for which
\be
T_{\m\n} =( \r +p) u_\m u_\n+p g_{\m\n}, \qquad u^2=-c^2,
\ee
where $\r$ and $p$ are respectively the energy and pressure densities of the fluid. In this case, $T^\m{}_\n$ is metric-independent, 
and this allows us to separate the metric and matter perturbations using the identity 
$T^{\sscr (1)}_{\m\n}=\bar g_{\m\r}T^{{\sscr (1)}\r}{}_\n+h_\m{}^{\r}\bar T_{\r\n}$. On-shell, the second term is proportional to the background curvature and therefore can be neglected again by the same argument as given above. We conclude that in 
the absence of matter perturbations, \eqref{eq:g} reduces to
\be
\bar{\square} 
\bar{h}_\ab \simeq 0.
\label{eq:smooth}
\ee

\subsubsection{WKB approximation}

For a general spacetime, the solution of Eq.~\eqref{eq:smooth} can be obtained in the WKB approximation. The underlying physical assumption is that the amplitude of the wave is slowly varying with respect to the frequency of the wave, and hence we write 
\be
\bar{h}_\mn(x) =\sum_p \Re \left[A_p(x) \epsilon^p_\mn(x) e^{iS(x)/\delta}\right]
\label{eq:WKB}
\ee
where  the sum is over polarisations $p$ with polarisation tensor $\epsilon_\mn$ satisfying $\epsilon_\mn \epsilon^\mn=1$ (recall that to this leading order in $h$, indices are raised and lowered with the background metric) and $A$ is the corresponding amplitude. The parameter  $\delta \rightarrow 0$, and we define
\be
k^\mu = \frac{\partial_\mu S}{\delta}.
\ee
Now substituting \eqref{eq:WKB} the de Donder condition Eq.~\ref{eq:LorenzC} becomes (dropping the $p$ sum for simplicity)
\be
 \bar{\nabla}^\mu \bar{h}_\mn = \left[ \bar{\nabla}^\mu(A\epsilon_\mn) + i A \epsilon_{\mn}\frac{\partial_\mu S}{\delta}\right] e^{iS(x)/\delta} = 0,
 \ee
 which to leading order in $\delta$ implies
 \be
 k^\mu \epsilon_\mn = 0
 \ee
and hence that $\epsilon_\mn$ is the transverse polarisation tensor. 
Substitution into the equation of motion Eq.~\eqref{eq:smooth} leads to terms in $\delta^{-2}$, $\delta^{-1}$ which are, respectively
\ba
\bar{g}^\mn k_\nu k_\mu &=& 0,
\label{eq:eikonal}
\\
2 \bar{\nabla}_\mu(A \epsilon_\ab) k^\mu + (A \epsilon_\ab) \bar{\nabla}_\mu k^\mu &=& 0.
\label{eq:inv}
\ea
 The first equation, the Eikonal equation (geometric optics limit), implies that GWs are massless with dispersion relation $\omega^2 = \vec{k}^2$ and propagate on null geodesics.  Contracting the second equation \eqref{eq:inv} with $\epsilon^{\ab}$, and using that $\epsilon_\mn \epsilon^\mn=1$ so that $(\bar{\nabla}^\alpha \epsilon_\mn) \epsilon^\mn = 0$, leads to 
\be
2(\bar{\nabla}_\mu A) k^\mu + A  \bar{\nabla}_\mu k^\mu  = 0 \qquad \Rightarrow \qquad  \bar{\nabla}_\mu(A^2 k^\mu) = 0.
\label{eq:amplitudeD}
\ee
This gives the decay of the GW amplitude $A$ along the null geodesics.
Finally, substituted back Eq.~\eqref{eq:amplitudeD} into Eq.~\eqref{eq:inv} gives
\be
k^\mu (\bar{\nabla}_\mu  \epsilon_\ab) = 0
\ee
which implies that the polarization tensor $\epsilon_\ab$ of the GW is parallel propagated along the null geodesics. 

To summarise, the solution of 
\be
\bar{g}^\mn \bar{\nabla}_\mu\bar{\nabla}_\nu\bar{h}_\ab \simeq 0
\label{eq:smoothbis}
\ee
in the WBK approximation, and in the Lorenz gauge, is $\bar{h}_\mn(x) = \Re \left[A(x) \epsilon_\mn(x) e^{ik_\mu x^\mu }\right]$ with
\ba
\bar{g}^\mn k_\nu k_\mu &=& 0
\\
k^\mu \epsilon_\mn &=& 0
\\
  \bar{\nabla}_\mu(A^2 k^\mu) &=& 0
  \\
  k^\mu (\bar{\nabla}_\mu  \epsilon_\ab) &=& 0
\ea
We now consider these equations in a FLRW metric.

\subsection{FLRW metric: background}

The flat Friedmann-Lemaitre-Robertson-Walker (FLRW) metric is 
\be
d\bar{s}^2 = -dt^2 + a^2(t) d\vec{x}^2 = -dt^2 + a^2(t)(dr^2 + r^2 d\Omega^2)
\label{eq:FLRW}
\ee
where $\vec{x}$ are comoving coordinates and $a(t)$ is the scale factor (normalised such that today, at $t=t_0$, $a(t_0)\equiv a_0=1$).  
For a perfect fluid source, Einsteins equations Eq.~\eqref{EE1} reduce to the Friedmann equations
\ba
H^2 &=& \frac{8\pi G}{3}\rho + \frac{\Lambda}{3}
\\
\frac{\ddot{a}}{a} &=& - \frac{8\pi G}{6}(\rho + 3P) + \frac{\Lambda}{3}
\ea
where $\rho$ and $P$ are respectively the energy density and pressure of the perfect fluid, and $H = \frac{\dot{a}}{a}$ is the Hubble parameter, whose value today is the Hubble constant $H_0$.  These two equations imply the conservation equation $\dot{\rho} + 3H(\rho + P)=0 = \nabla_\nu T^\mn$.  In terms of conformal time $\eta$ defined by $d\eta = dt/a(t)$, the metric in Eq.~\eqref{eq:FLRW} is conformally related to the Minkowski metric
\be
d\bar{s}^2 \equiv \bar{g}_{\mn} dx^\mu dx^\nu = a^2(\eta)[-d\eta^2 + dr^2 + r^2 d\Omega^2].
\label{eq:FLRWconformal}
\ee
Consider now a source (of photons or GWs) at fixed radial position $r=0$, and an observer at $r_o$. On a $(\eta,r)$ space-time diagram null radial geodesics propagate at 45 degrees. If two null geodesics are emitted at a conformal time interval $\delta \eta_s$ by the source, then they arrive at the observer with the same conformal time interval $\delta \eta_o = \delta \eta_s$. This implies the standard time-dilation relation 
\be
dt_o = \frac{a(t_o)}{ a(t_s)} dt_s \equiv (1+z) dt_s
\ee
where $z$ is the redshift of the source (the observer is at $t_o=t_0$), and equivalently that the emitted (or `source') frequency $f_s$ is related to the observed frequency $f_o$ by 
\be
f_o = \frac{f_s}{1+z}.
\ee
The radial comoving distance $R$ to an event with redshift $z$ is given by solving $ds^2=0$, thus $dr = dt/a(t)$, leading to
\be
R = \int dr = \int \frac{dt}{a(t)} = \int \frac{1}{a}\frac{dt}{da} \frac{da}{dz}dz =  \int_0^z dz' \frac{1}{H(z')}
\ee
where $H(z)= H_0 E(z) $ is the Hubble parameter expressed in terms of redshift, and from the  Friedmann equation 
\be
E(z) = \sqrt{\Omega_m(1+z)^3 + \Omega_r(1+z)^4 + \Omega_\Lambda}
\ee
where $\Omega_{r,m} = \frac{8\pi G \rho_{r,m}}{3 H_0^2}$, $\Omega_\Lambda = \frac{\Lambda}{3H_0^2}$ and $\Omega_r +\Omega_m + \Omega_\Lambda=1$.

A crucial quantity is the luminosity distance $d_L(z)$. This relates the EM luminosity of the source and the  luminosity measured by the observer. In the flat FRWL metric \eqref{eq:FLRW} it is given by
\be
d_L(z) =a(t_o) (1+z) R = (1+z) \int_0^z dz' \frac{1}{H(z')}.
\label{eq:dLdef}
\ee
As we will see, this same distance scale determines the GW amplitude in an expanding universe.

\subsection{FLRW metric: gravitational waveforms}

Consider a GW propagating radially outwards from the source at $r=0$ and redshift $z_s$ with $k_\mu = \omega(1,-1,0,0)$.  From Eq.~\eqref{eq:amplitudeD} it is possible to determine how the GW amplitude decreases along the null GW geodesic. In a FLRW metric (in conformal time) Eq.~\eqref{eq:amplitudeD}  becomes 
\be
\partial_\nu(\sqrt{-\bar{g}} A^2 k^\nu) = 0 = \partial_\nu(a(\eta)^2 A^2 r^2 k^\nu).
\ee
Thus $A(\eta,r) a(\eta) r$ remains conserved during the propagation, and
\be
A (\eta,r)=\left. \frac{\rm{const}}{a(\eta)r}\right|_{\eta-r={\rm const}}
\label{eq:amplitude}
\ee
The constant is fixed by the known amplitude of the wave in the wave-zone approximation, close to the source, where the Minkowski results are valid.  Then the remainder of the solution $\bar{h}_\mn$ is obtained by parallel transporting this solution from the source to the observer. We now carry out these steps.

Before doing so we note that in a flat FLRW universe and focusing on the {\it spatial} TT components only, then in fact \eqref{eq:g} reduces to 
\be
\bar{g}_\mn \bar{\nabla}^\mu\bar{\nabla}^\nu \bar{h}^{\rm{TT}}_{ij} =0 = h_{ij}^{{\rm TT}''} + 2 {\cal H} h_{ij}^{\rm{TT}'} + \partial_k \partial^k h_{ij}^{\rm{TT}}
\ee
since the spatial components of the Riemann and Ricci tensors vanish identically.  From here the scaling of the amplitude of GWs as $1/a(\eta)$ is also immediate.

We now consider a compact binary system on circular orbits, as discussed in section \ref{subsec:circular}, but this time in a FLRW background. In the wave-zone approximation and at a physical distance $R=a(t_s)r$ from the source as measured by time $t_s$ of the source clock, the plus and cross polarisations of the GW are given in \eqref{hplus} and \eqref{htimes}. Focusing on the cross polarisation,
\be
h_\times (t_s,\iota) = \frac{4}{R} \left(\frac{G{\cal M}}{c^2}\right)^{5/3} \left(\frac{\pi f_s(t_s^{\rm ret})}{c}\right)^{2/3} \cos \iota \sin(2\Phi_s(t^{\rm ret}_s))
\label{eq:step1}
\ee
where $t^{\rm ret}_s = t_s - t_c$ is the time to coalescence at $t_c$ and ${\cal M}$ is the  chirp mass \eqref{eq:chirpmass}. The time dependence of the frequency is given in Eq.~\eqref{eq:glo} namely
\be
\frac{df_s}{dt_s} = \frac{96}{5}\pi^{8/3}\left(\frac{G{\cal M}}{c^3}\right)^{5/3} f_s^{11/3}
\label{eq:fsourceevoln}
\ee
leading to 
\be
f_s(t^{\rm ret}_s) = \frac{1}{\pi}\left( \frac{5}{256 t^{\rm ret}_s} \right)^{3/8} \left(\frac{G{\cal M}}{c^3}\right)^{-5/8},
\ee
 so that the phase dependence is
\be
\Phi_s(t_s) = \Phi_c + 2\pi \int_{t_c}^{t_s} dt'_s f_s(t'_s) =  - 2 \left( \frac{t^{\rm ret}_s c^3}{5G{\cal M}}\right)^{5/8} + \Phi_c.
\ee

We now parallel transport this solution \eqref{eq:step1} along a null geodesic to the observer.  Along the geodesic the GW phase remains constant because the time dilation effects cancel the redshifting of the frequency. Thus at the observer whose clock measures $dt = dt_s (1+z)$, the observed GW frequency $f = f_s/(1+z)$ leading to $\Phi(t) =\Phi_s(t_s)$.  However, at the observer, the GW amplitude is changed.  From Eq.~\eqref{eq:amplitude}, and using \eqref{eq:step1}
\be
h_\times (t,\iota) = \frac{4}{a(t)R} \left(\frac{G{\cal M}}{c^2}\right)^{5/3} \left[\frac{\pi}{c} f(t^{\rm ret})(1+z)\right]^{2/3} \cos \iota \sin(2\Phi(t^{\rm ret}))
\label{eq:nearly}
\ee
where we have included the redshifting of frequency. Let us now define the {\it redshifted chirp mass}
\be
\boxed{{\cal M}_z = (1+z){\cal M}}
\label{eq:redshiftedM}
\ee
Then \eqref{eq:nearly} becomes
\ba
h_\times (t,\iota) &=& \frac{4}{a(t)R(1+z)} \left(\frac{G{\cal M}_z}{c^2}\right)^{5/3} \left(\frac{\pi f(t^{\rm ret})}{c}\right)^{2/3} \cos \iota \sin(2\Phi(t^{\rm ret}))
\label{eq:nearlybis}
\\
&=& \frac{4}{d_L(z)}  \left(\frac{G{\cal M}_z}{c^2}\right)^{5/3} \left(\frac{\pi f(t^{\rm ret})}{c}\right)^{2/3} \cos \iota \sin(2\Phi(t^{\rm ret}_o))
\label{eq:nearlyter}
\ea
where in the second line we have used  Eq.~\eqref{eq:dLdef} defining the luminosity distance to the source (today $a(t_o)=1$).
The dependence of the observed frequency on time $t$ is obtained by $f_s=(1+z)f$ into Eq.~\eqref{eq:fsourceevoln}:
\be
(1+z) \frac{d[f(1+z)]}{dt} = \frac{96}{5}\pi^{8/3}\left(\frac{G{\cal M}}{c^3}\right)^{5/3} f^{11/3}(1+z)^{11/3}.
\ee
{\it Assuming that changes in $z$ are negligible during the observation time}, then $z$ can be taken as constant\footnote{See \cite{Bonvin:2016qxr} for a discussion of where this assumption may lead to biases} leading to 
\be
\frac{df}{dt} = \frac{96}{5}\pi^{8/3}\left(\frac{G{\cal M}_z}{c^3}\right)^{5/3} f^{11/3},
\ee
namely the GW phase depends on the {\it redshifted chirp mass}, 
\be
\Phi(t^{\rm ret}) =  - 2 \left( \frac{t^{\rm ret} c^3}{5G{\cal M}_z}\right)^{5/8} + \Phi_c.
\ee

To summarize, the GW frequency depends on the redshifted chirp mass ${\cal M}_z$ which is therefore determined by measurements of the phase of an inspiral signal. The GW amplitude depends on both ${\cal M}_z$ and $d_L(z)$. Given that the former is determined from the phase, measurements of the amplitude of the signal determine $d_L(z)$.  Generally speaking therefore, GW observations from individual CBC events determine the luminosity distance $d_L(z)$ and the so-called `redshifted' masses,  
\be
m^{\rm detected}_{1,2} = (1+z) m_{1,2}
\label{msoursobs}
\ee
which are related to the `source' masses $m_{1,2}$ by the same factor of $1+z$ as in Eq.~\eqref{eq:redshiftedM}.\footnote{To determine each redshifted mass individually, rather than in the combination of the chirp mass, requires the waveform beyond the lowest order quadrupolar form discussed here}

It is important to note that while the redshifted masses and the luminosity
distance can  be deduced from the
waveform, the redshift $z$ of the individual CBC remains {\it undetermined}. To deduce this redshift, one
possibility is to assume a cosmological model --- such as $\Lambda$CDM, with given values of $H_0$, $\Omega_m$ etc, say from the Planck observations --- so that $z$ can be
read off the luminosity distance, see Eq.~\eqref{eq:dLdef}.  Another possibility is to find, and measure, effects in the waveform that
also depend on the {\it source-frame} masses as well as the redshifted masses. For binary black
holes there are no such effects (even to higher orders in the PN
expansion). {For binaries including neutron stars, tidal effects with these properties enter, however at higher PN order (see e.g.~\cite{DelPozzo:2015bna,Messenger:2011gi} and references within). Furthermore, the features depend on the  --- uncertain --- equation of state of the nuclear matter making up the neutron star.} In conclusion, to lowest order in the PN expansion for CBCs, {\it it is not possible to determine the redshift $z$ of the source from GW observations alone}: there is a perfect degeneracy between source masses, redshift, as well as spins.

Note that {\it if} $z$ is determined, then from the detected masses $m^{\rm detected}_{1,2}$ one can obtain the value of the source masses $m_{1,2}$ via Eq.~\eqref{msoursobs}.   This was done in \cite{LIGOScientific:2016aoc}, for example, assuming $\Lambda$CDM cosmological model with Planck values of $H_0$, $\Omega_m$ to find $z$: this is how  the source-frame values of the two black hole masses was determined.  However, the values of the cosmological parameters $H_0$ etc are in fact {\it not}  known precisely, and a source of tension in cosmology today, see e.g.~\cite{DiValentino:2021izs,Verde:2023lmm}. For these reasons, it can be interesting to use GW observations in a different way, namely as a new observable with which to measure cosmological parameters.

\subsection{Measuring cosmological parameters with GWs: outline of GW cosmology}

In this last brief subsection we outline how GW observations can be used to measure cosmological parameters. 

The luminosity distance $d_L(z)$ in $\Lambda$CDM is given in Eq.~\eqref{eq:dLdef}, and is a function of cosmological parameters such as $H_0$ and $\Omega_m$. At low redshifts $z \ll 1$, the domain of the O3 measurements of the LVK collaboration \cite{LIGOScientific:2021aug}, only $H_0$ enters since Eq.~\eqref{eq:dLdef} reduces to
\be
d_L \sim \frac{cz}{H_0} \qquad \Rightarrow \qquad H_0 \sim \frac{cz}{d_L}.
\label{eq:bbGWc}
\ee
Clearly, in order to {\it measure} $H_0$, not only is $d_L$ required (and obtained from GW observations, as we have discussed), but the redshift $z$ of the source is also needed.  However, as mentioned above, this cannot be determined from GW observations: extra non-gravitational information is necessary to determine $z$.  Such information could, for example, be electro-magnetic (EM).

Indeed, the most straightforward way to determine $z$ is use EM observations to uniquely identify the ``host galaxy'' of the GW signal, namely the galaxy in which GW event occurred. This was possible for GW event GW170817 which occurred on August 17th 2017 and which corresponded to the the merger of two neutron stars \cite{GW170817}.  This GW signal was observed the two LIGO and Virgo GWdetectors, and 1.7s following the GW merger, EM observers around the globe observed a subsequent gamma-ray burst as well as multiple EM signals in different frequency bands. Using this EM data it was possible to determine the host galaxy, namely NGC 4993, a galaxy in the Hydra constellation. This constellation is receding from us with a velocity $cz=3327 \pm 72$ km/s, due to the expansion of the universe. Combining this with the distance $d_L = 43.8^{+2.9}_{-6.9}$ Mpc inferred from the GW signal led, using Eq.~\eqref{eq:bbGWc}, to an estimated value for the Hubble constant of $H_0=70^{+12}_{-8}$ km/s/Mpc \cite{LIGOScientific:2017adf}. 
This result, using one GW event only, is consistent with other measurements but is of course less accurate because of its larger error bars. Its interest is that it shows that the idea works. The errors would be reduced (with a $\sim 1/\sqrt{N}$ scaling) if $N$ other measurements of this kind existed, but unfortunately GW170817 was an extremely rare event as since then no further GW events with associated EM counterparts (known as {\it standard sirens}) have been detected.

However, LVK has detected GWs from hundreds of BBHs and a few NS-BH, for each of which there is a measured $d_L$ and $m_{1,2}^{\rm detected}$ --- but no EM counterpart.  Even for these {\it dark sirens}, it is possible to obtain redshift information, and therefore measure $H_0$. Today, two pieces of information are used together to get a statistical redshift for GW events: (i) galaxy catalogues and (ii) astrophysical modelling of the formation channels of BBHs. We refer the reader to \cite{Mastrogiovanni:2023emh} for a review of these methods and results.


\section{Acknowledgements}

We are grateful to Abhay Ashtekar, Tom Bertheas, C\'edric Deffayet, Nathalie Deruelle, Eanna Flanagan, Alejandro Perez, Pierre Piovesan, Eric Poisson, Syksy R\"as\"anen, Carlo Rovelli, Robert Wald for discussions on the topics of this review, as well as multiple Master 2 and doctoral students whom we have had the pleasure to teach. We also thank ChatGPT for technical support (not for the calculations though). This work was supported in part by Perimeter Institute for Theoretical Physics. Research at Perimeter Institute is supported by the Government of Canada through the Department of Innovation, Science and Economic Development and by the Province of Ontario through the Ministry of Research, Innovation and Science.  This work was also supported in part by CERN.

\appendix
\setcounter{tocdepth}{1}

\section{Second order action for perturbations around any background solution}
\label{app:perturbed-action}

This appendix gives the details of the steps required to calculate the Einstein-Hilbert action of GR to second order in perturbations about a general background metric $\bar{g} _{\mu \nu}$:
\be
g_{\mu \nu} = \bar{g}_{\mu \nu}  + h_{\mu \nu}.
\label{eq:step1p}
\ee
The starting point is
\ba
S_{\rm EH} &\equiv & \int d^4 x \sqrt{-g} R 
= S^{(0)} +  S^{(1)} + S^{(2)} + \ldots
\label{eq:pertS}
\ea
where  $S^{(i)}$ is of ${\cal{O}}(h^i)$.  The extremisation of the second order action $S^{(2)}$ with respect to $h_{\mu \nu}$ will give the linearized equations of motion for $h_{\mu \nu}$, namely those we want to calculate.  The background metric $\bar{g}_{\mu \nu}$ satisfies the background Einstein equations which follow from $S^{(1)}$. 

To find $S^{(2)}$ there are two main steps: calculating the Ricci scalar $R$ to second order, and then the determinant of the metric and hence $\sqrt{-g}$ to second order.

\vspace{0.1cm}
\begin{center}
{\bf Perturbed Riemann tensor, Ricci tensor and scalar, Einstein tensor}
\end{center}

\noindent  $\bullet$ The inverse metric, or {\it contravariant metric tensor} corresponding to Eq.~\eqref{eq:step1p}  is given at second order by
\ba
g^{\mu \nu} &=& \bar{g}^{\mu \nu} - h^{\mu \nu}  +  h^{\mu \rho}h^\nu_\rho
\label{eq:step2p}
\ea
where indices of $h_{\mu \nu}$ are raised and lowered with $ \bar{g}_{\mu \nu}$.

\noindent $\bullet$  The perturbed {\it Christoffel symbols} are given by
\ba
\Gamma^\rho_{\mu \nu} &\equiv& 
\frac{1}{2} {g}^{\rho \lambda} \left({\partial}_\mu g_{\nu \lambda} + {\partial}_\nu g_{\mu \lambda} - {\partial}_\lambda g_{\mu \nu} \right)
=
\bar{\Gamma}^\rho_{\mu \nu} +   \Gamma^{\rho (1)}_{\mu \nu} + \Gamma^{\rho (2)}_{\mu \nu} .
\label{chris}
\ea
In the following we denote the covariant derivative with respect to $\bar{g}$ by $\bar{\nabla}$ (with of course $\bar{\nabla}_\mu \bar{g}_{\nu \alpha} = 0$). Substitution of 
Eqs.~\eqref{eq:step1p} and \eqref{eq:step2p} gives
\ba
\bar{\Gamma}^\rho_{\mu \nu} &=& \frac{1}{2} \bar{g}^{\rho \lambda} \left({\partial}_\mu \bar{g}_{\nu \lambda} + {\partial}_\nu \bar{g}_{\mu \lambda} - {\partial}_\lambda   \bar{g}_{\mu \nu} \right)
\nn
\\
\Gamma^{\rho (1)}_{\mu \nu} &=& \frac{1}{2} \bar{g}^{\rho \lambda} \left(\bar{\nabla}_\mu h_{\nu \lambda} + \bar{\nabla}_\nu h_{\mu \lambda} - \bar{\nabla}_\lambda h_{\mu \nu} \right)
\label{g1}
\\
\Gamma^{\rho (2)}_{\mu \nu}  &=&  - h_\beta^\rho  \Gamma^{\beta (1)}_{\mu \nu} .
\label{g2}
\ea

\noindent $\bullet$  Next we calculate the {\it Riemann tensor}.  Let 
\be
\Gamma^\rho_{\mu \nu} = \bar{\Gamma}^\rho_{\mu \nu} + \delta \Gamma^\rho_{\mu \nu}
\label{dG}
\ee
where, from (\ref{chris}), 
\be
\delta \Gamma^\rho_{\mu \nu} =  \Gamma^{\rho (1)}_{\mu \nu} + \Gamma^{\rho (2)}_{\mu \nu} .
\label{delta}
\ee
The definition of the Riemann tensor together with (\ref{dG}) gives
\ba
{R^{\mu}}_{ \nu \rho \sigma}&\equiv& {\partial}_\rho \Gamma^\mu_{\sigma \nu} + \Gamma^\mu_{\rho \lambda}\Gamma^\lambda_{\sigma \nu} - (\rho \leftrightarrow \sigma)
\label{Riemann1}
\\
&=& \bar{R^{\mu}}_{\nu \rho \sigma} + \bar{\nabla}_\rho (\delta \Gamma^\mu_{\sigma \nu}) - 
\bar{\nabla}_{\sigma} (\delta \Gamma^\mu_{\rho \nu}) + 
(\delta \Gamma^\mu_{\rho \lambda})(\delta \Gamma^\lambda_{\sigma \nu}) -
(\delta \Gamma^\mu_{\sigma \lambda})(\delta \Gamma^\lambda_{\rho \nu}) 
\label{rr}
\ea
On writing
\ba
{R^{\mu}}_{ \nu \rho \sigma}
&\equiv&{\bar{R}^{\mu}}_{ \; \; \nu \rho \sigma}+  {R^{ (1) \mu}}_{\nu \rho \sigma} + {R^{(2)\mu }}_{\nu \rho \sigma}
\nn
\ea
and using (\ref{delta}) one can read off the different orders of the Riemann tensor.  To first order
\ba
 {R^{(1)\mu }}_{\nu \rho \sigma} &=& \bar{\nabla}_\rho \Gamma^{\mu (1)}_{\sigma \nu} -  \bar{\nabla}_\sigma \Gamma^{\mu (1)}_{\rho \nu}
 \label{r1}
\\
&=& \frac{1}{2} \left[  
\left( \bar{\nabla}_\rho  \bar{\nabla}_\sigma   - \bar{\nabla}_\sigma  \bar{\nabla}_\rho   \right)
h_\nu^\mu +  \left( \bar{\nabla}_\rho  \bar{\nabla}_\nu h_\sigma^\mu -\bar{\nabla}_\sigma  \bar{\nabla}_\nu h_\rho^\mu \right)- \left( \bar{\nabla}_\rho  \bar{\nabla}^\mu h_{\sigma \nu}
  {-} \bar{\nabla}_\sigma  \bar{\nabla}^\mu h_{\rho \nu}\right) \right],
  \label{rr1}
\ea
where we have used Eq.~\eqref{g1}.
The second order term follows from (\ref{rr}) and (\ref{delta}) and reads
\begin{align}
& {R^{(2) \mu }}_{\nu \rho \sigma} = 
\left(  \bar{\nabla}_\rho  \Gamma^{\mu (2)}_{\sigma \nu} - 
\bar{\nabla}_{\sigma}  \Gamma^{\mu (2)}_{\rho \nu} \right)
+ \left( \Gamma^{\mu (1)}_{\rho \lambda}  \Gamma^{\lambda (1)}_{\sigma \nu} -
\Gamma^{\mu (1)}_{\sigma \lambda} \Gamma^{\lambda (1)}_{\rho \nu} \right).
\label{rr2}
\end{align}

\noindent $\bullet$  The {\it Ricci tensor} is then obtained by contraction:
\be
R_{\nu \sigma} \equiv {R^\mu}_{\nu \mu \sigma}  = \bar{R}_{\nu \sigma} +  R^{(1)}_{\nu \sigma}  + R^{(2)}_{\nu \sigma} .
\label{eq:Ricc}
\ee
From (\ref{rr1}) it follows that
\be
 R^{(1)}_{\nu \sigma} = \frac{1}{2}\left[ \bar{\nabla}_\mu  \bar{\nabla}_\sigma h_\nu^\mu + 
 \bar{\nabla}_\mu  \bar{\nabla}_\nu h_\sigma^\mu - \bar{\Box} h_{\nu \sigma} - 
\bar{\nabla}_\sigma  \bar{\nabla}_\nu h \right],
\label{ran}
\ee
whereas from (\ref{rr2}) 
\ba
R^{ (2) }_{\nu \sigma}
 &=&
\bar{\nabla}_\rho  \Gamma^{\rho (2)}_{\sigma \nu} - 
\bar{\nabla}_{\sigma}  \Gamma^{\rho (2)}_{\rho \nu} + 
\Gamma^{\rho (1)}_{\rho \lambda}  \Gamma^{\lambda (1)}_{\sigma \nu} -
\Gamma^{\rho (1)}_{\sigma \lambda} \Gamma^{\lambda (1)}_{\rho \nu}
\label{thus}
\ea
Its explicit form is not required below, but for completeness we give it here:
\ba
\bar{R}^{ (2) }_{\nu \sigma}
 &=& \frac{1}{4} \bar{\nabla}_\nu h_{\ab} \bar{\nabla}_\sigma h^{\ab} + \bar{\nabla}^{\beta} h_{\sigma}^\beta 
 \bar{\nabla}_{[\beta} h_{\alpha]\nu} 
+ \frac{1}{2} h^{\ab}\left( \bar{\nabla}_\nu \bar{\nabla}_\sigma h_\ab + \bar{\nabla}_\beta \bar{\nabla}_\alpha h_{\nu \sigma} - \bar{\nabla}_\beta \nabla_\sigma h_{\alpha \nu} - \bar{\nabla}_\beta \bar{\nabla}_\nu h_{\alpha \sigma}\right)
\nn
\\
 & -& 
 \frac{1}{2}(\bar{\nabla}_{\beta} h^{\ab} - \frac{1}{2}\bar{\nabla}^\alpha h)(\bar{\nabla}_\sigma h_{\nu\alpha} +  \bar{\nabla}_\nu h_{\alpha \sigma} - \bar{\nabla}_\alpha h_{\nu \sigma} ).
 \nn
 \ea 

\noindent $\bullet$ The {\it Ricci scalar} is obtained from  Eqs.~\eqref{eq:step2p} and \eqref{eq:Ricc} and is given by
\ba
R &\equiv &g^{\nu \sigma}R_{\nu \sigma} =( \bar{g}^{\nu \sigma} - h^{\nu \sigma}  +  h^{\nu \rho}h^\sigma_\rho)(\bar{R}_{\nu \sigma} +  R^{(1)}_{\nu \sigma}  + R^{(2)}_{\nu \sigma} )
= \bar{R} +  R^{(1)} +  R^{(2)},
\label{eq:expandedR}
\ea
where
\ba
R^{(1)} &=& \bar{g}^{\nu \sigma} R^{(1)}_{\nu \sigma} - \bar{R}^{\nu \sigma}  h_{\nu \sigma} 
= \bar{\nabla}_\mu  \bar{\nabla}_\nu h^{\mu \nu} - \Box h -  \bar{R}^{\nu \sigma}  h_{\nu \sigma} 
\label{ricci1}
\ea
(in the last line we have used (\ref{ran})), and
\ba
 R^{(2)} &=&  h^{\nu \alpha}h^{\sigma}_{\alpha}  \bar{R}_{\nu \sigma}  - h^{\nu \sigma} R^{(1)}_{\nu \sigma}  + \bar{g}^{\nu \sigma} R^{(2)}_{\nu \sigma}.
 \label{ricci2}
 \ea

\noindent $\bullet$ 
For completeness we also give the first order perturbed {\it Einstein tensor}
 \begin{align}
G^{\sscr (1)}_{\m\n} & = R^{\sscr (1)}_{\m\n} -\f12\bar R h_{\m\n} - \f12 \bar g_{\m\n} R^{\sscr (1)}\\ 
& = -\f12{\bar\square} h_{\m\n} +{\bar\na}_{(\m} {\bar\na}_\r h^\r{}_{\n)} -\f12 {\bar\na}_\m {\bar\na}_\n h \nn
-\f12\bar g_{\m\n}({\bar\na}_\m {\bar\na}_\n h^{\m\n} - {\bar\square} h) 
+ \bar G_{\r(\m} h_{\n)}{}^\r - (\bar R_{\m\r\n\s} -\f12\bar g_{\m\n}\bar R_{\r\s})h^{\r\s}.
\label{G1gen}
\end{align}

\vspace{0.1cm}
\begin{center}
{\bf Perturbed metric determinant}
\end{center}

To expand $\sqrt{-g}$ to second order, we write Eq.~\eqref{eq:step1p}
$g_{\alpha \beta} = \bar{g}_{\alpha \mu} (\delta^{\mu}_\beta  +  M^{\mu}_\beta)$ where $M^{\mu}_\beta = \bar{g}^{\mu \lambda} h_{\lambda \beta}$. 
Thus
\be
\det(g) = \det(\bar{g}) \det(\mathbf{1} +  \mathbf{M})
\label{eq:gddd}
\ee
where the matrix $\mathbf{M}$ has components $M^{\mu}_\beta$. To quadratic order
\be
\det(\mathbf{1} + \mathbf{M}) = 1 +  {\rm tr}{\mathbf{M}} + \frac{1}{2} \left( 2  ({\rm tr}{\mathbf{M}} )^2 -  {\rm tr}({\mathbf{M}}^2) \right)  + \ldots
\nn
\ee
Replacing ${\rm tr}{\mathbf{M}} = h$ and ${\rm tr}({\mathbf{M}}^2)= h_{\mu \nu}  h^{\mu \nu}$ in Eq.~\eqref{eq:gddd}, and then taking the square root, gives
\be
\sqrt{-\det (g_{\mu\nu})} =  \sqrt{-\det (\bar{g}_{\mu\nu})} \left[ 1 + \frac{1}{2} h  + \frac{1}{8}(h^2 - 2 h_{\mu \nu}^2) \right] .
\label{det}
\ee

\vspace{0.1cm}
\begin{center}
{\bf EH action to second order}
\end{center}

Substituting (\ref{eq:expandedR}) and (\ref{det}) into the perturbed Einstein Hilbert action Eq.~\eqref{eq:pertS} gives
\ba
S_{(0)} & =&  \int d^4 x \sqrt{-\bar{g}} \bar{R} 
\nn
\\
S_{(1)} &=&   \int d^4 x \sqrt{-\bar{g}}\left( R^{(1)}+ \frac{h}{2} \bar{R}\right) 
\nn
\\
S_{(2)} &=&
\int d^4 x \sqrt{-\bar{g}}\left( {R}^{(2)} + \frac{h}{2}R^{(1)} + \frac{\bar{R}}{8}(h^2 - 2 h_{\mu \nu}h^{\mu \nu}) \right).
\nn
\ea

\noindent $\bullet$ The {\it first order action} is
\ba
S_{(1)} &=&   \int d^4 x \sqrt{-\bar{g}}\left(   \bar{\nabla}_\mu  \bar{\nabla}_\nu h^{\mu \nu} - \Box h -  \bar{R}^{\nu \sigma}  h_{\nu \sigma} + \frac{h}{2} \bar{R}\right) 
\nn
\ea
The first two terms are total derivatives. After integration by parts and dropping the boundary terms
\ba
S_{(1)}&=&  \int d^4 x \sqrt{-\bar{g}}\left(  -  \bar{G}^{\nu \sigma}  h_{\nu \sigma} \right).
\ea
On including matter through the stress tensor, the variation of this gives the background Einstein equation.

\noindent $\bullet$ The {\it second order} part becomes, on substituting \eqref{ricci2}, 
\ba
S_{(2)} 
&=&  \int d^4 x \sqrt{-\bar{g}}\left( \bar{g}^{\nu \sigma} R^{(2)}_{\nu \sigma} + \frac{h}{2}R^{(1)}  - h^{\nu \sigma} R^{(1)}_{\nu \sigma}
 +  h^{\nu \alpha}h^{\sigma}_{\alpha}  \bar{R}_{\nu \sigma}  + \frac{\bar{R}}{8}(h^2 - 2 h_{\mu \nu}h^{\mu \nu}) \right)
\label{eq:S2}
\ea
where $R^{(1)}$ and $R^{(1)}_{\nu \sigma}$ are given in \eqref{ricci1} and \eqref{ran}  respectively.

The first term $\int d^4 x \sqrt{-\bar{g}} \bar{g}^{\nu \sigma} R^{(2)}_{\nu \sigma}$ splits into four parts on using (\ref{thus}). The first two parts are total derivatives and do not contribute.
The last two parts give
\ba
\int d^4 x \sqrt{-\bar{g}}\left( \bar{g}^{\nu \sigma} R^{(2)}_{\nu \sigma}\right) &=&  \int d^4 x \sqrt{-\bar{g}} \bar{g}^{\nu \sigma}\left( \Gamma^{\mu (1)}_{\rho \lambda}  \Gamma^{\lambda (1)}_{\sigma \nu} -
\Gamma^{\mu (1)}_{\sigma \lambda} \Gamma^{\lambda (1)}_{\rho \nu}
\right)
\nn
\\
&=& \int d^4 x \sqrt{-\bar{g}}\frac{1}{2} \left[  (\bar{\nabla}^\lambda h) \left(\bar{\nabla}^\nu h_{\lambda \nu} - \frac{1}{2}\bar{\nabla}_\lambda h\right)
- (\bar{\nabla}^\nu h^{\mu \lambda}) \left( \bar{\nabla}_\mu h_{\lambda \nu} -\frac{1}{2} \bar{\nabla}_\nu h_{\mu \lambda} \right) \right].
\nn
\ea
Collecting the expressions together into \eqref{eq:S2} gives
\ba
S_{(2)} &=& \frac{1}{2}  \int d^4 x \sqrt{-\bar{g}}\left( (\bar{\nabla}^\lambda h) \left(\bar{\nabla}^\nu h_{\lambda \nu} - \frac{1}{2}\bar{\nabla}_\lambda h\right) 
 - (\bar{\nabla}^\nu h^{\mu \lambda}) \left( \bar{\nabla}_\mu h_{\lambda \nu} -\frac{1}{2} \bar{\nabla}_\nu h_{\mu \lambda} \right) 
+ h \bar{\nabla}_\mu \bar{\nabla}_\nu h^{\mu \nu} - h \Box h \right.
\nn
\\ &&\left. + h^{\mu \nu} \Box h_{\mu \nu} - {2} h^{\nu \sigma} \bar{\nabla}^\mu \bar{\nabla}_\sigma h_{\mu \nu}  + (\bar{\nabla}_\sigma \bar{\nabla}_\nu h) h^{\sigma \nu}
+ 2 h^{\nu \alpha}h^{\sigma}_{\alpha}  \bar{R}_{\nu \sigma} - \bar{R}^{\mu \nu}h h_{\mu \nu}  + \frac{\bar{R}}{4}(h^2 - 2 h_{\mu \nu}h^{\mu \nu})
\right).
\nn
\ea
Finally, after integration by parts,
\ba
S_{(2)} &=&\frac{1}{2}  \int d^4 x \sqrt{-\bar{g}}\left(- (\bar{\nabla}^\lambda h) (\bar{\nabla}^\nu h_{\lambda \nu}) 
+ \frac{1}{2}(\bar{\nabla}_\lambda h) (\bar{\nabla}^\lambda h) \right.
-  \frac{1}{2} (\bar{\nabla}^\nu h^{\lambda \mu})(\bar{\nabla}_\nu h_{\lambda \mu}) + (\bar{\nabla}^\nu h^{\lambda \mu})(\bar{\nabla}_\mu h_{\lambda \nu}) 
\nn
\\
&& \qquad \qquad \qquad\left. + 2 h^{\nu \alpha}h^{\sigma}_{\alpha}  \bar{R}_{\nu \sigma} - \bar{R}^{\mu \nu}h h_{\mu \nu} + \frac{\bar{R}}{4}(h^2 - 2 h_{\mu \nu}h^{\mu \nu})
\right).
\nn
\ea
In flat space $\bar{g}_{\mu \nu}=\eta_{\mu \nu}$, the terms in square brackets all vanish and this reduces to the usual perturbed equations around Minkowski space.

In terms of the {\it trace reversed perturbation} 
\be
h_{\alpha \beta} = \bar{h}_{\alpha \beta}  - \frac{1}{2} \bar{g}_{\alpha \beta} \bar{h}
\ee
 this becomes
\ba
S_{(2)} &=&\frac{1}{2}  \int d^4 x \sqrt{-\bar{g}}\left(- \frac{1}{2} (\bar{\nabla}^\nu \bar{h}^{\lambda \mu})(\bar{\nabla}_\nu \bar{h}_{\lambda \mu})  +
\frac{1}{4} (\bar{\nabla}^\lambda \bar{h}) (\bar{\nabla}^\lambda \bar{h}) \right.
+ (\bar{\nabla}^\nu \bar{h}^{\lambda \mu})(\bar{\nabla}_\mu \bar{h}_{\lambda \nu}) 
\nn
\\
&& \qquad \qquad \qquad \left. + 2 \bar{h}^{\nu \alpha}\bar{h}^{\sigma}_{\alpha}  \bar{R}_{\nu \sigma} - \bar{R}^{\mu \nu}\bar{h} \bar{h}_{\mu \nu} - \frac{\bar{R}}{2}\bar{h}_{\mu \nu}\bar{h}^{\mu \nu} {+\frac{1}{4}\bar{R} \bar{h}^2}
\right).
\nn
\ea

\vspace{0.2cm}
\begin{center}
{\bf Linearised equations of motion}
\end{center}

\noindent $\bullet$ {\it Equations of motion}. Variation of action \eqref{eq:S2} with respect to $h^{\ab}$ gives
\ba
0 = \int d^4 x \sqrt{-\bar{g}}\delta h^{\ab}\left( -G_{\ab}^{(1)} - \frac{1}{2} h \bar{G}_{\ab} + h^{\mn}(\bar{g}_{\nu \beta}\bar{G}_{\alpha \nu} + \bar{g}_{\nu \alpha}\bar{G}_{\beta \nu})\right)
\nn
\ea
leading to the equations of motion
\be
 G_{\ab}^{(1)} = - \frac{1}{2} h \bar{G}_{\ab} + h^{\mn}(\bar{g}_{\nu \beta}\bar{G}_{\alpha \nu} + \bar{g}_{\nu \alpha}\bar{G}_{\beta \nu}).
 \ee

In terms of the trace reversed perturbation, after commuting covariant derivatives, for example,
\ba
\bar{\nabla}_\mu \bar{\nabla}_{\beta} \bar{h}^{\mu}_{\alpha}  &=&  \bar{\nabla}_{\beta}( \bar{\nabla}_\mu\bar{h}^{\mu}_{\alpha}  ) + \bar{R}_{\lambda \beta} h^{\lambda}_{\alpha} - \bar{R}^{\lambda}_{\; \; \alpha \mu \beta} h^{\mu}_\lambda
\ea
and then imposing the Lorenz gauge $\bar{\nabla}_\mu \bar{h}^{\mn}=0$ these read 
\be
{\bar\square} \bar{h}_{\m\n}  +2 \bar R_{\m\r\n\s} \bar{h}^{\r\s}-2 \bar G_{\r(\m} \bar{h}_{\n)}{}^\r 
-\bar g_{\m\n} (\bar R_{\r\s} \bar{h}^{\r\s}) =0
\ee
which can be rewritten as
\be
{ \bar{\Box} \bar{h}_{\alpha \beta}  + 2 \bar{R}^{\mu}_{\; \; \alpha \nu \beta} \bar{h}_{\mu}^{\nu} + \bar S_{\mu \alpha \nu\beta} \bar{h}^{\mu\nu} = 0,}
\label{eq:bloodyfinal}
\ee
where
\be
{\bar S_{\mu \alpha \nu\beta}   = 2\bar{G}_{\mu (\alpha}\bar g_{\beta) \nu}  - \bar{R}_{\mu \nu} \bar g_{\alpha \beta}.}
\label{eq:bloodyfinalS}
\ee

Including matter, and splitting its energy momentum tensor $T_{\m\n}=\bar T_{\m\n}+ T^{\sscr (1)}_{\m\n}$ into a background (determining $\bar g_{\m\n}$) plus first-order perturbation,
the first-order field equations are
\be
 \bar{\Box} \bar{h}_{\alpha \beta}  + 2 \bar{R}^{\mu}_{\; \; \alpha \nu \beta} \bar{h}_{\mu}^{\nu} + S_{\mu \alpha \nu\beta} \bar{h}^{\mu\nu} = 
-\f{16\pi G}{c^4} T^{\sscr (1)}_{\ab}.
\label{eq:bloodyfinalSO}
\ee

\section{Green's functions}
\label{app:Green}

Given a linear differential operator $\D$ acting on functions on $\mathbb{R}^n$, its corresponding Green's function is a function on $\mathbb{R}^n\times \mathbb{R}^n$ satisfying 
\be\label{defG}
\D G(x,x') = \d^{(n)}(x,x').
\ee
Green's functions are useful to solve differential equations in the presence of sources, since they allow one to write the solutions of 
$\D\Phi = J$
as
\be
\Phi(x) = \int d^nx' G(x,x') J(x') +\Phi^\circ(x),
\ee
where $\Phi^\circ$ is a solution of the homogeneous problem with $J=0$. For a given $\D$ the Green function is typically not unique, but a unique one can be selected via boundary conditions or other physical requirements.

For the Laplace equation, $n=3$ and $\D=\vec\p^2$, there is a unique solution of \eqref{defG} with vanishing boundary conditions at infinity, given by
\be\label{GLap}
G(\vec x,\vec x') = -\f{1}{4\pi |\vec x-\vec x'|}. 
\ee
Therefore
\be
\Phi(\vec x)=\f1{4\pi}\int d^3x'\f{J(\vec x')}{|\vec x-\vec x'|} +\Phi^\circ(\vec x).
\ee

For the d'Alembert equation, $n=4$ and $\D=\square$, fixing vanishing boundary conditions at spatial infinity is not enough to have a single solution: there are infinitely many solutions of the homogeneous equation that can be added to any given $G$ and still satisfy the defining equation.
Two notable examples are the 
retarded and advanced ones, which are uniquely characterised by vanishing for $x$  respectively in the past or the future of $x'$, and given by
\ba
G_{\pm}(x,x') &=& -\f{\d(t-t' \mp|\vec x-\vec x'|)}{4\pi |\vec x-\vec x'|} 
\nn
\\
&=& -\f1{2\pi}\Th\big(\pm(t-t')\big)\d\big((t-t')^2-|\vec x-\vec x'|^2\big),
\label{Gbox}
\ea
where $\Th$ is Heaviside's step function.
The retarded solution imposes no-incoming radiation boundary conditions, and it is the one relevant to study the emission of waves from a source.

\section{Landau-Lifshitz approach}
\label{appLL}

The Landau-Lifshitz formulation of Einstein's equations is a convenient approach to  perturbative theory around Minkowski, and it is widely used by the community working on the PN and PM expansions. We provide here a brief description of the approach from the perspective of the main text, and refer the reader to \cite{Blanchet:2006jqj,poisson} for more details. In the Landau-Lifshitz approach one uses a density-weighted inverse metric as fundamental variable, 
\be
{\mathfrak g}^{\m\n} := \sqrt{-g}g^{\m\n} = \bar g^{\m\n} -\bar h^{\m\n} +O(h^2).
\ee
The interest in doing so is that the quantity
\be
H^{\a\m\b\n}:={\mathfrak g}^{\a\b}{\mathfrak g}^{\m\n}-{\mathfrak g}^{\a\n}{\mathfrak g}^{\b\m}
\ee
is related to the Einstein tensor via ordinary derivatives. More precisely, 
\be\label{ppHid}
\p_\m\p_\n H^{\a\m\b\n} = 2(-g)(G^{\a\b}+\f{8\pi G}{c^4} t_{\sscr LL}^{\a\b})\eqons \f{16\pi G}{c^4}(-g)(T^{\a\b}+ t_{\sscr LL}^{\a\b}),
\ee
where $ t_{\sscr LL}^{\a\b}$ is a pseudo-tensor of density-weight two, given explicitly by some lengthy expression in terms of first derivatives of the metric.
Crucially, it is conserved on-shell $\p_\a(-gt_{\sscr LL}^{\a\b})\eqons 0$. The dimension-full numerical factor in front of it is included in its definition for convenience when going on-shell in the last equality above.

This arrangement of Einstein's equations manifestly breaks covariance.\footnote{It is like rearranging the covariant geodesic equation \eqref{geo} as in \eqref{geo2},  where neither side of the equation is covariant by itself.} Not only we have partial derivatives as opposed to covariant derivatives, but also tensor densities, aka pseudo-tensors, appearing. Its usefulness is limited to situations in which there are regions of spacetimes that are approximately flat, and where one can choose a Cartesian coordinate system, so that partial derivatives can be interpreted. This is precisely  the case when studying a perturbative approximation around Minkowski.
Within that context, the reformulation has two useful advantages.

The first is that it provides a prescription for a gravitational energy-momentum pseudo-tensor valid to all orders, given by $ t_{\sscr LL}^{\a\b}$. By analogy with the matter counterpart, one has a prescription to further split this quantity into contributions to energy, momentum and angular momentum. All these expression are gauge-dependent; in particular, we have the usual problem that the pseudo-tensor can be made to vanish at any given point, using a local inertial frame. But the logic in this approach is that we assume to have a preferred coordinate system, the Cartesian ones of the fiducial flat metric, and that is the gauge we stick to. Furthermore, since the left-hand side is a total derivative, the total energy, momentum and angular momentum can be expressed as surface integrals. This provides a prescription for these quantities that can be evaluated in a region far from the sources, where one can safely assume that spacetime is approximately flat and use Cartesian coordinates, and computed to all order in perturbation theory.

The second advantage of the formulation is that it allows to set up an iteration scheme for the perturbative resolution of the field equations in a very practical way.
To see that, we change variables to $\hat h^{\m\n}:=\eta^{\m\n}-{\mathfrak g}^{\m\n}$, where $\eta^{\m\n}$ is a fiducial background metric for which the coordinates are Cartesian. Then \eqref{ppHid} are equivalent to
\be\label{RFE}
\square\hat h^{\m\n} \eqons -\f{16\pi G}{c^4} \t^{\m\n}, \qquad \t^{\m\n}:=(-g)(T^{\m\n}+ t_{\sscr LL}^{\m\n}+t_{\sscr H}^{\m\n}+t_{\sscr NH}^{\m\n}), 
\ee
where $\square$ is the flat spacetime d'Alembertian, and $t_{\sscr H}^{\m\n}$ are the remaining terms in the LHS of  \eqref{ppHid}.
Specifically, $t_{\sscr H}^{\m\n}$ is quadratic in $h^{\m\n}$ and satisfies $\p_\m(-gt_{\sscr H}^{\m\n})\equiv 0$, whereas $t_{\sscr NH}^{\m\n}$ does not but contains only terms that vanish in harmonic gauge. It follows that in the harmonic gauge,
\be\label{HGC}
\p_\m\t^{\m\n}\eqons 0,
\ee
and this equation includes the conservation of the matter energy-momentum tensor, namely the matter dynamics.
We stress that \eqref{RFE} are the exact Einstein's equations, no approximation has been done yet. It is the introduction of a fiducial flat background and a fixed choice of Cartesian coordinates on it, that allows us to rewrite the exact equations in the form of a flat spacetime wave equation with a complicated source satisfying the matter dynamics through \eqref{HGC}. In particular, all non-linearities are recasted on the RHS of the equations.
To solve the equations perturbatively, we assume that $h$ is small, and proceed iteratively as explained in \eqref{gexpsecond}. The idea that makes this approach particularly convenient is to first solve the `relaxed field equations' \eqref{RFE} alone, in harmonic gauge, and afterwards impose the gauge consistency condition \eqref{HGC} on the matter dynamics. Notice that in this scheme, the $n$-th iteration of the gravitational potentials $h_{\m\n}$ are sourced by matter fields satisfying the equations of motion that use the $(n-1)$-th iteration of the gravitational field.
See \cite{Blanchet:2006jqj,poisson} for further details.

\bibliographystyle{JHEPs}
\bibliography{booksGW}

\end{document}